\documentclass[journal,12pt,onecolumn,draftclsnofoot,]{IEEEtran}
\usepackage{amsmath}
\usepackage{amssymb}
\usepackage{amsfonts}
\usepackage{amsthm}
\usepackage{graphicx}
\usepackage{mathtools}
\usepackage{balance}
\usepackage{multirow}
\usepackage{bm}
\usepackage{colortbl}
\usepackage{lipsum}      
\usepackage{changepage}
\usepackage{caption}
\usepackage{color}
\usepackage{pgf,tikz}
\usepackage{balance}
\usepackage{mathtools}
\usepackage{bbm}
\usepackage{array}
\usepackage{relsize}
\usepackage{cite}
\usepackage{amsthm}
\usepackage{verbatim}
\usepackage{epstopdf}
\usepackage{array}
\usepackage{url}
\usepackage{stfloats}
\usepackage[hidelinks]{hyperref}
\usepackage{url}
\usepackage[linesnumbered,ruled]{algorithm2e}
\usepackage{algpseudocode}
\newtheorem{theorem}{Theorem}
\newtheorem{corollary}{Corollary}
\newtheorem{proposition}{Proposition}
\newtheorem{lemma}{Lemma}
\newtheorem{remark}{Remark}
\newtheorem{definition}{Definition}
\newtheorem{condition}{Condition}
\newtheorem{assumption}{Assumption}
\newtheorem{fact}{Fact}
\newcommand{\eqdef}{\mathrel{\mathop:}=}
\newcommand*{\QEDclosed}{\hfill\ensuremath{\blacksquare}}

\newcommand*{\QED}{\hfill\ensuremath{\square}}
\DeclarePairedDelimiter\ceil{\lceil}{\rceil}
\DeclarePairedDelimiter\floor{\lfloor}{\rfloor}

\DeclareMathOperator*{\argmin}{arg\,min}

\newcommand{\minus}{\scalebox{0.65}[1.0]{$-$}}

\usepackage{tikz,amsmath}
\usetikzlibrary{arrows, decorations.markings}
\usetikzlibrary{positioning,fit,calc} 
\tikzset{block/.style={draw, text width=1.6cm, minimum height=1cm, align=center,node distance = 0.4cm},   
	input/.style    = {coordinate}, 
	output/.style   = {coordinate}, 
	line/.style={-latex,node distance = 0.25cm}   
}
\tikzset{block2/.style={draw, text width=0.48cm, minimum height=0.8cm, align=center,node distance = 0.4cm},   
	input/.style    = {coordinate}, 
	output/.style   = {coordinate}, 
	line/.style={-latex,node distance = 0.25cm}   
}

\tikzset{block2_large/.style={draw, text width=0.57cm, minimum height=0.8cm, align=center,node distance = 0.4cm},   
	input/.style    = {coordinate}, 
	output/.style   = {coordinate}, 
	line/.style={-latex,node distance = 0.25cm}   
}
\tikzset{block2_med/.style={draw, text width=0.52cm, minimum height=0.8cm, align=center,node distance = 0.4cm},   
	input/.style    = {coordinate}, 
	output/.style   = {coordinate}, 
	line/.style={-latex,node distance = 0.25cm}   
}

\tikzset{block3/.style={draw, text width=1.1cm, minimum height=0.8cm, align=center,node distance = 0.4cm},   
	input/.style    = {coordinate}, 
	output/.style   = {coordinate}, 
	line/.style={-latex,node distance = 0.25cm}   
}
\tikzset{block4/.style={draw, text width=1.4cm, minimum height=0.8cm, align=center,node distance = 0.4cm},   
	input/.style    = {coordinate}, 
	output/.style   = {coordinate}, 
	line/.style={-latex,node distance = 0.2cm}   
}

\tikzset{block4_small/.style={draw, text width=1.1cm, minimum height=0.6cm, align=center,node distance = 0.4cm},   
	input/.style    = {coordinate}, 
	output/.style   = {coordinate}, 
	line/.style={-latex,node distance = 0.25cm}   
}

\tikzset{be/.style={circle,draw, minimum width=0.25*\pgfkeysvalueof{/tikz/x radius} }}

\tikzset{blockA/.style={draw, text width=0.5cm, minimum height=0.95cm, align=center,node distance = 0.5cm},   
	input/.style    = {coordinate}, 
	output/.style   = {coordinate}, 
	line/.style={-latex,node distance = 0.25cm}   
}

\tikzset{blockW/.style={draw, text width=0.70cm, minimum height=0.95cm, align=center,node distance = 0.5cm},   
	input/.style    = {coordinate}, 
	output/.style   = {coordinate}, 
	line/.style={-latex,node distance = 0.25cm}   
}

\tikzset{blockB/.style={draw, text width=0.5cm, minimum height=0.95cm, align=center,node distance = 0.5cm},   
	input/.style    = {coordinate}, 
	output/.style   = {coordinate}, 
	line/.style={-latex,node distance = 0.25cm}   
}

\tikzset{blockChannel/.style={draw, text width=2.5cm, minimum height=3cm, align=center,node distance = 0.5cm},   
	input/.style    = {coordinate}, 
	output/.style   = {coordinate}, 
	line/.style={-latex,node distance = 0.25cm}   
}

\tikzstyle{vecArrow} = [decoration={markings,mark=at position
	1 with {\arrow[thick]{open triangle 60}}},
double distance=2.4pt, shorten >= 5.5pt,
preaction = {decorate},
postaction = {draw,line width=2.0pt, white,shorten >= 4.5pt}]

\tikzstyle{vecArrow2} = [decoration={markings,mark=at position
	1 with {\arrow[thick]{open triangle 60}}},
double distance=1.4pt, shorten >= 5.5pt,
preaction = {decorate},
postaction = {draw,line width=2.0pt, white,shorten >= 7pt}]

\begin{document}

\title{Blind Estimation of a Doubly Selective OFDM Channel: A Deep Learning Algorithm and Theory}

\author{Tilahun~M.~Getu,~\IEEEmembership{Member,~IEEE},~Nada~T.~Golmie,~\IEEEmembership{Fellow,~IEEE},~and~\\ ~David~W. Griffith   
	\thanks{\IEEEcompsocthanksitem T. M. Getu is with the Communications Technology Laboratory, National Institute of Standards and Technology (NIST), Gaithersburg, MD 20899, USA, and also with the Electrical Engineering Department, \'Ecole de Technologie Sup\'erieure (\'ETS), Montr\'eal, QC H3C 1K3, Canada (e-mail: tilahun.getu@nist.gov).}
	\thanks{\IEEEcompsocthanksitem N. T. Golmie is with the Communications Technology Laboratory, National Institute of Standards and Technology (NIST), Gaithersburg, MD 20899, USA (e-mail: nada.golmie@nist.gov). N. T. Golmie is also a Fellow of NIST.
	}
\thanks{\IEEEcompsocthanksitem D. W. Griffith is with the Communications Technology Laboratory, National Institute of Standards and Technology (NIST), Gaithersburg, MD 20899, USA (e-mail: david.griffith@nist.gov).
}

}

\maketitle

\begin{abstract}
Most works on wireless communications, signal processing, and networking employ frequency flat or frequency selective fading channel models. Despite these models' simplicity and tractability, high-speed broadband wireless communication channels manifest not only frequency-selectivity but also time-selectivity. Accounting for time-selectivity and frequency-selectivity, doubly selective fading channel models are realistic models that can capture the various realities of (ultra-)high-speed broadband wireless communications evident in several use-cases of the fifth-generation (5G) and sixth-generation (6G) wireless systems. Toward this end, we provide a new generation solution to the fundamental old problem of a doubly selective fading channel estimation for orthogonal frequency division multiplexing (OFDM) systems. For systems based on OFDM, we propose a deep learning (DL)-based blind doubly selective channel estimator. This estimator does require no pilot symbols -- unlike the corresponding state-of-the-art estimators -- even during the estimation of a deep fading doubly selective channel. This channel filters the transmitted OFDM signal and produces random inputs that would be fed to our proposed DL-based blind OFDM channel estimator. Accordingly, the performance quantification attempt of such a DL-based regression algorithm is often hampered by not only random inputs but also the inherent randomness of each per-layer activation function which would, in turn, complicate a DL model’s aggregated randomness. Overcoming this fundamental challenge in part, we provide an insightful performance quantification for our studied blind OFDM channel estimation using trained over-parameterized rectified linear unit feedforward neural networks (ReLU FNNs) while being inspired by the developments on the theory of deep learning. At the junction of the theory of deep learning, high-dimensional probability, wireless communications, and signal processing, we develop the first of its kind theory on the testing mean squared error (MSE) performance of our investigated blind OFDM channel estimator based on over-parameterized ReLU FNNs. Employing ReLU FNNs, we present extensive computer experiments that assess the performance of our proposed blind estimation algorithm. Quantifying this algorithm's testing MSE performance, the theoretical developments of this paper inspire fundamental research toward a unified theory of deep learning.     
\end{abstract}
\IEEEpeerreviewmaketitle

\section{Introduction}
\label{sec: DSFC_introduction}
\subsection{Motivation}
\label{subsec: DSFC_related_works}
Orthogonal frequency division multiplexing (OFDM) is the most widely adopted transmission technology. Due to this technology's spectral efficiency, flexibility and scalability, high multiple-input multiple-output (MIMO) compatibility, low transceiver complexity, and robustness to frequency-selectivity, OFDM has been adopted in numerous wired\footnote{OFDM-based wired communications standards: Asymmetric Digital Subscriber Line (ADSL); Very high-speed Digital Subscriber Line (VDSL); Digital Video Broadcasting -- Cable (DVB-C); DVB-C2; Multimedia Over Coax Alliance (MoCA) home networking; Data Over Cable Service Interface Specification (DOCSIS); and so on \cite{OFDM_Wiki}.} and wireless communications standards\footnote{OFDM-based wireless communications standards: the wireless local area network (WLAN) radio interfaces such as IEEE 802.11a, g, n, ac, ah, and HIPERLAN/2; Digital Video Broadcasting -- Terrestrial (DVB-T); Digital Video Broadcasting -- Handheld (DVB-H); evolved UMTS Terrestrial Radio Access (E-UTRA); the 4G downlink of the 3GPP Long Term Evolution (LTE) standard; the wireless MAN/broadband wireless access (BWA) standard IEEE 802.16e (Mobile-WiMAX); and alike \cite{OFDM_Wiki}.} \cite{Weinstein_OFDM_history,Jiang_OFDMA_WiMAX_07,Hwang_OFDM_Applications_09}, including 5G new radio (5GNR) \cite{Parkvall_5G_NR_17,Hirzallah_5GNR_20}. Despite such broad applicability and its efficient parallel transmission -- via orthogonal subcarriers  -- by transforming a frequency selective channel to a frequency-flat channel (per a sub-channel), OFDM is sensitive to a time-selectivity \cite{P_Sch_LCE_OFDM_04,Y_Mos_ICI_mit_05}. When a time-selectivity is compounded by a frequency-selectivity, a doubly selective fading channel \cite{GBG_BEMs_98,Giannakis_ST_BBWC_book_03} arises. During this channel's filtering of the transmitted OFDM symbols, orthogonality among the subcarriers can be destroyed to an extent that OFDM would suffer from a severe intercarrier interference (ICI). 

During (ultra-)high-speed broadband wireless communications evident in the fifth-generation (5G) and sixth-generation (6G) \cite{Saad_6G_Vision_20,Letaief_Edge_AI_Vision'22,Alwis_Survey_GG_Networks'21,Alsabah_6G_Wireless_Commun_Network'21,You_Towards_6G'21} wireless communication systems' use-cases such as autonomous vehicles, flying taxis, and ultra-high-speed trains, high data rates and multipath propagation give rise to the baseband equivalent  channel's frequency selectivity whereas carrier frequency offsets and mobility-induced Doppler shifts introduce time-selectivity \cite{X_MA_MDT_03,X_Ma_Opt_train_03}. Time-selectivity blended with frequency-selectivity, OFDM systems operating in (ultra-)high-speed broadband wireless communications scenarios will consequently suffer from severe ICI. To mitigate severe ICI efficiently, robust channel estimation is an inevitable signal processing paradigm required for robust equalization, robust symbol detection (recovery), and channel state information (CSI) feedback. Regarding CSI feedback, downlink CSI is estimated at the user equipment and fed back to the base station to eliminate the inter-user interference (via precoding) and increase channel capacity in massive MIMO frequency division duplex networks \cite{TW_DL_B_CSI_feed_2019,H_He_DL_B_CE_2018,Wen_M-MIMO_CSI_feedback_18}.
        
Devised for OFDM systems operating over a doubly selective channel, the state-of-the-art works -- inspired by statistical signal processing \cite{ZT_PA_TVCE_07,CE_PTP_DSCE_07,F_Qu_DSFC_estiim_10,PC_CE_for_OFDM_13,Liu_SC-FDMA_Comm_over_DSFC_20} and deep learning (DL) \cite{YYF_DL_DSFC_19} -- advocate various channel estimation algorithms. These algorithms count on -- despite being interesting -- the transmission of pilot (preamble) symbols and tend to waste the already scarce spectra. Saving scarce spectra that would otherwise be wasted, the development of a blind doubly selective channel estimation algorithm is important, as inspired by the following related works.               

\subsection{Related Works}
\label{subsec: related_works}
Aiming to address the fundamental problem of a doubly selective fading channel estimation for OFDM systems, the wireless communications and signal processing research communities propose many novel algorithms \cite{ZT_PA_TVCE_07,CE_PTP_DSCE_07,F_Qu_DSFC_estiim_10,PC_CE_for_OFDM_13,Liu_SC-FDMA_Comm_over_DSFC_20,YYF_DL_DSFC_19}. Beginning with the algorithms in \cite{ZT_PA_TVCE_07}, the authors propose doubly selective channel estimators devised to combat both out-of-band interference and noise by employing a basis expansion model (BEM) and a time-domain window. The authors of \cite{CE_PTP_DSCE_07} propose efficient doubly selective channel estimators by exploiting an efficient pilot tone placement scheme also proposed by the same authors. The authors of \cite{F_Qu_DSFC_estiim_10} develop a windowed least squares (LS) estimator for doubly selective channels. The authors of \cite{PC_CE_for_OFDM_13} propose a doubly selective channel estimation scheme based on the theory of distributed compressive sensing. The authors of \cite{Liu_SC-FDMA_Comm_over_DSFC_20} propose a joint channel estimation and data equalization scheme for a discrete Fourier transforms (DFT) spread OFDM system operating over a doubly selective channel.          

Notwithstanding the significant advancements made by the authors of the aforementioned works which are interesting in their own rights, there remain many challenges -- emanating from both time- and frequency-selectivity -- regarding the aforementioned fundamental problem. To this end, the shortcomings of the existing works are highlighted as follows.  
\begin{itemize}
	\item To reduce the parameters of the doubly selective channel estimation, most existing techniques \cite{ZT_PA_TVCE_07,PC_CE_for_OFDM_13,Liu_SC-FDMA_Comm_over_DSFC_20} rely on BEMs. However, BEMs cause a fitting (modeling) error \cite{ZT_PA_TVCE_07,PC_CE_for_OFDM_13,F_Qu_DSFC_estiim_10}. Aside from fitting error, BEMs require prior knowledge regarding upper bounds on the channel's delay spread and Doppler spread.    
	\item Some of the proposed schemes rely on time-domain windowing \cite{ZT_PA_TVCE_07} and also on time-domain de-windowing \cite{F_Qu_DSFC_estiim_10}.     
	\item In any attempt of employing pilot tones to estimate a time- and frequency-selective channel, the channel estimator itself suffers from ICI \cite{X_Cai_ICI_supp_03}. 
	\item Most of the proposed schemes entail prohibitive computational complexity -- e.g., performing an inverse operation of an (ultra-)large matrix.
	\item Almost all the proposed techniques \cite{ZT_PA_TVCE_07,CE_PTP_DSCE_07,F_Qu_DSFC_estiim_10,PC_CE_for_OFDM_13,Liu_SC-FDMA_Comm_over_DSFC_20} require huge number of pilot symbols whenever the filtering channel happens to be a doubly selective deep fading channel. 
	\item Most the state-of-the-art techniques are hardly data-driven techniques which will be inevitable in the upcoming 6G wireless systems \cite{Saad_6G_Vision_20,Letaief_Edge_AI_Vision'22,Alwis_Survey_GG_Networks'21,Alsabah_6G_Wireless_Commun_Network'21,You_Towards_6G'21} (also the references therein) envisioned to be artificial intelligence (AI)/machine learning (ML) enabled intelligent systems.    
\end{itemize}

In the context of the last shortcoming, most of the existing channel estimators employ mathematical channel models. Relying on models that attempt to represent the time- and/or frequency-variations of wireless communications channel, such a model-based approach can be effective for a few communicating devices connected to a base station or multi-antenna receivers. Nevertheless, depending only on mathematical channel models is bound to be infeasible \cite{TXU_DL_Int_Cancel_2018} for massive Internet of things (IoT) devices \cite{MRP_IoT_5G_era_16}, on the verge of being a norm in 5G and 6G \cite{Saad_6G_Vision_20,Letaief_Edge_AI_Vision'22,Alwis_Survey_GG_Networks'21,Alsabah_6G_Wireless_Commun_Network'21,You_Towards_6G'21}. For massive IoT devices employing OFDM in a high-speed broadband communication scenario, the parameter optimization schemes of \cite{HS_TCOM_Param_optim_99,HS_Param_Optimi_TWC_07} -- proposed for doubly selective channels -- would be computationally complex. Tackling computational complexity in a doubly selective channel setting, recent work by the authors of \cite{Campos_Low_Complexity_MC_System_16} employed the virtual decomposition of the channel to propose a modified OFDM system. Nevertheless, the entire modified OFDM system relies on a BEM that employs discrete prolate spheroidal (DPS) sequences \cite{Slepian_DPSS_1964}.       

Encompassing the major technological revolutions in AI/ML, DL \cite{PSSSYYTTRMMSS19,Lecun_DL_Nature_15,IGYAC16} has emerged as a powerful tool. As a tool driving the 21st-century science and technology revolution, DL (also deep reinforcement learning) has been applied in numerous research fields as diverse as computer vision, speech recognition, image processing, object recognition \cite{IGYAC16,Lecun_DL_Nature_15,LWXPJLM18}; clustering, information retrieval, dimensionality reduction, and natural language processing \cite{DL_methods_14,IGYAC16}; many-body quantum physics (e.g., quantum entanglement \cite{Giuseppe_Quan_MB_ANNs,Quan_Entag_DL_Arch_19}); gaming, finance, energy, and healthcare \cite{mnih-DQN-2015,Li2018_DRL}; and so forth. 

Aiming to overcome the limitations of the conventional model-based wireless communications, signal processing, and networking, DL has also emerged as a promising tool in wireless communications, signal processing, and networking \cite{QFQH_18,DRL_Applcs_19,CZPHH19,Chen_ANNs_Tutorial_19,Gunduz_ML_19,Huang_DL_PHY_20,OShea_TCCN_17,He_TSP_Mo_Driven_DL_20}. Along with these fields' latest trends, several DL-based channel estimation algorithms are proposed \cite{MSo_CommL_19,H_Ye_DLCE_2018,YYF_DL_DSFC_19,XMa_Lae_TVC_18,H_He_DL_B_CE_2018,Huang_SR_CE_18,Cheng_wcl_DL_CE_19,Chun_DL_MU-MIMO_19,Liao_MD_DL_M-MIMO'20,Liao_DL_TV_MIMO-OFDM'20,Bai_TCCN_DLCE'20,Balevi_JSAC'21,Gizzini_DL_for_IEEE_802.11p'20,Mthethwa_DL_CE_USTLD_20,Xiang_DL_JCE_DD'20}. Amongst these works, the authors of \cite{YYF_DL_DSFC_19} proposed a DL-based doubly selective fading channel estimator which employs a trained deep neural network (DNN). Employing DNNs, the authors of \cite{YYF_DL_DSFC_19} conduct pretraining and training using the outputs of an LS estimator and the doubly selective channel's previous estimates. The previous estimates are fed back to learn the time correlation of the time-varying frequency-selective channels. However, such feedback will result in intrinsic error propagation. As a result, the investigated DL-based algorithm of \cite{YYF_DL_DSFC_19} would suffer intrinsically from a performance loss, especially at low signal-to-noise ratio (SNR) regimes which would lead to an inaccurate feed of the previous channel estimates. Meanwhile, although the algorithm proposed in \cite{YYF_DL_DSFC_19} results in some improvements for a few pilot blocks (see \cite[Fig. 11]{YYF_DL_DSFC_19}), the proposed non-blind estimator relies on the assistance from the LS estimator that continuously feeds its estimates to the considered DNN. In this respect, the estimator of \cite{YYF_DL_DSFC_19} tends to serve as a refiner, with respect to (w.r.t.) the estimation of the LS, rather than learning autonomously the doubly selective channel characteristics. Moreover, the scheme of \cite{YYF_DL_DSFC_19} would require many pilot symbols, especially when the channel happens to be a deep fading doubly selective channel. Toward estimating such a channel, we deliver the following contributions.         

\subsection{Contributions}
\label{subsec: DSFC_contibs}
Aiming to tackle the fundamental challenges of a doubly selective channel estimation in OFDM systems, we propose a DL-based blind doubly selective channel estimation algorithm. This algorithm employs a deep rectified linear unit (ReLU) feedforward neural network (FNN) whose input-output mapping is derived from an optimization problem accounting for received OFDM signal samples. For this system setup, we develop a theory on the testing mean squared error (MSE) performance of our DL-based blind OFDM channel estimator. This estimator's performance is quantified through a blind OFDM channel estimation using a trained over-parameterized ReLU FNN while being inspired by the developments on the theory of deep learning \cite{Poggio_Theo_Issues_Dnets_2020,Zhu_GAN_deconstruction_20,Modern_DL_Math'21,Bartlett_DL_Stat_Viewpoint'21} pertaining to deep (also shallow) networks operating in an over-parameterized\footnote{Although over-parameterized deep and shallow networks should be \textit{overfitting} per the statistical wisdom of the classical bias–variance trade-off \cite{Belkin_PNAS_reconciling_MML'19,NIPS2014_Comp_Efficiency}, over-fitting does not usually occur with practical deep networks which are also over-parameterized deep networks \cite{Sun2019optimization,Sun_SPM_global_landscape'20}. This discrepancy has continued to be one of the fundamental research themes on the theory of DL \cite{Sun2019optimization,Sun_SPM_global_landscape'20}. Meanwhile, the authors of \cite{Chizat_lazy_training_2020} disseminated an implicit bias phenomenon dubbed lazy training that refers to the scenario when a non-linear parametric model behaves like a linear one. This happens implicitly under some choices of hyper-parameters governing normalization, initialization, and the number of iterations when the scale of the model becomes large \cite{Chizat_lazy_training_2020} (also applies to a large class of models beyond neural networks \cite{Sun2019optimization,Sun_SPM_global_landscape'20}). Meanwhile, over-parameterized deep networks need to move only a tiny distance and hence linearization is a good approximation, also in the context of the neural tangent kernel \cite{jacot2018neural}. However, since practical deep networks are not ultra-wide, their parameters will move a huge distance and thus likely to move out of the linearization regimes \cite{Sun2019optimization,Sun_SPM_global_landscape'20}.} regime \cite{allenzhu2018convergence,Zou2018_SGD,Zhang2019_Convergence,Soltanolkotabi_TIT_19,Allenzhu_Learing_and_Gen_2020,Chen_overparameterization_2020,Cao_generalization_error_bounds_2019,Oymak_Mod_Overparameterization_20}. This is the first of its kind performance quantification of a DL-based algorithm and such a theory has not been developed to date -- to the best of our knowledge -- for any of the various DL-based algorithms proposed for solving classification or regression AI/ML problems of wireless communications, signal processing, and networking. Apart from networking, signal processing, and wireless communications works, no work has also quantified the performance of any DL-based algorithm purposed for solving classification or regression AI/ML problems in the broader context of AI/ML research. The state-of-the-art AI/ML research on the theory of DL comprises interesting works on the generalization error bounds of deep (and shallow) networks using local Rademacher complexities, algorithmic stability, compression bounds, and algorithmic robustness \cite{Modern_DL_Math'21,Bartlett_DL_Stat_Viewpoint'21}. However, these bounds are not exactly performance quantification of a DL-based algorithm deployed -- after training -- for testing. Therefore, we shall report that this fundamental work develops the first of its kind theory at the crossover of the theory of DL, high-dimensional probability, wireless communications, and signal processing. 

Summing up, the fundamental contributions of this paper are itemized below.  
\begin{itemize}
\item We propose a DL-based blind channel estimator for a doubly selective OFDM channel.

\item For our blind channel estimator employing an over-parameterized ReLU FNN, we develop a theory on its testing MSE performance. Specifically, we derive theorems that quantify the asymptotic and non-asymptotic testing MSE performance of our proposed blind estimator. 

\item While quantifying our blind estimator's non-asymptotic testing MSE performance, we derive a new high-dimensional probability lemma on the concentration of the Euclidean norm of Gaussian random variables (RVs) with different variances.   
\end{itemize}
Apart from the itemized contributions, we assess the performance of our blind channel estimator by training and testing different ReLU FNNs using \textsc{Keras} with \textsc{TensorFlow} as a backend.     

With this introduction, the remainder of this paper is organized as follows. Section \ref{sec: notations_and_defs} outlines notation and definitions. Section \ref{sec: prelude} puts forward a prelude on the required paper preliminaries. Section \ref{sec: system_model} details the considered system model. Section \ref{sec: DL_architect_and_motivation} presents our DL-based blind OFDM channel estimation algorithm. Section \ref{sec: Per_Over_Param_ReLU_FNNs} documents our developed theory on the testing MSE performance of the proposed blind OFDM channel estimator. Section \ref{sec: Computer_experiments} reports corroborating computer experiments. Finally, Section \ref{sec: DSFC_conc_rem_and_res_outlook} provides concluding remarks and research outlook.     

\section{Notation and Definitions}
\label{sec: notations_and_defs}
\subsection{Notation}
\label{subsec: notations}
Scalars, vectors, matrices, and sets (and events) are represented by Italic letters, bold lowercase letters, bold uppercase letters, and calligraphic letters, respectively. $\mathbb{N}$, $\mathbb{Z}(\mathbb{Z}^+)$, $\mathbb{C}(\mathbb{R})$, $\mathbb{R}\backslash \{0\}$, $\mathbb{C}^{n}(\mathbb{R}^{n})$, and $\mathbb{C}^{m\times n}(\mathbb{R}^{m\times n})$ represent the set of natural numbers, the set of integer(positive integer) numbers, the set of complex(real) numbers, the set of real numbers excluding zero, the set of $n$-dimensional vectors of complex(real) numbers, and the set of $m\times n$ complex(real) matrices, respectively. $\mathcal{CN}(\mu, \sigma^2)$ represents a circularly symmetric complex Gaussian distribution with mean $\mu\in \mathbb{R}$ and variance $\sigma^2\in \mathbb{R}$. $\mathcal{N}(\mu, \sigma^2)$ stands for a Gaussian distribution with mean $\mu\in \mathbb{R}$ and variance $\sigma^2\in \mathbb{R}$. For a complex number $z=a+\jmath b$, where $\jmath = \sqrt{-1}$, its magnitude is given by $|z|=\sqrt{a^2+b^2}$ and its complex conjugate is denoted by $\overline{z}$ which is defined as $\overline{z}\eqdef a-\jmath b$. For $a\in\mathbb{R}$, its absolute value is denoted by $|a|$. For $z\in\mathbb{Z}^+$, its factorial and double factorial are denoted by $z!$ and $z!!$, respectively.  

$\ast$, $\sim$, $\forall$, $\delta(\cdot)$, $>>$, and $\textnormal{diag}(\cdot)$ stand for convolution, distributed as, for all, Dirac's delta function, much greater than, and a (block) diagonal matrix, respectively. $(\cdot)^{T}$, $(\cdot)^{H}$, $\otimes$, $\textnormal{Re}\{\cdot\}$, and $\textnormal{Im}\{\cdot\}$ represent transpose, Hermitian, Kronecker product, real part, and imaginary part, respectively. $\ell_2(\ell_p)$, $\textnormal{tr}(\cdot)$, $\textnormal{max}\{\cdot, \cdot\}$, $\textnormal{min}\{\cdot, \cdot\}$, and $\langle\cdot\rangle_{N}$ denote Euclidean norm($p$-th norm), trace, the maximum of two numbers, the minimum of two numbers, and modulo-$N$ operation\footnote{In MATLAB$^{\textregistered}$, $\langle m \rangle_{N}$ is computed via a function named \textit{mod}$(m,N)$ and equals to $m-\floor*{m./N}*N$.}, respectively. $\textnormal{vec}(\cdot)$, $\mathbb{P}(\cdot)$, $\mathbb{E}\{\cdot\}$, $\sigma_{\textnormal{min}}(\cdot)$, and $\sigma_{\textnormal{max}}(\cdot)$ stand for vectorization, probability, expectation, the minimum singular value, and the maximum singular value, respectively. $\textnormal{min}$, $\sup$, $\inf$, $\exp(x)$, and $\log x$ denote minimum, supremum, infimum, $e^x$, and natural logarithm, respectively. $\eqdef$, $|\cdot|$, $\floor*{\cdot}$, and $\ceil*{\cdot}$ represent equal by definition, cardinality, integer floor, and integer ceiling, respectively. $\mathbb{P}(\mathcal{A}|\mathcal{B})$ stands for the probability of an event $\mathcal{A}$ conditioned on an event $\mathcal{B}$. $\mathbb{P}(\mathcal{A}|\mathcal{B}, \mathcal{C})$ denotes the probability of an event $\mathcal{A}$ conditioned on events $\mathcal{B}$ and $\mathcal{C}$.    

The Euclidean norm of a vector $\bm{a}\in\mathbb{R}^m$ is denoted by $\|\bm{a}\|_2$ or $\|\bm{a}\|$. For $n\in\mathbb{N}$, we let $[n]\eqdef\big\{1, \ldots, n\big\}$. For $m, i\in\mathbb{N}$; $i\leq m$; and $\bm{a}\in\mathbb{R}^m$, we denote the $i$-th entry of $\bm{a}$ by ${(\bm{a})}_i$. We use $\rho(x)\eqdef \max \{0, x\}$ to denote a component-wise ReLU operation for $\bm{a}\in\mathbb{R}^m$ as $\rho(\bm{a})\eqdef \big[\rho({(\bm{a})}_1), \rho({(\bm{a})}_2), \ldots, \rho({(\bm{a})}_m) \big]^T$. $\rho'(x)$ denotes the derivative of $\rho(x)$. For $\mathbbm{1}\{\cdot\} $ being an indicator function that returns 1 when its argument is true (0 otherwise) and $\bm{a}\in\mathbb{R}^m$, we use $\mathbbm{1}\{\bm{a}>0\}$ to denote $\big[\mathbbm{1}\{(\bm{a})_1>0\}, \mathbbm{1}\{(\bm{a})_2>0\}, \ldots, \mathbbm{1}\{(\bm{a})_m>0\} \big]^T$. The inner product between two conformable vectors $\bm{a}$ and $\bm{b}$ is denoted by $\langle\bm{a},\bm{b}\rangle$. For $n\geq 2$ and $n\in\mathbb{N}$, the horizontal concatenation of $n$ conformable vectors and matrices is written as $[\bm{a}_1 \hspace{1mm} \bm{a}_2 \ldots \bm{a}_n ]$ and $[\bm{A}_1 \hspace{1mm} \bm{A}_2 \ldots \bm{A}_n ]$, respectively. $\bm{I}_{n}$, $\bm{0}_{m\times n}$, and $\bm{0}$ denote an $n\times n$ identity matrix, an $m\times n$ zero matrix, and a zero matrix whose dimension coincides with the stated context, respectively. $\bm{F}_N$ denotes an $N\times N$ DFT matrix defined via ${(\bm{F}_N)}_{k,l}\eqdef 1/\sqrt{N} e^{-j \frac{2\pi (k-1)(l-1)}{N}}$ \cite{X_Cai_ICI_supp_03} and ${\bm{F}}_N^{H}$ denotes the inverse DFT (IDFT) matrix ($\bm{F}_N$ is a unitary matrix, i.e., $\bm{F}_N{\bm{F}}_N^{H}={\bm{F}}_N^{H}\bm{F}_N=\bm{I}_N$ \cite{Giannakis_ST_BBWC_book_03}.). For a matrix $\bm{A}\in\mathbb{R}^{n \times n}$, its spectral norm, Frobenius norm, and inverse are denoted by $\|\bm{A}\|_2$, $\|\bm{A}\|_F$, and $\bm{A}^{-1}$, respectively. The $(i,j)$-th element of $\bm{A}$ is denoted by $(\bm{A})_{i,j}$. We use $\bm{A}\succeq \bm{0}$ to denote that each element of $\bm{A}$ is greater than or equal to 0. For $\bm{W}\eqdef [\bm{W}_1 \hspace{1mm} \bm{W}_2  \ldots \bm{W}_K ]$, we let $\|\bm{W}\|_2=\max_{k\in [K]} \|\bm{W}_k\|_2$ and $\|\bm{W}\|_F=\big( \sum_{k=1}^K \|\bm{W}_k\|_F^2  \big)^{1/2}$. We denote the gradient of $F(\bm{W})$ by $\nabla F(\bm{W})$.     

For a function $f(x)$ and a function $g(x)$, $\Omega\big(g(x)\big)$ implicates the asymptotic lower bounds of $f(x)$ evaluated at $x\geq x_0$, i.e., $f(x)=\Omega\big(g(x)\big)$ if and only if (iff) there exists a real constant $c>0$ and an integer constant $x_0\geq 1$ such that $f(x)\geq cg(x)$ for all integer $x\geq x_0$. The Landau notation $O(\cdot)$ describes the limiting behavior of a function whenever the underlying argument tends towards a particular value or infinity, i.e., $f(x) = O\big(g(x)\big)$ iff there exists a real constant $c>0$ and an integer constant $x_0\geq 1$ such that $\displaystyle f(x) \leq cg(x)$ for all integer $x\geq x_0$. On the other hand, for $f(x)$ and $g(x)$ being functions that map positive integers to positive real numbers, we write that $f(x)=\Theta (g(x))$ iff $f(x)=\Omega\big(g(x)\big)$ and $f(x) = O\big(g(x)\big)$, i.e., $f(x)=\Theta (g(x))$ iff there exists real constants $c_1,c_2$ and an integer $x_0\geq 1$ such that $c_1g(x)\leq f(X)\leq c_2g(x)$ for all integer $x\geq x_0$. We use $\tilde{\Omega}(\cdot)$ to hide polynomial terms in $\Omega(\cdot)$. We use $\textnormal{poly}(\cdot)$ to imply polynomial dependency. Following these notation, we follow up with definitions. 

\subsection{Definitions}
\label{subsec: ndefs}
For the sake of clarity and completeness, we hereinafter provide definitions on double factorial, error function, Gamma function, $L$-Lipschitz function, a positive semidefinite matrix, a symmetric Bernoulli distribution, extreme singular values, Frobenius norm, and spectral norm.    
\begin{definition}[{Double factorial \cite{Wolfram_Double_Fact}}]
\label{Double_factorial_def}
The double factorial of $n$ is denoted by $n!!$ and defined as 
\begin{equation}
\label{double_fact_real}
n!!=\begin{cases}
n.(n-2) \ldots 5.3.1 & \text{if $n>0$ and $n$ is odd} \\
n.(n-2) \ldots 6.4.2 & \text{if $n>0$ and $n$ is even} \\
1 & \text{if $n=-1, 0$.}
\end{cases}
\end{equation} 	 
\end{definition}

\begin{definition}[{Error function \cite[p. 887]{ISGI07}}]
\label{Error_func_def}
The error function is denoted by $\textnormal{erf}(\cdot)$ and   
\begin{equation}
\label{erf_func_def}
\textnormal{erf}(x)=\Phi(x)=\frac{2}{\sqrt{\pi}}\int_{0}^x e^{-t^2}dt,
\end{equation}
where $\Phi(x)$ is the probability integral.   
\end{definition}

\begin{definition}[{Gamma function \cite[p. 892]{ISGI07}}]
\label{Gamma_func_def_all_cases}
The Gamma function is denoted by $\Gamma(\cdot)$ and  
\begin{equation}
\label{gamma_func_def}
\Gamma(z)=\int_{0}^{\infty} t^{z-1}e^{-t} dt.	
\end{equation}
For a positive integer $z$, $\Gamma(z)=(z-1)!$. 		
\end{definition} 
\begin{definition}[{$L$-Lipschitz function \cite{Boucheron_Con_eqs_13}}]
	\label{L_Lipschitz_func_def}
	Let $f: \mathbb{R}^n\to\mathbb{R}$ denote an $L$-Lipschitz function. Then, there exists a constant $L>0$ such that for all $x,y\in\mathbb{R}^n$ \cite[p. 125]{Boucheron_Con_eqs_13}, 
	\begin{equation}
	\label{L_Lipschitz_relation}
	|f(x)-f(y)|\leq L\|x-y\|.
	\end{equation}  	
\end{definition}

\begin{definition}[{Positive semidefinite matrix \cite{RHCR13}}]
\label{Pos_semi_definite_def}
If $\bm{A}\in\mathbb{C}^{n\times n}$ is a Hermitian matrix, $\bm{A}=\bm{A}^H$. Such Hermitian matrix $\bm{A}$ is positive semidefinite if \cite[p. 429]{RHCR13} 
\begin{equation}
\label{Pos_semi_definite_matrix}
\bm{x}^H \bm{A}\bm{x} \geq 0 \hspace{2mm}\textnormal{for all nonzero} \hspace{1mm} \bm{x}\in\mathbb{C}^n.
\end{equation}
\end{definition}

\begin{definition}[{Symmetric Bernoulli distribution \cite{vershynin_2018}}]
\label{Symm_Bernoulli_distribution}
A RV $X$ has \textit{symmetric Bernoulli} distribution (also known as \textit{Rademacher} distribution) provided that it takes values -1 and 1 with probabilities 1/2 each, i.e., $\mathbb{P}(X=1)=\mathbb{P}(X=-1)=1/2$.   
\end{definition}

\begin{definition}[{Extreme singular values \cite{Rudelson_min_Sing_Vals_2006,Rudelson_invertibility_2005,Tatarko_upper_bound_2018,Rudelson_Extreme_SVals_10}}]
\label{Ext_singular_values_def}
For $\bm{A}\in\mathbb{R}^{N\times n}$, its singular values are the eigenvalues of $|\bm{A}|=\sqrt{\bm{A}^H\bm{A}}$. Meanwhile, the maximum and minimum singular values of $\bm{A}$ are defined mathematically as \cite{Rudelson_min_Sing_Vals_2006,Rudelson_invertibility_2005,Tatarko_upper_bound_2018,Rudelson_Extreme_SVals_10} 
\begin{equation}
\label{max_min_sing_values}
\sigma_{\textnormal{max}}(\bm{A})=\sup_{\|\bm{x}\|=1}\|\bm{A}\bm{x}\| \hspace{2mm}\textnormal{and}\hspace{2mm} \sigma_{\textnormal{min}}(\bm{A})=\inf_{\|\bm{x}\|=1}\|\bm{A}\bm{x}\|.  
\end{equation}
For all $\bm{x}\in\mathbb{R}^n$, the extreme singular values of $\bm{A}$ satisfy the inequality \cite[eq. (4.5)]{vershynin_2018}
\begin{equation}
\label{Euclidean_norm_constraint}
\sigma_{\textnormal{min}}(\bm{A}) \|\bm{x}\|\leq \|\bm{A}\bm{x}\|\leq \sigma_{\textnormal{max}}(\bm{A})\|\bm{x}\|.
\end{equation}
\end{definition}

\begin{definition}[{Frobenius and spectral norms \cite{CDM00}}]
\label{Frob_and_spec_norma}	
For a real matrix $\bm{A}\in\mathbb{R}^{m\times n}$, its Frobenius and spectral norms are defined as $\|\bm{A}\|_F \eqdef \sqrt{\sum_{i=1}^m \sum_{j=1}^m \big(\bm{A}\big)_{i,j}^2}$ and $\|\bm{A}\|_2\eqdef \sigma_{\textnormal{max}}(\bm{A})$, respectively. For a complex matrix $\bm{A}\in\mathbb{C}^{m\times n}$, its Frobenius norm is defined as $\|\bm{A}\|_F \eqdef \sqrt{\sum_{i=1}^m \sum_{j=1}^m \big|\big(\bm{A}\big)_{i,j}\big| ^2}$.      		
\end{definition}

Following Definitions \ref{Double_factorial_def}-\ref{Frob_and_spec_norma}, we follow with a prelude.
	
\section{Prelude}
\label{sec: prelude}
This section presents preliminaries on wireless doubly selective fading channel and classical and high-dimensional probability.   

\subsection{Classical and High-Dimensional Probability Preliminaries}
\label{subsec: HDP_results}
In the following, we state the main classical \cite{DPJN08} and high-dimensional probability \cite{vershynin_2018,wainwright_2019} (also concentration inequalities \cite{Boucheron_Con_eqs_13}) results that have guided the theoretical developments of this paper. To begin with, we state a classical result in probability theory known as the \textit{total probability theorem} \cite[p. 28]{DPJN08} (also known as the \textit{law of total probability} \cite{Grimmett_Prob_Random_Process'01}). 
\begin{theorem}[{Total probability theorem \cite{DPJN08}}]
\label{Total_Prob_Thm}
Let $\mathcal{A}_1, \ldots, \mathcal{A}_n$ be disjoint events that constitute a partition of the sample space (i.e., each probable outcome is included in exactly one of the events $\mathcal{A}_1, \ldots, \mathcal{A}_n$). Then, for any event $\mathcal{B}$,
\begin{equation}
\label{Total_Prob_Thm_Expression}
\mathbb{P}(\mathcal{B})=\sum_{j=1}^n \mathbb{P}(\mathcal{A}_j \cap \mathcal{B})=\sum_{j=1}^n \mathbb{P}(\mathcal{A}_j)\mathbb{P}(\mathcal{B} |\mathcal{A}_j).
\end{equation}
\end{theorem}

The upper bound of an upper tail probability exhibited by any RV is characterized via a classical result known as \textit{Markov's inequality} \cite{vershynin_2018,Boucheron_Con_eqs_13,wainwright_2019} which is stated beneath. 
\begin{proposition}[{Markov's inequality \cite[Proposition 1.2.4]{vershynin_2018}}]
\label{Prop_Markov_inequality}
For a non-negative RV $X$ and $t>0$, 
\begin{equation}
\label{Markov_inequality}
\mathbb{P}\big(X\geq t\big)\leq \frac{\mathbb{E}\{X\}}{t}.
\end{equation} 
\end{proposition} 
Concerning the right-hand side (RHS) of (\ref{Markov_inequality}), we state the \textit{integral identity} \cite[p. 7]{vershynin_2018} below. 
\begin{lemma}[{Integral identity \cite[Lemma 1.2.1]{vershynin_2018}}]
\label{Lemma_integral_identity}
Let $X$ be a non-negative RV. Then, 
\begin{equation}
\label{integral_identity}
\mathbb{E}\{X\}=\int_{0}^{\infty}\mathbb{P}\big(X> t\big)dt.                 
\end{equation}
The two sides of (\ref{integral_identity}) are either finite or infinite at the same time.  
\end{lemma}

Characterizing the expectation of an RV that might be computed through the integral identity, \textit{Jensen’s inequality} \cite{Tropp_Mat_Con_InEq_15,wainwright_2019,vershynin_2018} that describes how averaging interacts with convexity or concavity is stated beneath.   
\begin{proposition}[{Jensen’s inequality \cite[eq. (2.2.2)]{Tropp_Mat_Con_InEq_15}}]
\label{Prop_Jensen_inequality}
Let $\bm{X}$ be a random matrix and let $h$ be a real-valued function on matrices. Then,  
\begin{subequations}
\begin{align}
\label{Jensen_inequality_concave}
\mathbb{E}\{h(\bm{X})\}&\leq h(\mathbb{E}\{\bm{X}\}) \hspace{3mm} \textnormal{when $h$ is concave, and}     \\
\label{Jensen_inequality_convex}
\mathbb{E}\{h(\bm{X})\}&\geq h(\mathbb{E}\{\bm{X}\}) \hspace{3mm} \textnormal{when $h$ is convex}.
\end{align}
\end{subequations} 
\end{proposition}

Apart from the upper bound of an upper tail probability manifested by a given RV, the lower bound of an upper tail probability plays an important role in the random matrix theory. In this vein, we state the \textit{Paley--Zygmund inequality} \cite{Litvak_smallest_sing_value_05} as follows. 
\begin{lemma}[Paley--Zygmund inequality {\cite[Lemma 3.5]{Litvak_smallest_sing_value_05}}]
	\label{PZ_identity}
	Let $p\in(1,\infty)$ and $q=p/(p-1)$. Suppose $f>0$ be an RV with $\mathbb{E}\big\{ f^{2p} \big\}\leq \infty$. Then, for all $0\leq \lambda \leq \sqrt{\mathbb{E}\{ f^2 \}}$, 
	\begin{equation}
	\label{PZ_identity_expression}
	\mathbb{P}(f>\lambda)\geq \frac{\big(\mathbb{E}\{ f^2 \}-\lambda^2\big)^q}{\big(\mathbb{E}\big\{ f^{2p} \big\}\big)^{q/p}}.
	\end{equation}           
\end{lemma}

Extreme singular values (via extreme eigenvalues) are often used to characterize the performance of detection algorithms applicable in wireless communications, signal processing, and wireless networking. Toward this end, we state a fundamental classical result -- due to Gordon \cite{Gordon_inequalities_1985,Gordon_majorization_1992} -- on the extreme singular values of Gaussian matrices.   
\begin{theorem}[{Extreme singular values of Gaussian matrices \cite[Theorem 2.6]{Rudelson_Extreme_SVals_10}}]
	\label{Thm_Gordon_inequality}
	Let $\bm{A}$ be an $N\times n$ matrix with independent entries $(\bm{A})_{i,j}\sim\mathcal{N}(0,1)$ for $1\leq i\leq N$ and $1\leq j\leq n$. Then, 
	\begin{equation}
	\label{Gordon_inequality}
	\sqrt{N}-\sqrt{n}\leq \mathbb{E}\{\sigma_{\textnormal{min}}(\bm{A})\}\leq\mathbb{E}\{\sigma_{\textnormal{max}}(\bm{A})\}\leq\sqrt{N}+\sqrt{n}.
	\end{equation}    
\end{theorem}

At last, we state a fundamental result on the \textit{concentration of measure} in the
Gaussian space \cite{Boucheron_Con_eqs_13,wainwright_2019,Peng:Thesis:2015}. 
\begin{theorem}[{Gaussian concentration inequality \cite[Theorem 5.6]{Boucheron_Con_eqs_13}}]
\label{Thm_Gaussian_conc_inequality}
Suppose $X=[X_1, \ldots, X_n]$ be a vector of $n$ independent standard normal RVs, i.e., $X_i\sim\mathcal{N}(0,1)$ for $1\leq i \leq n$. Let $f: \mathbb{R}^n\to\mathbb{R}$ be an $L$-Lipschitz function. Then, for all $t>0$, 
\begin{equation}
\label{Gaussian_conc_inequality}
\mathbb{P}\big(f(X)-\mathbb{E}\{f(X)\}\geq t\big) \leq e^{-t^2/(2L^2)}.
\end{equation} 
	
\end{theorem}	
It is worthwhile underscoring that the RHS of (\ref{Gaussian_conc_inequality}) does not depend on the dimension $n$ \cite{Boucheron_Con_eqs_13}. Meanwhile, we proceed to preliminaries on wireless doubly selective fading channel.

\subsection{Wireless Doubly Selective Fading Channel Preliminaries}
\label{subsec: DSC_prelims}
Consider the transmission of a baseband signal $s(t)$ after modulation with a carrier frequency $f_c$ as $s_p(t)=\textnormal{Re}\big\{s(t) e^{\jmath 2\pi f_ct} \big\}$. This baseband signal is filtered at the transmitter by a linear filter denoted by $h^{\textnormal{tr}}(t)$, distorted by the possibly time-varying physical wireless channel which is also a linear filter denoted by $h^{\textnormal{ch}}(t; \tau)$, contaminated by the additive noise $v(t)$, and filtered at the receiver also by a linear filter denoted by $h^{\textnormal{rec}}(t)$. To this end, the received baseband signal prior the sampling operation is expressed as \cite{GBG_BEMs_98,G_Leus_OMA_03}   
\begin{equation}
\label{BB_reception}
y(t)=\sum_{n=-\infty}^{\infty}s(n)h(t; t-nT_s)+z(t), 
\end{equation}
where $T_s$ is the symbol period, $z(t)=h^{\textnormal{rec}}(t)\ast v(t)$ is the filtered noise, and $h(t; \tau)$ is the continuous-time baseband equivalent channel. The continuous-time baseband equivalent channel is obtained as the convolution of $h^{\textnormal{tr}}(t)$, $h^{\textnormal{ch}}(t; \tau)$, and $h^{\textnormal{rec}}(t)$ -- all linear filters -- that leads to the expression \cite[eq. (9.3)]{Giannakis_ST_BBWC_book_03}\cite[eq. (1)]{G_Leus_OMA_03}    
\begin{equation}
\label{DSFC_model_1}
h(t; \tau)\eqdef \int_{-\infty}^{\infty}\int_{-\infty}^{\infty} h^{\textnormal{rec}}(s)h^{\textnormal{tr}}(\tau-\theta-s)h^{\textnormal{ch}}(t-s; \theta) dsd\theta,  
\end{equation} 
where the time-varying impulse response of the physical wireless channel at the baseband can be expressed as a superposition of a cluster of rays whose overall aggregation can be written as \cite[eqs. (9.1) and (9.2)]{Giannakis_ST_BBWC_book_03} 
\begin{equation}
\label{TV_PHY_channel}
h^{\textnormal{ch}}(t;\tau)=\sum_{p}\sum_{m} \alpha_{m, p}(t)e^{\jmath[\phi_{m,p}(t)+2\pi f_{m,p}(t)t]}\delta(\tau-\tau_p(t)), 
\end{equation}
where $\tau_p(t)$ is the common delay of the $p$-th cluster (path) of rays arriving coherently at the same time, $\alpha_{m, p}(t)$ is the amplitude of the $m$-th ray in the $p$-th cluster with delay $\tau_p(t)$, $f_{m,p}(t)$ denotes a frequency offset (e.g., Doppler frequency and/or transmit-receive oscillator drifts), $\phi_{m,p}(t)\in[-\pi, \pi)$ is the phase shift of the $m$-th ray in the $p$-th cluster, and $\delta(\tau-\tau_p(t))=1$ iff $\tau=\tau_p(t)$. 
   
Taking into account how fast $h(t; \tau)$ varies in $t$ and $\tau$ relative to the symbol duration, the carrier frequency, and the coherence time of the channel, (\ref{DSFC_model_1}) yields simplifications that model different channels \cite{Giannakis_ST_BBWC_book_03}. Among the channels, a doubly selective fading channel is modeled by (\ref{DSFC_model_1}) when the physical channel variations are extremely fast to cause time and frequency selectivity (even when the delay spread is small) \cite{Giannakis_ST_BBWC_book_03}. To simplify such double selectivity, the following assumption is often employed \cite[A9.1]{Giannakis_ST_BBWC_book_03}.
\begin{assumption}
\label{Assump_receive_TV_filter}
During the span of the receive filter $h^{\textnormal{rec}}(t)$, the time-variant impulse response $h^{\textnormal{ch}}(t;\tau)$ remains invariant in $t$.  
\end{assumption}           
Note that Assumption \ref{Assump_receive_TV_filter} is practically plausible because the receive filter is typically shorter than the coherence time of the channel \cite{Giannakis_ST_BBWC_book_03}, especially for a broadband receive filters whose nonzero support is very short. Meanwhile, using Assumption \ref{Assump_receive_TV_filter}, (\ref{DSFC_model_1}) simplifies to \cite[eq. (9.4)]{Giannakis_ST_BBWC_book_03}
\begin{multline}
\label{DSFC_model_2}
h(t; \tau)= \int_{-\infty}^{\infty} h^{\textnormal{ch}}(t; \theta) \bigg( \int_{-\infty}^{\infty} h^{\textnormal{rec}}(s)h^{\textnormal{tr}}(\tau-\theta-s) ds \bigg) d\theta 
=\int_{-\infty}^{\infty} h^{\textnormal{ch}}(t; \theta) \psi(\tau-\theta) d\theta,    
\end{multline}
where $\psi(t)\eqdef h^{\textnormal{tr}}(t)\ast h^{\textnormal{rec}}(t)$ and $\psi(\tau-\theta) =\displaystyle\int_{-\infty}^{\infty} h^{\textnormal{rec}}(s)h^{\textnormal{tr}}(\tau-\theta-s) ds$. Substituting (\ref{TV_PHY_channel}) into (\ref{DSFC_model_2}) leads to the following continuous-time baseband equivalent channel \cite[eq. (9.5)]{Giannakis_ST_BBWC_book_03} 
\begin{equation}
\label{DSFC_model_3}
h(t; \tau)=\sum_{p} \psi(\tau-\tau_p(t)) \sum_{m} \alpha_{m, p}(t)e^{\jmath[\phi_{m,p}(t)+2\pi f_{m,p}(t)t]}. 
\end{equation} 
Concerning the symbol period, the amplitude, delay, phase, and frequency offset of each ray vary slowly as functions of time \cite{Giannakis_ST_BBWC_book_03}. Accordingly, the following assumption \cite[A9.2]{Giannakis_ST_BBWC_book_03} is justified.  
\begin{assumption}
\label{Invariance_mult_params}
For the sampling period $T_s$ seconds chosen to be equal to the symbol period (also denoted by $T_s$), $\tau_p(t)$, $\alpha_{m, p}(t)$, $\phi_{m,p}(t)$, and $f_{m,p}(t)$ remain invariant over $\tilde{N}$ symbol periods.     
\end{assumption}
Without loss of generality, consequently, we consider transmissions in bursts (blocks) of duration $[0, \tilde{N}T_s)$ and discard the time index $t$ from the RHS parameters of (\ref{DSFC_model_3}):  
\begin{equation}
\label{DSFC_model_4}
h(t; \tau)=\sum_{p} \psi(\tau-\tau_p) \sum_{m} \alpha_{m, p}e^{\jmath(\phi_{m,p}+2\pi f_{m,p}t)}, \hspace{2mm} \forall t\in [0, \tilde{N}T_s).  
\end{equation}          
If the wireless channel exhibits a rich scattering, the model in (\ref{DSFC_model_4}) could be irrelevant for an accurate channel estimation since the possible number of paths and rays can be astronomically large (theoretically infinite). To alleviate this complexity, deterministic BEMs \cite{GBG_BEMs_98,Giannakis_ST_BBWC_book_03} were derived -- by employing (\ref{DSFC_model_4}) -- for $h[n;l]\eqdef h(nT_s;lT_s)$ which is the time-$n$ response to an impulse applied at time $n-l$ \cite{P_Sch_LCE_OFDM_04,Proakis_5th_ed_08}. Meanwhile, BEMs are motivated by the observation that the temporal ($n$) variation of $h[n;l]$ is rather smooth due to the channel’s limited Doppler spread \cite{GMa_Fundam_TVC_2011}. A limited Doppler spread's effect can be captured by using a fixed set of functions, where BEMs can be seen as low-rank approximations of the channel's singular value decomposition (SVD) \cite{GMa_Fundam_TVC_2011}. In such approximations, the following assumptions -- leading to less complex models -- are used to derive BEMs \cite{G_Leus_OMA_03,IBar_TV_FIR_Eq_05,X_MA_MDT_03,X_Ma_Opt_train_03}. 
\begin{assumption}
\label{Assum_max_delay}
The maximum delay spread of $h^{\textnormal{ch}}(t; \tau)$ is bounded by $\tau_{\textnormal{max}}$.	
\end{assumption}	  
\begin{assumption}
\label{Assum_max_Doppler}
The maximum Doppler spread\footnote{The maximum Doppler frequency (spread) $f_{\textnormal{max}}$ is given by $f_{\textnormal{max}}=\frac{f_cv_{\textnormal{max}}}{c}$ \cite[p. 839]{Proakis_5th_ed_08} for $f_c$, $v_{\textnormal{max}}$, and $c$ being the carrier frequency, \textquotedblleft virtual" mobile velocity, and the speed of light in the medium of transmission, respectively. In case of a moving terminal, the \textquotedblleft virtual” velocity equals the terminal velocity; whereas in a moving scatterer, it would be twice the scatterer velocity \cite{ZT_PA_TVCE_07}.} of $h^{\textnormal{ch}}(t; \tau)$ is bounded by $f_{\textnormal{max}}$.	
\end{assumption}
Under Assumptions \ref{Assum_max_delay} and \ref{Assum_max_Doppler}, the deterministic doubly selective fading channel $h[n;l]$ can be modeled by the complex exponential BEM (CE-BEM)\footnote{For the sake of completeness, we shall mention that the following BEMs have also been proposed: polynomial BEM (P-BEM) \cite{Borah_P-BEM_99}, discrete Karhuen–Loève BEM (DKL-BEM) \cite{AFT96}, generalized CE-BEM (GCE-BEM) \cite{Leus_GCE-BEM_04}, and DPS-BEM \cite{Zemen_DPS_BEM_05}.} as \cite[eq. (1)]{X_Ma_Opt_train_03}   
\begin{equation}
\label{BEM_equation}
h[n;l] \eqdef \sum_{q=0}^{Q}  h_q[ \lfloor{n/\tilde{N}\rfloor}; l]e^{\jmath \frac{2\pi (q-Q/2)n}{\tilde{N}}},  
\end{equation}  
where $l \in [0, L]$; $\tilde{N}$ is the block size; and $L, Q\in\mathbb{Z}^+$ satisfy the following condition \cite{X_Ma_Opt_train_03}: 
\begin{condition}
\label{cond_L_Q}
$L\eqdef\floor*{ \tau_{\textnormal{max}}/T_s}$\footnote{If one considers a channel with a total of $L$ taps (i.e., $l\in[0, L-1]$), then $(L-1)T_s\leq \tau_{\textnormal{max}}\Leftrightarrow L=\ceil*{ \tau_{\textnormal{max}}/T_s}$ would be the maximum discrete-time delay \cite{GMa_Fundam_TVC_2011,WCom_TVC_11}. If one rather considers a total of $L+1$ channel taps (i.e., $l\in[0, L]$), then $LT_s\leq \tau_{\textnormal{max}}\Leftrightarrow L=\floor*{ \tau_{\textnormal{max}}/T_s}$ would be the maximum discrete-time delay.} and $Q\eqdef 2\lceil{f_{\textnormal{max}}\tilde{N}T_s \rceil}$.
\end{condition}
To make the channel estimation well-posed, meanwhile, the following constraint is presumed on the delay-Doppler spread factor that bounds the channel's degrees of freedom \cite{X_Ma_Opt_train_03}.
\begin{assumption}
\label{tau_max_and_f_max}
$\tau_{\textnormal{max}}$ and $f_{\textnormal{max}}$ are known and satisfy the constraint $2f_{\textnormal{max}}\tau_{\textnormal{max}}<1$. 		
\end{assumption}
Note that the constraint $2f_{\textnormal{max}}\tau_{\textnormal{max}}<1$ asserts an underspead channel \cite[p. 30]{WCom_TVC_11} which does not cause severe time and frequency dispersion. Moreover, for wireless channel manifesting a rich scattering and no dominant line-of-sight propagation, one can evoke the central limit theorem to validate the underneath assumption on the CE-BEM coefficients \cite{X_Ma_Opt_train_03}. 
\begin{assumption}
\label{CE_BEM_coefficients}
The CE-BEM coefficients $h_q[ \lfloor{n/\tilde{N}\rfloor}; l]$ are zero mean circularly symmetric complex Gaussian RVs with variance $\sigma_{q,l}^2$.		
\end{assumption}
In light of Assumption \ref{CE_BEM_coefficients} and the CE-BEM per (\ref{BEM_equation}), the following remarks are in order.
\begin{remark}
\label{CE_BEM_modeling_rem}
Pursuant to Assumption \ref{CE_BEM_coefficients} and (\ref{BEM_equation}), $(Q+1)(L+1)$ Gaussian distributed CE-BEM coefficients -- constant per a block -- would be multiplied with time-varying exponentials to model a doubly selective channel. Modeling a doubly selective channel using a CE-BEM, thus, needs $(Q+1)(L+1)<<N(L+1)$ parameters. 
\end{remark}    
\begin{remark}
\label{Rem_block_size}
When a doubly selective OFDM channel is modeled using the CE-BEM given by (\ref{BEM_equation}), a block duration $\tilde{N}T_s$ equals the OFDM symbol duration $NT_s$ plus the cyclic prefix (CP) duration $N_{cp}T_s$, i.e., $\tilde{N}=N+N_{cp}$.     
\end{remark}
Accommodating the CE-BEM described above, our system model follows. 

\section{System Model}
\label{sec: system_model}
\begin{figure*}[hbt!]
\centering  
\newcommand{\suma}{\Large\bm{$+$}}
\hspace*{-0.15cm}\begin{tikzpicture}[line width=0.6pt, node distance=0.5cm, auto]  
\node (a) {$d^i[k]$\hspace{-1mm} };  
\node[block2,right=of a] (b') {S/P};
\draw[line] (a)-- (b');
\node[block2_med,right=of b'] (b) {$\bm{F}_N^H$}; 
\draw[vecArrow] (b') to (b);
\node[block2_large,right=of b] (c') {$\bm{A}_{cp}$};
\draw[vecArrow] (b) to (c');
\node[block2,right=of c'] (c'') {P/S};
\draw[vecArrow] (c') to (c'');
\node[block3,right=of c''] (c) {$h^i[n;l]$}; 
\draw[line] (c'')-- (c);
\node[be,node distance=0.4cm, scale=0.4, right=of c] (d'){\suma};
\draw[line] (c)-- (d');
\node[block2,right=of d'] (d) {S/P};  
\draw[line] (d')-- (d);
\node[block2_large,right=of d] (e') {$\bm{D}_{cp}$};
\draw[vecArrow] (d)-- (e');
\node[block2,right=of e'] (e) {$\bm{F}_N$}; 
\draw[vecArrow] (e')-- (e);
\node[inner sep=0,minimum size=0,right of=e] (k) {}; 
\node[block4,right=of k] (g) {Detector};
\node[block,below=0.60cm of g] (k2) {Channel \\ Estimator};
\draw[vecArrow] (e)-- (g);
\draw[vecArrow] (k) |- (k2);
\draw[vecArrow] (k2)-- (g);

\node[block4_small, below = 0.95cm of d'](f) {Noise}; 
\draw[line] (f) -- (d');

\node[block2,right=of g] (h) {P/S};
\draw[vecArrow] (g)-- (h);
\node[node distance=0.4cm, right=of h] (a) {\hspace{-1mm}$\hat{d}^i[k]$};
\draw[line] (h)-- (a);

\end{tikzpicture}   \\ [2mm]
\caption{Baseband model of an OFDM transceiver: $\bm{A}_{cp}\eqdef \big[ [\bm{0}_{N_{cp} \times (N-N_{cp})}, \bm{I}_{N_{cp}} ]^{T}, \bm{I}_{N} \big]^{T}$ is a CP addition matrix; $\bm{D}_{cp}\eqdef \big[\bm{0}_{N\times N_{cp}}, \bm{I}_{N} \big]$ is a CP deletion matrix; and $i,k\in\mathbb{N}$.}
\label{OFDM_transceiver_fig}
\end{figure*}
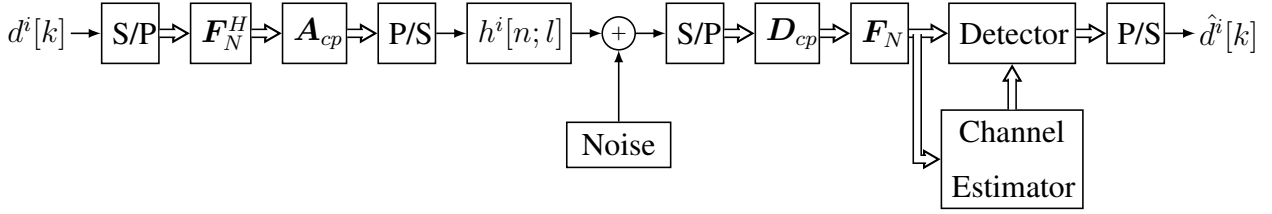

The baseband model of an OFDM transceiver is depicted in Fig. \ref{OFDM_transceiver_fig} which illustrates the chain of signal processing on the transmission and reception of an OFDM symbol.      
To transmit the $i$-th OFDM symbol, symbols $d^i[k]$ chosen from a finite constellation $\mathcal{Q}$ \cite{T_Cui_RD_OFDM_07} are collected to form a sequence $\big\{d^i[k]\big\}_{k=0}^{N-1}$. This sequence is fed to the serial-to-parallel (S/P) converter which yields $\bm{d}^i=\big[d^{i}[0], \ldots, d^{i}[N-1]\big]^{T}$. To modulate these symbols into $N$ orthogonal subcarriers\footnote{To transmit $P$ symbols over $N$ orthogonal subcarriers parallelly while avoiding aliasing, $N-P$ (virtual) subcarriers at the spectrum edges are not used in practice \cite{MM_CE_OFDM_01}. Similarly, $G$ (out of $N$) subcarriers are used at the center and edge as direct current and edge guard band, respectively \cite{ZZ_Learn_OFDM_18}.}, an $N$-point IDFT\footnote{An $N$-point (I)DFT is efficiently implemented via  (inverse)fast Fourier transforms ((I)FFT) \cite{P_Sch_LCE_OFDM_04}.} is applied to $\bm{d}^i$  to produce the $i$-th OFDM symbol $\bm{s}^i$ given by 
\begin{equation}
\label{vec_IFFT_operation}
\bm{s}^i={\bm{F}}_N^{H}\bm{d}^i, 
\end{equation}	
where $\bm{s}^{i}=\big[s^{i}[0], \ldots, s^{i}[N-1]\big]^T\in\mathbb{C}^{N}$ for $s^i[n]=1/\sqrt{N} \sum_{k=0}^{N-1} d^i[k] e^{\jmath \frac{2\pi kn}{N}}$. A CP is then prepended\footnote{The size of CP is chosen to be longer than the channel impulse response (maximum delay spread) \cite{P_Sch_LCE_OFDM_04,MM_CE_OFDM_01,T_Cui_RD_OFDM_07}. If $T$ is the duration of an OFDM symbol extended by CP, $T_s=T/(N+N_{cp})$ would be the signaling interval (sampling period) \cite{Book_OFDM_and_CDMA_05}.} to $\bm{s}^i$ -- to avoid an inter-OFDM symbol interference -- via an $(N+N_{cp})\times N$ CP addition matrix $\bm{A}_{cp}\eqdef \big[ [\bm{0}_{N_{cp} \times (N-N_{cp})}, \bm{I}_{N_{cp}} ]^{T}, \bm{I}_{N} \big]^{T}$ \cite{PC_CE_for_OFDM_13} which amounts to copying the last part of $\bm{s}^i$ and prepending it to the beginning of the extended $i$-th OFDM symbol $\tilde{\bm{s}}^i$ given by   
\begin{equation} 
\label{with_CP_vec_IFFT_operation}
\tilde{\bm{s}}^i= \bm{A}_{cp}{\bm{F}}_N^{H}\bm{d}^i, 
\end{equation}
where $\tilde{\bm{s}}^i=\big[ s^{i}[-N_{cp}], \ldots, s^{i}[-1], s^{i}[0], \ldots, s^{i}[N-1] \big]^{T}\in\mathbb{C}^{\tilde{N}}$ for $s^i[-N_{cp}+j]\eqdef s^{i}[N-N_{cp}+j]$ and $0\leq j < N_{cp}$. This extended $i$-th symbol is then transmitted -- after a parallel-to-serial (P/S) conversion -- via a doubly selective channel $h^{i}[n;l]$ modeled by the CE-BEM of (\ref{BEM_equation})\footnote{As stated in Assumptions \ref{Assum_max_delay}, \ref{Assum_max_Doppler}, and \ref{tau_max_and_f_max}, the CE-BEM of (\ref{BEM_equation}) requires knowledge of the upper bounds of the channel's delay spread and Doppler spread. Although we use (\ref{BEM_equation}) for the generation of a doubly selective fading channel matrix, our upcoming channel estimator does not require such knowledge.} as   
\begin{equation}  
\label{DT_CIR}
h^{i}[n;l] \eqdef h[i\tilde{N}+n; l] \hspace{2mm} \textnormal{for} \hspace{2mm} -N_{cp}\leq n < N, 
\end{equation}
where $h^{i}[n;l]=0$ for $l\not\in[0, L-1]$.\footnote{Without loss of generality, we consider a channel with $L$ taps. Thus, $(L-1)T_s\leq \tau_{\textnormal{max}}\Leftrightarrow L=\ceil*{ \tau_{\textnormal{max}}/T_s}$ \cite{GMa_Fundam_TVC_2011}.} Transmitted through this channel and contaminated by an additive white Gaussian noise (AWGN) $z^{i}[n]\sim \mathcal{CN}(0, \sigma^2)$, the received OFDM signal samples are given by  
\begin{equation}  
\label{Rec_sig_samples}
y^{i}[n]=\sum_{l=0}^{L-1} h^{i}[n;l]\tilde{s}^{i}[n-l]+z^{i}[n],      
\end{equation}
where $y^{i}[n]=y[i\tilde{N}+n]$ for $-N_{cp}\leq n <N$ and $y[n]$ being the received OFDM signal sample at the $n$-th sampling period; $\tilde{s}^{i}[n]\eqdef {(\tilde{\bm{s}}^i)}_{n}$ and $z^{i}[n]\eqdef {(\bm{z}^{i})}_{n}$ for $\bm{z}^{i}=\big[z^{i}[0], \ldots, z^{i}[\tilde{N}-1] \big]^T\in\mathbb{C}^{\tilde{N}}$; and $L$ propagation paths are considered in line with the modeling of \cite{P_Sch_LCE_OFDM_04,PC_CE_for_OFDM_13,T_Cui_RD_OFDM_07}.

Once $\tilde{N}$ received OFDM signal samples are produced, the S/P converter transforms $\big\{y^{i}[n]\big\}_{n=0}^{\tilde{N}-1}$ into $\bm{y}^{i}=\big[y^{i}[0], \ldots, y^{i}[N+N_{cp}-1] \big]^{T} \in\mathbb{C}^{(N+N_{cp})}$. Thereafter, $\bm{y}^i$ is transformed by an $N \times \tilde{N}$ CP deletion matrix $\bm{D}_{cp}\eqdef \big[\bm{0}_{N\times N_{cp}}, \bm{I}_{N} \big]$ \cite{PC_CE_for_OFDM_13} followed by an $N$-point DFT matrix $\bm{F}_N$ to yield the $i$-th post-DFT received OFDM signal vector $\tilde{\bm{y}}^i$ given by        
\begin{subequations}
\begin{align}
\label{DFT_received_samples_2_1}
\tilde{\bm{y}}^i&=\bm{F}_N\bm{D}_{cp}\bm{y}^i	   \\
\label{DFT_received_samples_2}
&=\bm{F}_N\bm{H}^i\bm{F}_N^H\bm{d}^{i}+\bm{F}_N\bm{D}_{cp}\bm{z}^{i}, 	     
\end{align} 
\end{subequations}
where $\tilde{\bm{y}}^i=\big[\tilde{y}^i[0], \ldots, \tilde{y}^i[N-1] \big]^T\in\mathbb{C}^N$ and $\bm{H}^i$ stands for the (time-domain) doubly selective fading channel matrix -- for the $i$-th OFDM symbol -- including the effect of CP insertion and removal. Specifically, $\bm{H}^i$ is a \textit{pseudo-circulant matrix} \cite{ZT_PA_TVCE_07,PC_CE_for_OFDM_13} and is defined via \cite{P_Sch_LCE_OFDM_04} 
\begin{equation}
\label{H_T_time_dom}
(\bm{H}^{i})_{n,l} \eqdef h^{i}[n; \langle n-l \rangle_{N}], \hspace{2mm} n, l\in[0, N-1],  
\end{equation}
where $h^{i}[n;l]=0$ for $l\not\in[0, L-1]$. Employing the $i$-th subcarrier coupling matrix defined as
\begin{equation}
\label{H_sc_matrix}
\tilde{\bm{H}}^{i}\eqdef\bm{F}_N\bm{H}^{i}\bm{F}_N^H,
\end{equation}
and $\tilde{\bm{z}}^{i}\eqdef \bm{F}_N\bm{D}_{cp}\bm{z}^i\sim \mathcal{CN}(\bm{0}, \sigma^2\bm{I}_N)$ (since $\bm{F}_N$ is a unitary matrix), (\ref{DFT_received_samples_2}) can be expressed as 
\begin{equation}
\label{DFT_received_samples_3} 
\tilde{\bm{y}}^{i}=\tilde{\bm{H}}^{i}\bm{d}^{i}+\tilde{\bm{z}}^{i}, 
\end{equation}
where $(\tilde{\bm{H}}^i)_{p+d,p} \eqdef\frac{1}{N}\sum_{n=0}^{N-1}\sum_{l=0}^{L-1}h^{i}[n; l]e^{-j\frac{2\pi(pl+dn)}{N}}$ \cite[eq. (6)]{PC_CE_for_OFDM_13}. Employing the model in (\ref{DFT_received_samples_3}), we follow with the operations of our DL-based blind OFDM channel estimation.    
               
\section{Blind OFDM Channel Estimation: A Deep Learning Algorithm}
\label{sec: DL_architect_and_motivation}
Since $h^{i}[n, l]$ varies over $n$ and $l$, the pseudo-circulant matrix $\bm{H}^i$ -- defined via (\ref{H_T_time_dom}) -- can not be diagonalized by (post- and) pre-multiplication by an (I)DFT matrix. Accordingly, (\ref{DFT_received_samples_3}) would manifest severe ICI that can complicate channel estimation and symbol detection, especially at low SNR. To mitigate a severe ICI, the existing ICI mitigation techniques start with channel estimation which tends to rely on many pilot symbols which would result in a spectrally inefficient transmission. To overcome such an inefficiency, we herein propose a DL-based blind channel estimation algorithm whose architectural motivation is provided below.    

\subsection{Architectural Motivation}
\label{subsec: DL_architecture}
If $\bm{d}^{i}$ were known, it follows from (\ref{DFT_received_samples_3}) that        
\begin{equation}
\label{DFT_samples_manip_1} 
\tilde{\bm{y}}^{i}(\bm{d}^{i})^H=\tilde{\bm{H}}^{i}\bm{D}^{i}+\tilde{\bm{z}}^{i}(\bm{d}^{i})^H,
\end{equation} 
where $\bm{D}^{i}=\bm{d}^{i}(\bm{d}^{i})^H$. Note that $\bm{D}^{i}$ would have a full rank provided that none of its two columns is identical. In this regard, as the probability of a given symbol being chosen is $|\mathcal{Q}|^{-1}$, the probability of any two columns of $\bm{D}^{i}$ being the same is $|\mathcal{Q}|^{-N}$. This probability is almost zero since $N$ is often large. As a result, $\bm{D}^{i}$ would have a full rank and hence invertible with a very high probability. To this end,   
\begin{equation}
\label{DFT_samples_manip_2} 
\tilde{\bm{y}}^{i}(\bm{d}^{i})^H(\bm{D}^{i})^{-1}=\tilde{\bm{H}}^{i}+\tilde{\bm{z}}^{i}(\bm{d}^{i})^H(\bm{D}^{i})^{-1}.
\end{equation}
If we let $\tilde{\bm{w}}^{i}\eqdef(\bm{d}^{i})^H(\bm{D}^{i})^{-1}\in\mathbb{C}^{1\times N}$, (\ref{DFT_samples_manip_2}) reduces to 
\begin{equation}
\label{DFT_samples_manip_3} 
\tilde{\bm{y}}^{i}\tilde{\bm{w}}^{i}=\tilde{\bm{H}}^{i}+\tilde{\bm{z}}^{i}\tilde{\bm{w}}^{i}.
\end{equation}
Should we then assume no noise, it can be deduced from (\ref{DFT_samples_manip_3}) that   
\begin{equation}
\label{DFT_samples_manip_4} 
\tilde{\bm{H}}^{i}=\tilde{\bm{y}}^{i}\tilde{\bm{w}}^{i}.  
\end{equation} 
Using the identities $\textnormal{Re}\{ab\}=\textnormal{Re}\{a\}\textnormal{Re}\{b\}-\textnormal{Im}\{a\}\textnormal{Im}\{b\}$, $\textnormal{Im}\{ab\}=\textnormal{Im}\{a\}\textnormal{Re}\{b\}+\textnormal{Re}\{a\}\textnormal{Im}\{b\}$, and $\textnormal{vec}(\bm{ABC})=(\bm{C}^T\otimes \bm{A})\textnormal{vec}(\bm{B})$ \cite[eq. (5), p. 5]{Magnus_07}, (\ref{DFT_samples_manip_4}) can also be equated as  
\begin{equation}
\label{DFT_samples_manip_5}
\textnormal{vec}\Bigg(\begin{bmatrix}
[\textnormal{Re}\{\tilde{\bm{H}}^{i}\}]^T   \\
[\textnormal{Im}\{\tilde{\bm{H}}^{i}\}]^T 
\end{bmatrix}^T\Bigg)=\tilde{\bm{W}}^{i}\begin{bmatrix}
\textnormal{Re}\{\tilde{\bm{y}}^{i}\}   \\
\textnormal{Im}\{\tilde{\bm{y}}^{i}\} 
\end{bmatrix}, 
\end{equation}
where 
\begin{equation}
\label{constraints}
\begin{aligned}
\tilde{\bm{W}}^i&= [\tilde{\bm{W}}_1^i \hspace{2mm} \tilde{\bm{W}}_2^i]\in\mathbb{R}^{2N^2\times 2N}  \\
\tilde{\bm{W}}_1^i & = \big[ \big[ \textnormal{Re}\{\tilde{\bm{w}}^i\} \hspace{2mm} \textnormal{Im}\{\tilde{\bm{w}}^i\} \big] \otimes \bm{I}_N \big]^T\in\mathbb{R}^{2N^2\times N}  \\ 
\tilde{\bm{W}}_2^i & = \big[ \big[ -\textnormal{Im}\{\tilde{\bm{w}}^i\} \hspace{2mm} \textnormal{Re}\{\tilde{\bm{w}}^i\} \big] \otimes \bm{I}_N \big]^T\in\mathbb{R}^{2N^2\times N}.
\end{aligned}
\end{equation}

If the transmitted symbols were known at the receiver and no AWGN were contaminating them, the exact channel matrix estimates would be characterized by (\ref{DFT_samples_manip_4})-(\ref{constraints}). Nevertheless, concerning our goal of devising a blind channel estimation technique, we do not know $\bm{d}^{i}$ (and hence $\tilde{\bm{w}}^{i}$). Besides, AWGN is inevitable. Consequently, (\ref{DFT_samples_manip_4})-(\ref{constraints}) would be infeasible. If we now assume a very high SNR scenario, the following condition follows from (\ref{DFT_samples_manip_3}):   
\begin{equation}
\label{High_SNR_condition}
\|\tilde{\bm{z}}^{i}\tilde{\bm{w}}^i\|_2^2=\| \tilde{\bm{y}}^{i}\tilde{\bm{w}}^i-\tilde{\bm{H}}^{i} \|_2^2 \to 0.
\end{equation}
In the context of (\ref{High_SNR_condition}), assuming the availability of channel labels contained in $\tilde{\bm{H}}^{i}$ that comprises a complex dataset $\tilde{\mathcal{S}}\eqdef\big\{ (\tilde{\bm{y}}^{i}, \tilde{\bm{H}}^{i} ) \big\}_{i\in [\tilde{n}]}$ for $\tilde{n}\eqdef |\tilde{\mathcal{S}}|$, a model-based channel estimation solution -- valid for high SNR regimes -- can then be cast via (\ref{DFT_received_samples_3}), (\ref{DFT_samples_manip_5}), (\ref{constraints}), and (\ref{High_SNR_condition}) as 
\begin{equation}
\label{est_eqn_1}
\textnormal{vec}\Bigg(\begin{bmatrix}
[\textnormal{Re}\{\hat{\tilde{\bm{H}}}^{i}\}]^T   \\
[\textnormal{Im}\{\hat{\tilde{\bm{H}}}^{i}\}]^T 
\end{bmatrix}^T\Bigg)=\bm{W}^{*}\begin{bmatrix}
\textnormal{Re}\{\tilde{\bm{y}}^{i}\}   \\
\textnormal{Im}\{\tilde{\bm{y}}^{i}\} 
\end{bmatrix}, 
\end{equation}
where $\hat{\tilde{\bm{H}}}^{i}$ denotes the estimated subcarrier coupling matrix for $i\geq\tilde{n}+1$ and $\bm{W}^{*}\in\mathbb{R}^{2N^2\times 2N}$ represents an optimal weight matrix given by  
\begin{equation}
\label{Bl_est_optimization}
\begin{aligned}
\bm{W}^{*}=\displaystyle \argmin_{\{\bm{W}^i\}_{i\in [\tilde{n}]}} \quad & \frac{1}{2} \sum_{i=1}^{\tilde{n}}\bigg\|\bm{W}^i\begin{bmatrix}
\textnormal{Re}\{\tilde{\bm{y}}^{i}\}   \\
\textnormal{Im}\{\tilde{\bm{y}}^{i}\} 
\end{bmatrix}- 
\textnormal{vec}\bigg(\begin{bmatrix}
[\textnormal{Re}\{\tilde{\bm{H}}^{i}\}]^T   \\
[\textnormal{Im}\{\tilde{\bm{H}}^{i}\}]^T 
\end{bmatrix}^T\bigg)\bigg\|_2^2  \\ 
\textrm{s.t.} \quad & \tilde{\bm{y}}^{i}=\tilde{\bm{H}}^{i}\bm{d}^{i}+\tilde{\bm{z}}^{i}    \\ 
\bm{W}^i&= [\bm{W}_1^i \hspace{2mm} \bm{W}_2^i]\in\mathbb{R}^{2N^2\times 2N}  \\
\bm{W}_1^i & = \big[ \big[ \textnormal{Re}\{\bm{w}^i\} \hspace{2mm} \textnormal{Im}\{\bm{w}^i\} \big] \otimes \bm{I}_N \big]^T\in\mathbb{R}^{2N^2\times N}  \\ 
\bm{W}_2^i & = \big[ \big[ -\textnormal{Im}\{\bm{w}^i\} \hspace{2mm} \textnormal{Re}\{\bm{w}^i\} \big] \otimes \bm{I}_N \big]^T\in\mathbb{R}^{2N^2\times N},  
\end{aligned}
\end{equation}
where the square loss is chosen as our loss function. However, the non-convex optimization problem given by (\ref{Bl_est_optimization}) is known to be an NP-hard problem. To overcome this problem, deep FNNs can be trained and deployed in regard to the blind OFDM channel estimation algorithm detailed below. 

\subsection{Blind OFDM Channel Estimation Algorithm}
\label{subsec: blind_estimation_algorithm}
In accordance with the above-presented motivation, it is shown in \cite{Getu_error_bounds_20} that the affine transformation $\bm{a}^i=\bm{W}^i[\textnormal{Re}\{\tilde{\bm{y}}^i\}^T \hspace{2mm}   \textnormal{Im}\{\tilde{\bm{y}}^i\}^T]^T$ can be represented by a $K$-layer ReLU FNN for $K\geq2$ \cite[Proposition 1]{Getu_error_bounds_20}. Thus, (\ref{Bl_est_optimization}) can be cast as minimizing the overall error produced by the representation of a $K$-layer ReLU FNN fed with $\{[\textnormal{Re}\{\tilde{\bm{y}}^i\}^T \hspace{2mm} \textnormal{Im}\{\tilde{\bm{y}}^i\}^T]^T\}_{i=1}^{\tilde{n}}$ and trained via a supervision w.r.t. the channel labels $\{\textnormal{vec}\big([\textnormal{Re}\{\tilde{\bm{H}}^{i}\} \hspace{2mm}  \textnormal{Im}\{\tilde{\bm{H}}^i\}]\big)\}_{i=1}^{\tilde{n}}$. In this regard, the blind OFDM channel estimation problem boils down to online learning \cite[p. 128]{Haykin_NNs_09} problem addressed using the back-propagation algorithm (BackProp) \cite[p. 139-141]{Haykin_NNs_09} that employs stochastic optimizers such as stochastic gradient descent (\texttt{SGD}). Thus, the weights of a $K$-layer ReLU FNN will be updated via BackProp for the presentation of every training pair until convergence.\footnote{BackProp is considered to have converged when one of the following criteria is satisfied: 1) \textquotedblleft when the Euclidean norm of the gradient vector reaches a sufficiently small gradient threshold''; 2) \textquotedblleft when the absolute rate of change in the average squared error per epoch is sufficiently small'' \cite[p. 139]{Haykin_NNs_09}.}

Despite its simplicity and high model update frequency that can sometimes render effective solutions to large-scale pattern classification problems (by tracking small changes in the training data) \cite{Haykin_NNs_09}, online learning (also \texttt{SGD}\footnote{As a well-known optimizer frequently employed by several DL papers, \texttt{Adam} \cite{Adam_ICLR_15} can be used -- instead of \texttt{SGD} -- as an optimizer during the training of our studied blind OFDM channel estimator based on a ReLU FNN.} with a mini-batch size of one) suffers from computational complexity. Online learning's noisy learning process can make it hard for the learning model to settle on good local minima for a given model. To alleviate these downsides, mini-batch learning which updates the ReLU FNN weights once per a mini-batch size is often preferred in practice. Specifically, a practitioner can set the mini-batch size as a hyperparameter of a learning algorithm and optimizes it while training a DL model. Concerning model training followed by testing, our ReLU FNN based blind channel estimation algorithm comprises a testing phase preceded by a training phase. During a training phase, feedforward and backward computations are carried out. Toward the latter end, the local gradient computations are outlined in Algorithm \ref{Blind_DSCE_Alg} per the BackProp summarized in \cite[p. 139-141]{Haykin_NNs_09}. Once a BackProp converges heralding the completion of a ReLU FNN training, forward computations via Algorithm \ref{Blind_DSCE_Alg_cont} will be performed to execute blind channel estimation using the received OFDM signal samples as the only inputs.

In compliance with the aforementioned learning problem formulation and ReLU FNN based blind channel estimation, the channel labels and the network inputs are derived from the subcarrier coupling matrix and the received OFDM signal samples given by (\ref{H_sc_matrix}) and (\ref{DFT_received_samples_3}), respectively. Accordingly, the training set $\mathcal{D}\eqdef \big\{(\bm{x}_i, \bm{y}_i^{*})\big\}_{i=1}^{\breve{n}}$ is made of   
\begin{subequations}
\begin{align}
\label{training_input_i}
\bm{x}_i&\eqdef\big[\Xi_i \textnormal{Re}\{\tilde{\bm{y}}^{i}\}^T \hspace{2mm} \Xi_i\textnormal{Im}\{\tilde{\bm{y}}^{i}\}^T \hspace{2mm} 1/\sqrt{2}\big]^T \in \mathbb{R}^{d_x}  \\
\label{training_label_vector_1} 
\bm{y}_i^{*}&\eqdef\textnormal{vec}\big([ \textnormal{Re}\{\tilde{\bm{H}}^{i} \} \hspace{2mm}   \textnormal{Im}\{ \tilde{\bm{H}}^{i} \} ] \big) \in \mathbb{R}^{d_y},
\end{align}
\end{subequations}	
where $(\bm{x}_i)_{2N+1}=1/\sqrt{2}$ is a bias term\footnote{Without loss of generality, introducing $(\bm{x}_i)_{2N+1}=1/\sqrt{2}$ is equivalent to introducing a (random) bias term \cite{allenzhu2018convergence}. This specific constant does not alter nor affect our upcoming analyses.},  $\Xi_i\eqdef\Big(\sqrt{2\big[\|\textnormal{Re}\{\tilde{\bm{y}}^{i}\}\|^2+\|\textnormal{Im}\{\tilde{\bm{y}}^{i}\}\|^2\big]}\Big)^{-1}$ is a normalization factor that would make $\|\bm{x}_i\|=1$, $d_x=2N+1$, and $d_y=2N^2$. $\mathcal{D}$ will then be split into a training set and a validation set consistent with customary DL training practices \cite{Bengio_Prac_Recomm_2012,Glorot_Understanding_10}. Moreover, the testing set $\mathcal{T}\eqdef \{(\check{\bm{x}}_i, \check{\bm{y}}_i^{*}) \}_{i=\breve{n}+1}^{\breve{n}+\check{n}}$ would consist of        
\begin{subequations}
	\begin{align}
	\label{check_x_i}
	\check{\bm{x}}_i&\eqdef\big[\check{\Xi}_i \textnormal{Re}\{\tilde{\bm{y}}^{i}\}^T \hspace{2mm} \check{\Xi}_i\textnormal{Im}\{\tilde{\bm{y}}^{i}\}^T \hspace{2mm} 1/\sqrt{2}\big]^T \in \mathbb{R}^{d_x}  \\
	\label{test_label_vector_1} 
	\check{\bm{y}}_i^{*}&\eqdef\textnormal{vec}\big([ \textnormal{Re}\{\tilde{\bm{H}}^{i} \} \hspace{2mm} \textnormal{Im}\{ \tilde{\bm{H}}^{i} \} ] \big) \in\mathbb{R}^{d_y},
	\end{align}
\end{subequations}
where $(\check{\bm{x}}_i)_{2N+1}=1/\sqrt{2}$, $\check{\Xi}_i=\Big(\sqrt{2\big[\|\textnormal{Re}\{\tilde{\bm{y}}^{i}\}\|^2+\|\textnormal{Im}\{\tilde{\bm{y}}^{i}\}\|^2\big]}\Big)^{-1}$ is a normalization factor that would set $\|\check{\bm{x}}_i\|=1$, $d_x=2N+1$, $d_y=2N^2$, and $\check{n}\eqdef|\mathcal{T}|$.

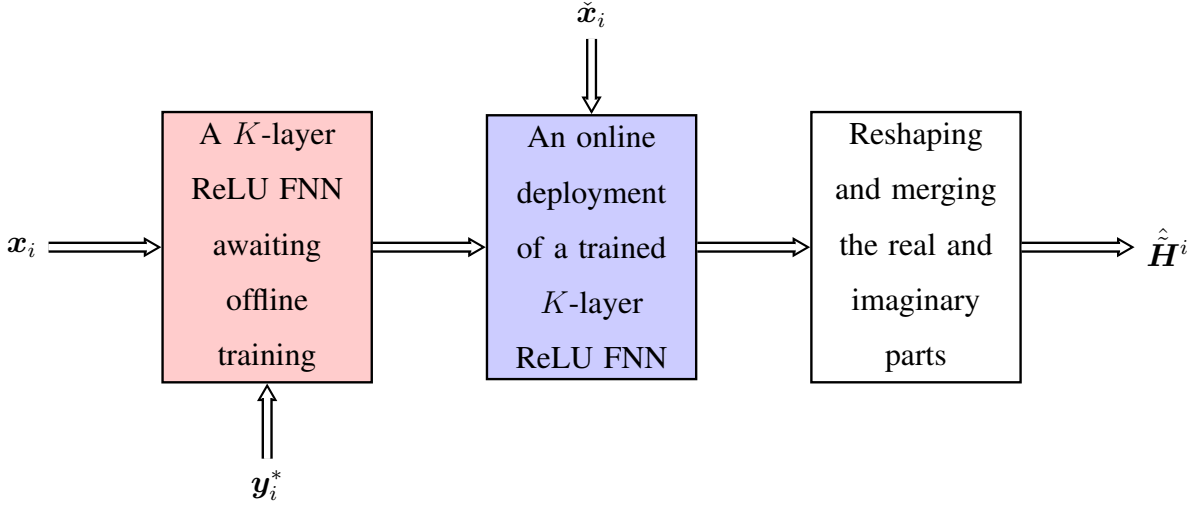
\begin{figure}
	\centering
	\begin{tikzpicture}[line width=0.9pt]   
	\node (a) {$\bm{x}_i$};  
	\node[blockChannel,right=1.5cm of a,fill=red!20] (b) {A $K$-layer ReLU FNN awaiting offline training};
	\draw[vecArrow] (a)-- (b);
	\node[below = 1cm of b](b') {$\bm{y}_i^{*}$}; 
	\draw[vecArrow] (b') -- (b);
	\node[blockChannel,right=1.5cm of b,fill=blue!20] (c') {An online deployment of a trained $K$-layer ReLU FNN};
	\draw[vecArrow] (b) to (c');
	%
	
	\node[above=1cm of c'] (a1) {$\check{\bm{x}}_i$}; 
	\draw[vecArrow] (a1) -- (c');

	\node[blockChannel,right=1.5cm of c'] (c) {Reshaping and merging the real and imaginary parts}; 
	\draw[vecArrow] (c') to (c);
	\node[right=1.5cm of c] (c1) {$\hat{\tilde{\bm{H}}}^i$}; 
	\draw[vecArrow] (c) to (c1);	
	\end{tikzpicture}   \\ [2mm]
	\caption{A schematic on the proposed blind OFDM channel estimator: $\hat{\tilde{\bm{H}}}^i$ is the estimate of $\tilde{\bm{H}}^i$.}
	\label{proposed_blind_channel_estimator}
\end{figure}

\begin{algorithm}
	\SetKwData{Left}{left}
	\SetKwData{This}{this}
	\SetKwData{Up}{up}
	\SetKwFunction{Union}{Union}
	\SetKwFunction{FindCompress}{FindCompress}
	\SetKwInOut{Input}{Input}
	\SetKwInOut{Output}{Output}
	\caption{DL-Based Blind Doubly Selective OFDM Channel Estimation (Training)}
	\label{Blind_DSCE_Alg}
	\Input{ $\mathcal{D}=\{(\bm{x}_i, \bm{y}_i^{*})\}_{i=1}^{\breve{n}}$}
	\Output{set of trained network weights $\big\{ \bm{W}_k^{(T)}\big\}_{k\in [K]}$}
	\BlankLine
	\emph{Compiled $K$-layer ReLU FNN with \texttt{SGD with momentum} $\big($neurons of the $j$-th layer $N_j$; mini-batch size $B$; learning rate $\eta$; momentum constant $\alpha$; and initialized weights $\big\{ \bm{W}_k^{(0)}\big\}_{k\in [K]}\big)$ }\\
	\emph{Start training: \emph{$t=0$}} \\
	\While{until convergence}{
		$c\leftarrow \textnormal{choose randomly } i\in[\breve{n}/B]$\;
		\For{$b\leftarrow 1$ \KwTo $B$}{\nllabel{forins}
			$\bm{x}_0\leftarrow \bm{x}_{(c-1)B+b}$\;
			\For{$k\leftarrow 1$ \KwTo $K$}{\nllabel{forins}	
				$\bm{v}_k^{(t)}\leftarrow \bm{W}_k^{(t)}\bm{x}_{k-1}      $\;
				$\bm{x}_k^{(t)}\leftarrow \rho \big(\bm{v}_k^{(t)}\big) $\;
				
			}
			$ \bm{e}_b^{(t)}\leftarrow \bm{y}_{(c-1)B+b}^{*} -\bm{x}_K^{(t)}$\;
		}
		\emph{ $\bm{e}^{(t)} \leftarrow \frac{1}{B}\sum_{b=1}^B \bm{e}_b^{(t)}$}\;
		\For{$l\leftarrow K$ \KwTo $1$}{\nllabel{forins}
			\For{$j\leftarrow 1$ \KwTo $N_j$}{\nllabel{forins}	
				\lIf{$l=K$}{\\ $\big(\bm{\delta}_l^{(t)}\big)_{j}\leftarrow \big(\bm{e}^{(t)}\big)_j\rho'\big( \big(\bm{v}_K^{(t)}\big)_{j} \big)$}
				\lElse{\\ $\big(\bm{\delta}_l^{(t)}\big)_{j}\leftarrow \rho'\big( \big(\bm{v}_l^{(t)}\big)_{j} \big) $  $\sum_{k=1}^{N_{j+1}} \big(\bm{\delta}_{l+1}^{(t)}\big)_{k} \big(\bm{W}_{l+1}^{(t)}\big)_{k,j}$}
			}
			$\big(\bm{W}_l^{(t+1)}\big)_{j,i} \leftarrow \big(\bm{W}_l^{(t)}\big)_{j,i}+ \alpha \big(\Delta\bm{W}_l^{(t-1)}\big)_{j,i} + \eta\big(\bm{\delta}_l^{(t)}\big)_{j} \big(\bm{x}_{l-1}^{(t)}\big)_{i} $
		}
		$t\leftarrow t+1$;	
	}
\emph{$\big\{ \bm{W}_k^{(T)}\big\}_{k\in [K]} \leftarrow \big\{ \bm{W}_k^{(t)}\big\}_{k\in [K]}$}
	
\end{algorithm}
\begin{algorithm}
	\SetKwData{Left}{left}
	\SetKwData{This}{this}
	\SetKwData{Up}{up}
	\SetKwFunction{Union}{Union}
	\SetKwFunction{FindCompress}{FindCompress}
	\SetKwInOut{Input}{Input}
	\SetKwInOut{Output}{Output}
	\caption{DL-Based Blind Doubly Selective OFDM Channel Estimation (Testing)}
	\label{Blind_DSCE_Alg_cont}
	\Input{set of trained network weights $ \big\{ \bm{W}_k^{(T)}\big\}_{k\in [K]}$ and $\mathcal{T}= \{(\check{\bm{x}}_i, \check{\bm{y}}_i^{*}) \}_{i=\breve{n}+1}^{\breve{n}+\check{n}}$ $\big($$\{\check{\bm{y}}_i^{*}\}_{i=\breve{n}+1}^{\breve{n}+\check{n}}$ is used only for MSE performance assessment$\big)$}
	\Output{set of estimated OFDM channel matrices $\{\hat{\tilde{\bm{H}}}^i\}_{i=\breve{n}+1}^{\breve{n}+\check{n}}$ and MSE}
	\BlankLine
	\emph{Learned weights: $\check{\bm{W}}_k\leftarrow\bm{W}_k^{(T)}$, $\forall k\in [K]$}\;
	\For{$i\leftarrow \breve{n}+1$ \KwTo $\breve{n}+\check{n}$}{
		$\bm{x}_0\leftarrow \check{\bm{x}}_i$\;
		\For{$k\leftarrow 1$ \KwTo $K$}{\nllabel{forins}
			$\bm{v}_k\leftarrow \check{\bm{W}}_k\bm{x}_{k-1} $\;
			$\bm{x}_k\leftarrow \rho (\bm{v}_k) $\;
			
		}
		$\bm{e}_i \leftarrow \bm{x}_K-\check{\bm{y}}_i^{*}$\;
		$\textnormal{Re}\big\{\hat{\tilde{\bm{H}}}^i\big\} \leftarrow \texttt{reshape}\big(\bm{x}_K(1:N^2), N, N\big)$\;
		$\textnormal{Im}\big\{\hat{\tilde{\bm{H}}}^i\big\} \leftarrow \texttt{reshape}\big(\bm{x}_K(N^2+1:2N^2), N, N\big)$\;
		$\hat{\tilde{\bm{H}}}^i\leftarrow \textnormal{Re}\big\{\hat{\tilde{\bm{H}}}^i\big\}+\jmath \textnormal{Im}\big\{\hat{\tilde{\bm{H}}}^i\big\}$\;	
	}
	$\textnormal{MSE} \leftarrow \frac{1}{2\check{n}N^2} \sum_{i=\breve{n}+1}^{\breve{n}+\check{n}} \| \bm{e}_i \|^2$ 	
\end{algorithm}

Using the training set $\mathcal{D}$ generated via (\ref{training_input_i})-(\ref{training_label_vector_1}) and the testing inputs generated via (\ref{check_x_i}), a schematic of the proposed blind channel estimator is shown in Fig. \ref{proposed_blind_channel_estimator}. Concerning this blind estimator, Algorithm \ref{Blind_DSCE_Alg} outlines the offline training algorithm -- for the proposed DL-based blind channel estimation -- using mini-batch learning, a square loss, BackProp, and the generalized delta rule\footnote{Avoiding the danger of instability, the generalized delta rule is a simple method of increasing the learning rate by including a momentum term \cite[p. 137]{Haykin_NNs_09}.} \cite[eq. (4.41)]{Haykin_NNs_09}. Algorithm \ref{Blind_DSCE_Alg_cont} then executes the online deployment of the trained deep ReLU FNN whose estimation performance is quantified in terms of the MSE computed in the last routine of Algorithm \ref{Blind_DSCE_Alg_cont}.     

In light of Algorithms \ref{Blind_DSCE_Alg} and \ref{Blind_DSCE_Alg_cont}, the following remark is in order.
\begin{remark}
Since we employ known channel statistics as labels for the training of ReLU FNNs as in Algorithm \ref{Blind_DSCE_Alg}, our proposed OFDM channel estimator may not be strictly blind (though we deploy no pilot symbols) during training. However, this estimator is strictly blind during testing since it does not need any pilot symbols nor knowledge of the channel statistics.
\end{remark} 
In line with this remark, our developed theory on the testing MSE performance of the detailed blind OFDM channel estimation algorithm ensues.
 
\section{Theory on the MSE Performance of a DL-Based Blind Channel Estimator}
\label{sec: Per_Over_Param_ReLU_FNNs}
We hereinafter develop a theory on the testing MSE performance of the proposed blind channel estimator -- for a doubly selective fading OFDM channel -- based on an over-parameterized ReLU FNN. For an over-parameterized ReLU FNN-based blind OFDM channel estimator, we are going to quantify its testing MSE performance by deriving theorems on its asymptotic and non-asymptotic testing MSE performance. Toward this performance quantification pertaining to a testing phase following the training of an over-parameterized ReLU FNN using \texttt{SGD}, we consider the convergence of \texttt{SGD} per \cite[Theorem 2]{allenzhu2018convergence}. Meanwhile, we begin the entire theoretical development by stating our assumptions and definitions. 
   
\subsection{Assumptions and Definitions}
\label{subsec: assumptions_and_definitions}
Concerning the performance quantification of an over-parameterized ReLU FNN-based blind OFDM channel estimator, we deploy the underneath assumptions and definitions. 
\begin{assumption}
\label{CE_BEM_coeffcients_normalization}	
The variances of the CE-BEM Gaussian coefficients are normalized to satisfy the constraint $\sum_{l=0}^{L-1} \sum_{q=0}^Q \sigma_{q,l}^2=1$.	
\end{assumption}
Assumption \ref{CE_BEM_coeffcients_normalization} ascertains the generation constraint pertaining to the CE-BEM coefficients used to model the considered doubly selective OFDM channel. Following the stated assumption on this channel's modeling constraint, we proceed to an assumption on the inputs to a ReLU FNN.      
\begin{assumption}
\label{normalization_assumption}
Without loss of generality, the training and testing inputs are assumed to be normalized so that $\|\bm{x}_i\|=1$ and $\|\check{\bm{x}}_j\|=1$ for all $i \in [\breve{n}]$ and all $j\in \{\breve{n}+1, \ldots, \breve{n}+\check{n} \}$.
\end{assumption}
Assumption \ref{normalization_assumption} is already satisfied by our training inputs [see (\ref{training_input_i})] and testing inputs [see (\ref{check_x_i})]. We then make the beneath non-degeneracy (or separateness) assumption.   
\begin{assumption}
\label{non-degeneracy_assumption}
For every $i, j\in [\breve{n}]$ and $\check{i}, \check{j}\in \{\breve{n}+1, \ldots, \breve{n}+\check{n} \}$, we presume that $\|\bm{x}_i-\bm{x}_j\|, \|\check{\bm{x}}_{\check{i}}-\check{\bm{x}}_{\check{j}}\| \geq \delta$ for some $\delta>0$.   
\end{assumption}
Regarding the training set made with (\ref{training_input_i})-(\ref{training_label_vector_1}), the testing set comprised with (\ref{check_x_i})-(\ref{test_label_vector_1}), and the received OFDM signal model given by (\ref{DFT_received_samples_3}), Assumption \ref{non-degeneracy_assumption} is satisfied with a high probability, especially, for large $N$ due to the inevitability of AWGN making any two received OFDM signal samples different.        

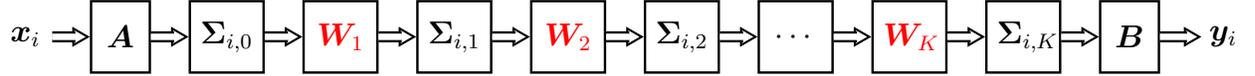
\begin{figure*}[htb!]
	\centering
	\hspace{-0.2mm}\begin{tikzpicture}[line width=0.9pt,node distance=0.55cm, auto]   
		\node (a) {$\bm{x}_i$\hspace{-2mm}};  
		\node[blockA,right=of a] (b) {$\bm{A}$};
		\draw[vecArrow] (a)-- (b);
		\node[blockW,right=of b] (c') {$\bm{\Sigma}_{i,0}$};
		\draw[vecArrow] (b) to (c');
		\node[blockW,right=of c'] (c) {$\textcolor{red}{\bm{W}_1}$}; 
		\draw[vecArrow] (c') to (c);  
		\node[blockW,right=of c] (d) {$\bm{\Sigma}_{i,1}$};
		\draw[vecArrow] (c) to (d);
		\node[blockW,right=of d] (e) {$\textcolor{red}{\bm{W}_2}$}; 
		\draw[vecArrow] (d) to (e);  
		\node[blockW,right=of e] (f) {$\bm{\Sigma}_{i,2}$};
		\draw[vecArrow] (e) to (f);
		\node[blockW,right= of f] (g) {$\cdots$}; 
		\draw[vecArrow] (f) to (g);
		%
		\node[blockW,right=of g] (h) {$\textcolor{red}{\bm{W}_K}$}; 
		\draw[vecArrow] (g) to (h);  
		\node[blockW,right=of h] (i) {$\bm{\Sigma}_{i,K}$};
		\draw[vecArrow] (h) to (i);
		
		\node[blockB,right=of i] (j) {$\bm{B}$};
		\draw[vecArrow] (i) to (j);
		\node[right=of j](k) {\hspace{-2mm} $\bm{y}_i$};
		\draw[vecArrow] (j) to (k);
	\end{tikzpicture}   \\ [4mm]
	\caption{Matrix model of a ReLU FNN: the considered trainable weights are in red; $\bm{\Sigma}_{i,k}=\textnormal{diag}\big( \{0,1\}^{1 \times m} \big)$; and $\bm{W}_k\in\mathbb{R}^{m\times m}$ for all $k\in[K]$.}
	\label{ReLU_FNN_matrix_model}
\end{figure*}

We consider a ReLU FNN with $K$ hidden layers -- depth-$(K+2)$ network per Fig. \ref{ReLU_FNN_matrix_model} -- to be trained for an $\ell_2$-regression task (channel estimation) and we assume that $\bm{y}_i^{*} \in \mathbb{R}^{d_y}$ is a label vector assumed to be available. Without loss of generality, all hidden layers are presumed to have $m$ neurons. From this case, it is worth underlining that scenarios with a different number of hidden neurons can be generalized \cite{allenzhu2018convergence}.   

Let $\bm{A}$, $\bm{W}_k$, and $\bm{B}$ be the weight matrices for the first layer, the $k$-th hidden layer, and the output layer, respectively, of the ReLU FNN modeled via Fig. \ref{ReLU_FNN_matrix_model}. This network's feedforward computations can then be expressed as 
\begin{subequations}
\begin{align}
\label{FF_network_1}
 \bm{x}_{i,k}&=\rho\big(\bm{W}_k\bm{x}_{i,k-1}\big)     \\ 
\label{FF_network_2}
\bm{y}_i&=\bm{B}\bm{x}_{i,K}, 
\end{align}
\end{subequations}
where $i\in[\breve{n}]$, $k\in\{0, 1, \dots, K\}$, $\bm{x}_{i,-1}=\bm{x}_i$, $\bm{W}_0=\bm{A}\in\mathbb{R}^{m\times d_x}$, $\bm{W}_k\in\mathbb{R}^{m\times m}$ for $k\in[K-1]$, and $\bm{B}\in\mathbb{R}^{d_y\times m}$. To re-write (\ref{FF_network_1}) and (\ref{FF_network_2}), we are going to deploy the following definition.     
\begin{definition}
\label{diagonal_sign_matrix_def}	
A diagonal sign matrix is denoted by $\bm{\Sigma}_{i,k}$ defined as $(\bm{\Sigma}_{i,k})_{j,j}=\mathbbm{1}\{(\bm{W}_k\bm{x}_{i,k-1})_{j}\geq 0\}$, $\forall j\in[m]$; $(\bm{\Sigma}_{i,k})_{j,l}=0$, $\forall l\neq j$. For the ReLU FNN matrix model of Fig. \ref{ReLU_FNN_matrix_model}, 
\begin{subequations}
\begin{align}
\label{diag_mat_defn_1}
\bm{\Sigma}_{i,0}&=\textnormal{diag}\big(\mathbbm{1}\big\{\bm{W}_0\bm{x}_{i}=\bm{A}\bm{x}_{i}\geq 0\big\}\big)    \\
\label{diag_mat_defn_2}
\bm{\Sigma}_{i,k}&=\textnormal{diag}\Big(\mathbbm{1}\Big\{\bm{W}_k\Big(\prod_{l=1}^{k-1}  \bm{\Sigma}_{i,k-l}\bm{W}_{k-l}\Big)\bm{\Sigma}_{i,0}\bm{A}\bm{x}_i\geq 0\Big\}\Big),
\end{align}	
\end{subequations}
where (\ref{diag_mat_defn_1})-(\ref{diag_mat_defn_2}) are valid $\forall i\in[\breve{n}]$ and $\forall k\in[K]$.	  	
\end{definition}
 
Employing (\ref{FF_network_2}) and Definition \ref{diagonal_sign_matrix_def} while applying (\ref{FF_network_1}), (\ref{diag_mat_defn_1}), and (\ref{diag_mat_defn_2}) recursively lead to    
\begin{subequations}
\begin{align}
\label{intermediate_network_output}
\bm{x}_{i,k}&=\bm{\Sigma}_{i,k}\bm{W}_k\bm{x}_{i,k-1} \\
\label{FF_network_output}
\bm{y}_i&=\bm{B}\bm{x}_{i,K}\stackrel{(a)}{=}\bm{B} \Big( \prod_{l=k}^{K} \bm{\Sigma}_{i,K-l+k} \bm{W}_{K-l+k} \Big) \bm{x}_{i,k-1}\stackrel{(b)}{=}\bm{B} \Big( \prod_{k=0}^{K} \bm{\Sigma}_{i,K-k} \bm{W}_{K-k} \Big)\bm{x}_i, 
\end{align}
\end{subequations} 
where $(a)$ and $(b)$ hold for $k\geq 0$ and $\bm{W}_0=\bm{A}$, respectively. On initializing the non-diagonal matrices of the RHS of (\ref{FF_network_output}), we adopt the following definition on \textit{random initialization}.     
\begin{definition}
\label{Def_random_initialization} 
We state that $\bm{A}$, $\bm{W}\eqdef [\bm{W}_1 \hspace{1mm} \bm{W}_2, \ldots \hspace{1mm} \bm{W}_K ]$, and $\bm{B}$ are at random initialization provided that 
\begin{itemize}
\item $(\bm{A})_{i,j}\sim \mathcal{N}(0, 2/m)$ for every $(i, j)\in [m]\times [d_x]$;  	
\item $(\bm{W}_k)_{i,j}\sim \mathcal{N}(0, 2/m)$ for every $i, j\in [m]$ and $k\in [K]$; and     
\item $(\bm{B})_{i,j}\sim \mathcal{N}(0, 1/d_y)$ for every $(i, j)\in [d_y]\times [m]$.   
\end{itemize} 
\end{definition}
   
Concerning over-parameterization, meanwhile, we follow the underneath assumption \cite{allenzhu2018convergence}.       
\begin{assumption}
\label{over-param_assumption}
For some sufficiently large polynomial, we assume $m\geq \Omega\big(\textnormal{poly}(\breve{n}, K, \delta^{-1}).d_y\big)$.    
\end{assumption}          

Under Assumptions \ref{CE_BEM_coeffcients_normalization}, \ref{normalization_assumption}, \ref{non-degeneracy_assumption}, and \ref{over-param_assumption}, moreover, we consider the training of hidden layer weights $\big\{ \bm{W}_k\big\}_{k=1}^K$ while leaving $\bm{A}$ and $\bm{B}$ at random initialization in line with Fig. \ref{ReLU_FNN_matrix_model}. This training scenario's results extend to the setting where all weight matrices are jointly trained \cite{allenzhu2018convergence}. For the training scenario Fig. \ref{ReLU_FNN_matrix_model}, accordingly, we proceed to characterize the \texttt{SGD} convergence.       
 
\subsection{\texttt{SGD} Convergence}
\label{subsec: SGD_convergence}
For over-parameterized ReLU FNNs, the convergence of \texttt{SGD} has already been characterized by the authors of \cite{allenzhu2018convergence}. Hence, we adopt the following theorem \cite[p. 33]{allenzhu2018convergence} on over-parameterized ReLU FNNs trained with \texttt{SGD} under the satisfaction of Assumptions \ref{normalization_assumption}, \ref{non-degeneracy_assumption}, and \ref{over-param_assumption}.\footnote{The work in \cite{allenzhu2018convergence} makes Assumptions \ref{normalization_assumption} and \ref{non-degeneracy_assumption} w.r.t. a training input. Nevertheless, we make Assumptions \ref{normalization_assumption} and \ref{non-degeneracy_assumption} w.r.t. a training input and a testing input.} 
\begin{theorem}[\texttt{SGD} convergence {\cite[Theorem 2]{allenzhu2018convergence}}]
\label{SGD_over_parameterized_net_convergence}
Let $K, d_y, \breve{n}, m\in\mathbb{N}$. Suppose $\varepsilon\in(0, 1]$, $b\in[\breve{n}]$ be a mini-batch size, $\delta\in (0, O(1/K)]$, $m\geq \tilde{\Omega}\Big( \frac{\textnormal{poly}(\breve{n}, K, \delta^{-1}).d_y}{b} \Big)$, and $\eta\eqdef\Theta\Big( \frac{b\delta d_y}{\textnormal{poly}(\breve{n},K)m\log^2 m}\Big)$ be a learning rate. Let $\bm{W}^{(0)}=\bm{W}$, $\bm{A}$, and $\bm{B}$ are at random initialization per Definition \ref{Def_random_initialization}; $F\big(\bm{W}^{(t)}\big) \eqdef  \frac{1}{2}\sum_{i=1}^{\breve{n}} \big\|\bm{B} \big( \prod_{k=0}^{K-1} \bm{\Sigma}_{i,K-k} \bm{W}_{K-k}^{(t)} \big)\bm{\Sigma}_{i,0} \bm{A}\bm{x}_i-\bm{y}_i^{*} \big\|^2$; and training with only trainable hidden layers  per Fig. \ref{ReLU_FNN_matrix_model} is assumed. Assuming that we begin at $\bm{W}^{(0)}$, optimization is carried out using \texttt{SGD} as
\begin{equation*}
\label{SGD_relation}
\bm{W}^{(t+1)}=\bm{W}^{(t)}-\eta\times \frac{\breve{n}}{|S_t |} \sum_{i\in S_t} \nabla F\big(\bm{W}^{(t)}\big), 
\end{equation*}      
where $t=0, 1, \ldots, T-1$; $\bm{W}^{(t)}\eqdef \big[\bm{W}_1^{(t)} \hspace{1mm} \bm{W}_2^{(t)} \ldots \hspace{1mm} \bm{W}_K^{(t)} \big]\in\mathbb{R}^{m\times mK}$; and $S_t\subset [\breve{n}]$ is a random subset of fixed cardinality $b$. Then, with probability at least $1-e^{-\Omega(\log^2 m)}$ over the randomness of $S_1, S_2, \ldots, S_T$:  
\begin{equation*}
\label{SGD_convergence_result}
F\big(\bm{W}^{(T)}\big) \leq \varepsilon,  \hspace{2mm}  \forall T=\Theta\Big(\frac{\textnormal{poly}(\breve{n},K)\log^2 m}{b\delta^2}\log\frac{n\log m}{\varepsilon}  \Big).   
\end{equation*}        
\end{theorem} 
As a consequence of Theorem \ref{SGD_over_parameterized_net_convergence}, the underneath corollary -- also comprising the proof of Theorem \ref{SGD_over_parameterized_net_convergence} detailed in \cite[p. 33-36]{allenzhu2018convergence} -- ensues.      
\begin{corollary}[Optimization constraint of \texttt{SGD} {\cite{allenzhu2018convergence}}]
\label{SGD_convergence_prop}
Let $K, d_y, \breve{n}, m \in\mathbb{N}$. Let $b\in[\breve{n}]$ be a mini-batch size, $\delta\in (0, O\big(\frac{1}{K}\big)]$, $m\geq \tilde{\Omega}\Big( \frac{\textnormal{poly}(\breve{n}, K, \delta^{-1}).d_y}{b} \Big)$, and $\bm{W}^{(t)}\eqdef \big[\bm{W}_1^{(t)} \hspace{1mm} \bm{W}_2^{(t)} \ldots \hspace{1mm} \bm{W}_K^{(t)} \big]\in\mathbb{R}^{m\times mK}$. Throughout an optimization of an over-parameterized ReLU FNN by \texttt{SGD}, the following constraint is satisfied with a high probability for every $t=0, 1, \ldots, T-1$: 
\begin{equation}
\label{Conv_constraint}
\big\|\bm{W}^{(t)}-\bm{W}^{(0)} \big\|_{F}\leq\omega, 
\end{equation}   
where $\omega= O\bigg(\frac{\breve{n}^{3.5}\sqrt{d_y}}{\delta \sqrt{bm}}\log m \bigg)$ and $\bm{W}^{(0)}=\bm{W}\eqdef [\bm{W}_1 \hspace{1mm} \bm{W}_2, \ldots \hspace{1mm} \bm{W}_K ]\in\mathbb{R}^{m\times mK}$.   
\end{corollary}
Building on Corollary \ref{SGD_convergence_prop} and Theorem \ref{SGD_over_parameterized_net_convergence} which characterize the convergence of an over-parameterized ReLU FNNs trained using \texttt{SGD}, we hereunder provide the testing MSE performance quantification of an over-parameterized deep ReLU FNN-based blind OFDM channel estimator.

\subsection{Testing MSE Performance Quantification of a DL-based Blind OFDM Channel Estimator}
\label{subsec: theory_on_MSE_bounds}
For an over-parameterized ReLU FNN initialized randomly per Definition \ref{Def_random_initialization} and operating under the satisfaction of Assumptions \ref{CE_BEM_coeffcients_normalization}, \ref{normalization_assumption}, \ref{non-degeneracy_assumption}, and \ref{over-param_assumption}, we henceforth develop a theory. This theory concerns the asymptotic and non-asymptotic testing MSE performance of an over-parameterized ReLU FNN-based blind OFDM channel estimator per Algorithm \ref{Blind_DSCE_Alg_cont}. Per Algorithm \ref{Blind_DSCE_Alg_cont} and Fig. \ref{ReLU_FNN_matrix_model}, the ReLU FNN's output during testing can be expressed as 
\begin{equation}
\label{ReLU_FNN_output_testing}
\check{\bm{y}}_i=\bm{B}\check{\bm{x}}_{i,K},
\end{equation}  
where (\ref{ReLU_FNN_output_testing}) is valid for all $i\in\{\breve{n}+1, \breve{n}+2, \ldots, \breve{n}+\check{n}\}$ and 
\begin{equation}
\label{bm_x_i_K_definition}
\check{\bm{x}}_{i,K}=\check{\bm{\Sigma}}_{i,K}\check{\bm{W}}_{K}\check{\bm{x}}_{i,K-1} = \Big(\prod_{k=0}^{K-1} \check{\bm{\Sigma}}_{i,K-k} \check{\bm{W}}_{K-k} \Big) \check{\bm{\Sigma}}_{i,0}\bm{A}\check{\bm{x}}_i, 	
\end{equation}
where $\check{\bm{\Sigma}}_{i,k}=\textnormal{diag}\big( \{0,1\}^{1 \times m} \big)$ for $k=0$ and all $k\in[K]$ (in line with Definition \ref{diagonal_sign_matrix_def}), and $\check{\bm{x}}_{i,K-1}$ is the output of the ReLU FNN's $(K-1)$-th layer. Considering a testing set $\mathcal{T}\eqdef \{(\check{\bm{x}}_i, \check{\bm{y}}_i^{*}) \}_{i=\breve{n}+1}^{\breve{n}+\check{n}}$ made of (\ref{check_x_i})-(\ref{test_label_vector_1}) and normalization by $d_y\check{n}$, we define the MSE exhibited by the presented blind OFDM channel estimator as   
\begin{subequations}
\begin{align}
\label{MSE_cost_2}
F(\check{\bm{W}})  \eqdef& \frac{1}{d_y\check{n}} \sum_{i=\breve{n}+1}^ {\breve{n}+\check{n}}  \|\bm{B}\check{\bm{x}}_{i,K}-\check{\bm{y}}_i^{*} \|^2    \\
\label{MSE_cost_3}
\stackrel{(a)}{=}&\frac{1}{d_y\check{n}} \sum_{i=\breve{n}+1}^{\breve{n}+\check{n}}\big\|\bm{B} \big( \prod_{k=0}^{K-1} \check{\bm{\Sigma}}_{i,K-k} \check{\bm{W}}_{K-k} \big) \check{\bm{\Sigma}}_{i,0}\bm{A}\check{\bm{x}}_i-\check{\bm{y}}_i^{*} \big\|^2, 
\end{align}
\end{subequations}
where $d_y=2N^2$; $(a)$ follows from (\ref{bm_x_i_K_definition}); and $\check{\bm{W}}=\big[\check{\bm{W}}_1 \hspace{1mm} \check{\bm{W}}_2  \ldots  \check{\bm{W}}_K\big]\in\mathbb{R}^{m\times mK}$ denote the hidden layers' trained weights that can be characterized through Corollary \ref{SGD_convergence_prop}. Employing Corollary \ref{SGD_convergence_prop} and (\ref{MSE_cost_3}), meanwhile, the asymptotic testing MSE performance quantification of the investigated blind OFDM channel estimator follows.  

\subsubsection{Asymptotic Testing MSE Performance Quantification}  
\label{Asymptotic_per_characterization}
in various statistical signal processing and wireless communications works, infinite received signal samples have been presumed -- so that the central limit theorem can be evoked to ascertain Gaussianity -- by the disseminated asymptotic analyses. In our upcoming asymptotic analysis, however, we consider infinite subcarriers by letting $N\to\infty$ without resorting to any Gaussian assumption since we have $2N+1$ inputs for $N$ being the number of subcarriers. For this scenario, the asymptotic expectation of the testing MSE exhibited by the discussed blind OFDM channel estimator is quantified below.     
\begin{theorem}
\label{Thm_lim_expectation_MSE}
Let $m,K\in\mathbb{N}$ and $m,K<\infty$. Upon the convergence of training using \texttt{SGD}, let $\check{\bm{W}}=\big[\check{\bm{W}}_1 \hspace{1mm} \check{\bm{W}}_2 \ldots \check{\bm{W}}_K\big]\in\mathbb{R}^{m\times mK}$ be the trained hidden layer weights pertaining to an over-parameterized ReLU FNN of Fig. \ref{ReLU_FNN_matrix_model} such that $\check{\bm{W}}_k\in\mathbb{R}^{m\times m}$ for all $k\in[K]$. Then,     
\begin{equation}
\label{lim_expectation_MSE}
\lim_{N\to\infty} \mathbb{E}\{ F(\check{\bm{W}}) \}=0.  
\end{equation}
\proof The proof\footnote{The write-up of our proofs is aimed at clarity, completeness, and rigor.} is deferred to Appendix \ref{proof_Thm_lim_expectation_MSE}.  
\end{theorem}

To infer an insightful remark from Theorem \ref{Thm_lim_expectation_MSE}, we prove the following lemma which asserts Parseval's theorem \cite[p. 18]{Proakis_5th_ed_08}, \cite[p. 60]{AORS10}, as also noted in Remark \ref{On_Parseval's theorem} (see Appendix \ref{proof_Thm_lim_expectation_MSE}).     
\begin{lemma}
\label{lem_TD_FD_MSE_relation}
Let $\hat{\tilde{\bm{H}}}^{i}$ and $\hat{\bm{H}}^{i}$ be the estimated frequency-domain and time-domain doubly selective fading OFDM channel matrices, respectively. If $\hat{\tilde{\bm{H}}}^{i}$ is estimated by the trained over-parameterized ReLU FNN of Fig. \ref{ReLU_FNN_matrix_model}, its exhibited MSE is given by       
\begin{equation}
\label{MSE_alt_defn}
F(\check{\bm{W}})\eqdef \frac{1}{2\check{n}N^2}\sum_{i=\breve{n}+1}^ {\breve{n}+\check{n}} \big\| \hat{\tilde{\bm{H}}}^{i}-\tilde{\bm{H}}^{i} \big\|_{F}^2, 
\end{equation}
where $\tilde{\bm{H}}^{i}$ is given by (\ref{H_sc_matrix}). Then,     
\begin{equation}
\label{MSE_eq_defn}
F(\check{\bm{W}})= \frac{1}{2\check{n}N^2}\sum_{i=\breve{n}+1}^ {\breve{n}+\check{n}} \big\| \hat{\bm{H}}^{i}-\bm{H}^{i} \big\|_{F}^2. 
\end{equation}
\proof The proof is provided in Appendix \ref{proof_lem_TD_FD_MSE_relation}.
\end{lemma}
          
Consequently, Theorem \ref{Thm_lim_expectation_MSE}, Lemma \ref{lem_TD_FD_MSE_relation}, and (\ref{H_i_matrix_expr_1}) lead to the following remark.     
\begin{remark}
\label{prob_lower_bound}
It follows directly from Theorem \ref{Thm_lim_expectation_MSE}, Lemma \ref{lem_TD_FD_MSE_relation}, and (\ref{H_i_matrix_expr_1}) that $\big\| \hat{\bm{H}}^{i}-\bm{H}^{i} \big\|_{F}^2=\Theta(NL).$             
\end{remark}

The asymptotic performance as in Theorem \ref{lem_TD_FD_MSE_relation} is mathematically plausible regarding an infinite number of subcarriers. However, as the number of subcarriers increases, the peak-to-average power ratio (PAPR) increases significantly for the overall OFDM transceiver \cite{Jiang_TBC_PAPR_Red_Overview_08,Yoshizawa_PAPR_redu_15}. Such a PAPR increment will make the high power amplifier introduce inter-modulation  between  the  different  subcarriers  and  introduce  additional interference into the OFDM system \cite{Jiang_TBC_PAPR_Red_Overview_08}. Accordingly, the underneath non-asymptotic testing MSE performance quantification -- for an arbitrary number of subcarriers -- would be important.  

\subsubsection{Non-Asymptotic Testing MSE Performance Quantification}  
\label{Non_asymptotic_per_characterization}
we herein present a theorem on the non-asymptotic probabilistic lower bound pertaining to the testing MSE exhibited by the proposed blind OFDM channel estimator. 
 
\begin{theorem}
\label{Thm_probabilistic_MSE_lower_bound}
Suppose $m,K, N\in\mathbb{N}$; $m,K, N<\infty$; and $t\in\mathbb{R}$. Upon convergence of training using \texttt{SGD}, let $\check{\bm{W}}=\big[\check{\bm{W}}_1 \hspace{1mm} \check{\bm{W}}_2 \ldots \check{\bm{W}}_K\big]\in\mathbb{R}^{m\times mK}$ be the trained hidden layer weights corresponding to an over-parameterized ReLU FNN modeled through Fig. \ref{ReLU_FNN_matrix_model} such that $\check{\bm{W}}_k\in\mathbb{R}^{m\times m}$ for all $k\in[K]$. Let $t>(\sqrt{m}+\sqrt{2}N)^4$, $C_{\sigma}(m, N)=\sqrt{2}N\big[(\sqrt{m}+\sqrt{2}N)+\sqrt{N}\big(\sqrt{m}+\sqrt{2}N\big)^{-1}\big]\in \mathbb{R}$, and $C_{\sigma} \in \mathbb{R}\backslash \{0\}$. If $\displaystyle \prod_{k=0}^{K-1}\sigma_{\textnormal{min}}\big(\check{\bm{\Sigma}}_{i,K-k}\check{\bm{W}}_{K-k}\big)  \sigma_{\textnormal{min}}\big(\check{\bm{\Sigma}}_{i,0}\bm{A}\big)\geq C_{\sigma} > C_{\sigma}(m, N)$, then   
\begin{equation}
\label{Prob_Exp_instant_MSE_bound}
\mathbb{P}\bigg( F(\check{\bm{W}}) \geq \frac{t}{2N^2} \bigg) \geq \Bigg(\frac{\big( f_1^2- f_2^2 \big)^{2}}{2^{2L(Q+1)}} \bigg[1+ \textnormal{erf}\bigg( \frac{1}{\sqrt{2(Q+1)}} \bigg) \bigg]\Bigg)^{\check{n}},  
\end{equation}
where $f_1=\frac{\sqrt{m}-\sqrt{2}N}{\sqrt{m}+\sqrt{2}N}$ and $f_2=\frac{\sqrt{2}NC_{\sigma}^{-1}(\sqrt{t}+ \sqrt{N})}{\sqrt{m}+\sqrt{2}N}$. 

\proof The proof is deferred to Appendix \ref{proof_Thm_probabilistic_MSE_lower_bound}.  
\end{theorem}

Inspired by Theorem \ref{Thm_probabilistic_MSE_lower_bound}, the underneath remarks are in order.   
\begin{remark}
\label{remark_open_problems}
Theorem \ref{Thm_probabilistic_MSE_lower_bound} is the first of its kind non-asymptotic theory that quantifies the performance of a DL algorithm used to solve a regression AI/ML problem. However, this theory is not a unified theory of DL, as embodied in the following open problems\footnote{Regarding the open problems of Remark \ref{remark_open_problems}, the authors of \cite{Akemann_Mat_Products_2015} review recent results for the complex eigenvalues and singular values of finite products of finite size random matrices that are taken from ensembles of independent real, complex, or quaternionic Ginibre matrices, or truncated unitary matrices. Besides, a recent work reported in \cite{Hanin_2019_Mat_Products} provides results on products of random matrices when the number of terms and the size of the matrices tends to infinity at the same time.} w.r.t. the model of Fig. \ref{ReLU_FNN_matrix_model}: 
\begin{enumerate}
\item Deriving the distribution of $\prod_{k=0}^{K-1}\sigma_{\textnormal{min}}\big(\check{\bm{\Sigma}}_{i,K-k}\check{\bm{W}}_{K-k}\big) \sigma_{\textnormal{min}}\big(\check{\bm{\Sigma}}_{i,0}\bm{A}\big)$;
\item Deriving the lower bound for $\hspace{0.5mm} \mathbb{E} \big\{ \prod_{k=0}^{K-1}\sigma_{\textnormal{min}}\big(\check{\bm{\Sigma}}_{i,K-k}\check{\bm{W}}_{K-k}\big) \sigma_{\textnormal{min}}\big(\check{\bm{\Sigma}}_{i,0}\bm{A}\big)  \big\} $.
\end{enumerate}  
\end{remark}

\begin{remark}
After convergence is established per Theorem \ref{SGD_over_parameterized_net_convergence}, the studied blind OFDM channel estimator manifests a probabilistic testing MSE quantified per Theorem \ref{Thm_probabilistic_MSE_lower_bound}. This is because the real and imaginary testing inputs -- made of (\ref{check_x_i}) -- are Gaussian random vectors evoking randomness onto the $K$-layer ReLU FNN.\footnote{The Gaussian random inputs sampled from the received wireless signals can also evoke randomness in the trained network even if the network is a linear FNN (see \cite{Arora2018_Convergence} on a convergence analysis of deep linear FNNs).} Quantifying the effect of such randomness, Theorem \ref{Thm_probabilistic_MSE_lower_bound} inspires fundamental research toward the non-asymptotic performance quantification of DL-based algorithms as applied to wireless communications, signal processing, and networking.  
\end{remark}

Minimizing the left-hand side (LHS) of (\ref{Prob_Exp_instant_MSE_bound}), meanwhile, the following corollary arises.        
\begin{corollary}
\label{Coro_min_lower_probab_bound}
Under the conditions of Theorem \ref{Thm_probabilistic_MSE_lower_bound},  
\begin{equation}
\label{min_proba_bound_exp}
\min \mathbb{P}\bigg( F(\check{\bm{W}}) \geq \frac{t}{2N^2} \bigg) \geq 0.
\end{equation}
\proof The proof is given in Appendix \ref{proof_Coro_min_lower_probab_bound}.

\end{corollary}
  
Moreover, inspired by Lemma \ref{Lem_dep_Gaussian_norm} which constitutes Theorem \ref{Thm_probabilistic_MSE_lower_bound}'s proof detailed in Appendix \ref{proof_Thm_probabilistic_MSE_lower_bound}, the following lemma -- as our contribution to high-dimensional probability\footnote{Using Gaussian concentration inequality (see Theorem \ref{Thm_Gaussian_conc_inequality}), the existing concentration inequality for a Gaussian $\ell_p$-norm was derived only for comprising independent standard normal RVs \cite[p. 126-127]{Boucheron_Con_eqs_13}.} -- follows.    
\begin{lemma}[Concentration of the Euclidean norm of Gaussian RVs with different variances]
\label{Lem_Conc_Gau_norm_diff_variance}
Let $\bm{x}=[X_1, \ldots, X_{2n}]^T\in\mathbb{R}^{2n}$ be a Gaussian random vector made of $2n$ independent Gaussian RVs $X_i\sim\mathcal{N}(0, \sigma_i^2)$. Let $\sigma^2=\sum_{i=1}^{2n} \sigma_i^2$ and $t\geq 0$. Then, 
\begin{equation}
\label{Lem_Conc_Gau_norm_diff_variance_1}
\mathbb{P}\big( \|\bm{x}\|- \mathbb{E}\big\{ \|\bm{x}\| \big\}  > t\big)\leq 1-2^{-2n} \bigg[ 1 +\textnormal{erf}\bigg(\frac{t}{\sigma\sqrt{2}} \bigg)  \bigg].
\end{equation} 
\proof The proof is relegated to Appendix \ref{proof_Lem_Conc_Gau_norm_diff_variance}. 
\end{lemma}  
With this ending of our theoretical developments, we proceed to our computer experiments.

\section{Computer Experiments}
\label{sec: Computer_experiments}
This section documents extensive computer experiments on the performance assessment of the investigated DL-based blind OFDM channel estimator: Sections \ref{subsec: Gen_DSF_channel}, \ref{subsec: training_and_testing_data_generation}, \ref{subsec: deep_ReLU_FNN_training}, and \ref{subsec: results} present generation of a doubly selective fading channel, generation of training and testing sets, training of over-parameterized (deep) ReLU FNNs, and training/testing results, respectively.

\begin{table*}	
	\caption{Generation parameters of the training and testing sets.}		
	\label{table: sim_parameters}
	\centering
	\begin{tabular}{ | l | l | l | } 
		\hline      
		\textbf{Parameters} & \textbf{Type/Value} & \textbf{Remark}     \\  \hline  
		Modulation  & quadrature phase shift keying (QPSK)  & chosen as $d^i[k]=\frac{1}{\sqrt{2}}[s_{I}^{i}[k]+js_{Q}^{i}[k] ]$  \\      
		&   & for $\big\{ s_{I}^{i}[k], s_{Q}^{i}[k] \big\}  \in \big\{-1, 1 \big\}^2$ \\  \hline 
		Carrier frequency ($f_c$) & $3.55$ GHz      & chosen Citizens Broadband Radio     \\    
		&    & Service (CBRS) band \cite{Lees_DL_Classification_19}       \\   \hline      
		Bandwidth ($B$) & $10.24$ MHz       & chosen        \\   \hline   
		Number of subcarriers ($N$) & $N \in \{32, 64, 128\}$  & chosen     \\   \hline    
		Subcarrier spacing ($\Delta f$)   & $\Delta f \in \{320, 160, 80\}$   kHz               & $\Delta f=\frac{B}{N}$            \\   \hline
		Useful OFDM symbol duration ($T_u$)& $T_u \in \{3.125, 6.25, 12.5\}$ $\mu s$    & $T_u=\frac{1}{\Delta f}$  \\  \hline   
		Symbol duration ($T_s$) & $T_s \in \{97.6562, 97.6562, 97.6562\}$ $\eta$s & $T_s=\frac{T_u}{N}$   \\   \hline  
		Maximum delay spread ($\tau_{\textnormal{max}}$)  & 500 $\eta$s      &  Assumed  \\ \hline
		Channel order ($L$) & $L \in \{6, 6, 6\}$  & $L=\ceil*{ \tau_{\textnormal{max}}/T_s}$   \\ \hline   
		Cyclic prefix length ($N_{cp}$) &   $N_{cp} \in \{10, 10, 10\}$   & $N_{cp}\geq L-1$  \\ \hline                                  
		Maximum mobile velocity ($v_{\textnormal{max}}$) & 160 km/hr    & Assumed     \\   \hline  
		Speed of light ($c$) & $c=3\times 10^8$ m/s    & Known value  \\   \hline    
		Maximum Doppler spread ($f_{\textnormal{max}}$) & 525.9259 Hz    & $f_{\textnormal{max}}=f_c v_{\textnormal{max}}/c$         \\  \hline    
		$\tilde{N}$ & $\tilde{N}\in\{42, 74, 138\}$  & $\tilde{N}=N+N_{cp}$     \\ \hline 
		$Q$                  &   $Q\in\{2, 2, 2\}$      & $Q=2\lceil{f_{\textnormal{max}}\tilde{N}T_s \rceil}$     	     \\    \hline 
	\end{tabular}
\end{table*}

\begin{algorithm}
	\SetKwData{Left}{left}
	\SetKwData{This}{this}
	\SetKwData{Up}{up}
	\SetKwFunction{Union}{Union}
	\SetKwFunction{FindCompress}{FindCompress}
	\SetKwInOut{Input}{Input}
	\SetKwInOut{Output}{Output}
	\caption{Generation of the subcarrier coupling matrix w.r.t. the doubly selective fading channel of the $i$-th OFDM symbol }
	\label{OFDM_DSFC_generation}
	\Input{ $f_c$, $v_{\textnormal{max}}$, $\tau_{\textnormal{max}}$, $N$, $B$, $i$   }
	\Output{frequency-domain OFDM channel matrix $\tilde{\bm{H}}^i$}
	\BlankLine
	\emph{Compute $L$ as $L=\ceil*{ \tau_{\textnormal{max}}/T_s}$  }\\
	\emph{Choose $N_{cp}$ as $N_{cp}\geq L-1$}  \\
	\emph{Calculate $\tilde{N}$ as $\tilde{N}=N+N_{cp}$}  \\
	\emph{Compute $f_{\textnormal{max}}$ as $f_{\textnormal{max}}=\frac{f_c v_{\textnormal{max}}}{c}$  $\big($$c=3\times 10^8$ m/s$\big)$} \\
	\emph{Compute $Q$ as $Q=2\lceil{f_{\textnormal{max}}\tilde{N}T_s \rceil}$   }  \\
	\emph{Determine $\Delta f$ as $\Delta f=\frac{B}{N}$ }  \\
	\emph{Compute $T_u$ as $T_u=\frac{1}{\Delta f}$}  \\
	\emph{Determine $T_s$ as $T_s=\frac{T_u}{N}$}  \\
	\For{$l\leftarrow 0$ \KwTo $L-1$}{\nllabel{forins}
		\For{$q\leftarrow 0$ \KwTo $Q$}{\nllabel{forins}	
			Compute $\sigma_{q, l}^2$ using (\ref{sigma_q_l_def}) [via (\ref{S_H_tau}), (\ref{S_H_nu}), and (\ref{gamma_def})]\;
			Determine $h_q[i; l]$ as $h_q[i; l]=\frac{\sigma_{q, l}}{\sqrt{2}}\big[ \texttt{randn}+\jmath \texttt{randn} \big]$\;
			
		}
		\For{$n\leftarrow 0$ \KwTo $N-1$}{\nllabel{forins}
			Compute $h^{i}[n;l]$ as $h^{i}[n;l]=h[i\tilde{N}+n; l]=\bm{h}_{i\tilde{N}+n,l}\bm{v}_{i\tilde{N}+n}$ for $\bm{v}_{i\tilde{N}+n}=\big[e^{\jmath \frac{2\pi (0-Q/2)(i\tilde{N}+n)}{\tilde{N}}}, \ldots, e^{\jmath \frac{2\pi (Q/2)(i\tilde{N}+n)}{\tilde{N}}}\big]^T$ and $\bm{h}_{i\tilde{N}+n,l}=\big[ h_0[i; l], \ldots, h_Q[i; l] \big] $ -- see (\ref{DT_CIR}) and (\ref{BEM_equation_simpli_2})-(\ref{BEM_coefficient_relation_2_1})\;
			
		}
	}
	$\bm{H}^i=\texttt{zeros}(N,N)$\;
	\For{$n\leftarrow 0$ \KwTo $N-1$}{\nllabel{forins}
		\For{$l\leftarrow 0$ \KwTo $L-1$}{\nllabel{forins}
			$\big(\bm{H}^{i}\big)_{n,l} = h^{i}[n; \langle n-l \rangle_{N}]$ if $\langle n-l \rangle_{N}\in[0, L-1]$, otherwise $\big(\bm{H}^{i}\big)_{n,l} = 0$\; 
		}
	}
	Determine $\bm{F}_N$ via ${\big[\bm{F}_N\big]}_{k,l}\eqdef 1/\sqrt{N} e^{-j \frac{2\pi (k-1)(l-1)}{N}}$ for every $(k,l) \in [N] \times [N]$\;
	Compute $\tilde{\bm{H}}^i$ as $\tilde{\bm{H}}^i=\bm{F}_N\bm{H}^i\bm{F}_N^H$ [see (\ref{H_sc_matrix})]\; 
\end{algorithm}

\subsection{Generation of a Doubly Selective Fading Channel}
\label{subsec: Gen_DSF_channel}
Practitioners usually model a doubly selective fading wireless communications channel as a wide-sense stationary uncorrelated scattering (WSSUS) channel \cite{X_Cai_ICI_supp_03}. For a WSSUS channel, a popular channel model uses a separable scattering function $\mathcal{S}_h(\tau; \nu)=\frac{1}{\rho_h^2}\tilde{\phi}_h(\tau)\tilde{\mathcal{S}}_h(\nu)$ \cite{WCom_TVC_11} with an exponential power delay profile (PDP) $\tilde{\phi}_h(\tau)$ and Jakes Doppler power profile (Doppler power spectrum)\footnote{As depicted in \cite[Figure 13.1--8]{Proakis_5th_ed_08}, this spectrum is also known as the \textquotedblleft bathtub-shaped'' Doppler spectrum which follows from the assumption of a uniformly distributed angle of arrival \cite{Jakes:1994:MMC:561302,GMa_Fundam_TVC_2011,WCom_TVC_11}.} \cite[p. 839]{Proakis_5th_ed_08} $\tilde{\mathcal{S}}_h(\nu)$ that are, respectively, defined as \cite[p. 27]{GMa_Fundam_TVC_2011}: $\tilde{\phi}_h(\tau) \eqdef \frac{\rho_h^2}{\tau_0} e^{-\frac{\tau}{\tau_0}}$ for $\tau\geq 0$ and $\tilde{\phi}_h(\tau)\eqdef 0$ otherwise; $\tilde{\mathcal{S}}_h(\nu) \eqdef \frac{\rho_h^2}{\pi \sqrt{f_{\textnormal{max}}^2-\nu^2}}$ for $\left|\nu\right|<f_{\textnormal{max}}$ and $\tilde{\mathcal{S}}_h(\nu) \eqdef 0$ otherwise -- given that $\tau\in [0, \tau_{\textnormal{max}}]$, $\nu\in(-f_{\textnormal{max}}, f_{\textnormal{max}})$, $\rho_h^2$, and $\tau_0$ denote delay, Doppler spread, path gain, and the root mean square (RMS) delay spread (multipath spread), respectively. Concerning these definitions and the aforementioned separable scattering function of a WSSUS channel, we opt for the PDP and the Jakes Doppler power spectrum given by  
\begin{subequations}
	\begin{align}
	\label{S_H_tau}
	\phi_h(\tau)&=
	\begin{cases}
	e^{-\tau/(L-1)T_s},     & \text{$\tau\geq 0$}\\
	0,  & \text{$\tau<0$}
	\end{cases}     \\
	\label{S_H_nu}
	\mathcal{S}_h(\nu)&=
	\begin{cases}
	\Big( \pi\sqrt{f_{\textnormal{max}}^2-\nu^2} \Big)^{-1},  & \text{$\left|\nu\right|<f_{\textnormal{max}}$}\\
	0,  & \text{$\left|\nu\right|>f_{\textnormal{max}}$}, 
	\end{cases} 
	\end{align}
\end{subequations}
where $\phi_h(\tau)=\tilde{\phi}_h(\tau)\big|_{\rho_h^2=\tau_0=(L-1)T_s}$ and $\mathcal{S}_h(\nu)=\frac{1}{\rho_h^2}\tilde{\mathcal{S}}_h(\nu)$. To these ends, without loss of generality and similar to the doubly selective fading channel generation technique mentioned in \cite[Sec. VII]{X_Ma_Opt_train_03}, we define the variance of $ h_q[ \lfloor{n/\tilde{N}\rfloor}; l]$ denoted by $\sigma_{q, l}^2$ as 
\begin{equation}
\label{sigma_q_l_def}
\sigma_{q, l}^2\eqdef\gamma\phi_h(lT_s)\mathcal{S}_h\big(qf_{\textnormal{max}}/(Q+1)\big), 
\end{equation}
where $l\in [0, L-1]$, $q\in [0, Q]$, and $\gamma$ is a normalization factor defined as 
\begin{equation}
\label{gamma_def}
\gamma\eqdef   \bigg( \sum_{l=0}^{L-1}\sum_{q=0}^Q \phi_h(lT_s)\mathcal{S}_h\big(qf_{\textnormal{max}}/(Q+1)\big)\bigg)^{-1}.
\end{equation}
Regarding this generation technique and the definition on the CE-BEM of (\ref{BEM_equation}), it shall be recalled that the variances remain the same per an OFDM symbol duration. Meanwhile, the generation of the subcarrier coupling matrix w.r.t. the doubly selective fading channel of the $i$-th OFDM symbol is executed per Algorithm \ref{OFDM_DSFC_generation}\footnote{Per MATLAB$^{\textregistered}$, \texttt{randn} and $\texttt{zeros}(m,n)$ generate a zero mean Gaussian RV with unit variance and an $m\times n$ zero matrix, respectively.} using the parameters of Table \ref{table: sim_parameters}. Thereafter, we follow up with the generation of our training and testing sets.

\begin{algorithm}
	\SetKwData{Left}{left}
	\SetKwData{This}{this}
	\SetKwData{Up}{up}
	\SetKwFunction{Union}{Union}
	\SetKwFunction{FindCompress}{FindCompress}
	\SetKwInOut{Input}{Input}
	\SetKwInOut{Output}{Output}
	\caption{Generation of the training and testing sets}
	\label{Training_testing_datasets_generation}
	\Input{ $\breve{n}$, $\check{n}$     }
	\Output{training set $\mathcal{D}=\big\{(\bm{x}_i, \bm{y}_i^{*}) \big\}_{i=1}^{\breve{n}}$; testing set $\mathcal{T}=\big\{ (\check{\bm{x}}_i, \check{\bm{y}}_i^{*}) \big\}_{i=\breve{n}+1}^{\breve{n}+\check{n}}$ }
	\BlankLine
	\For{$i\leftarrow 1$ \KwTo $\breve{n}$}{\nllabel{forins}	
			Generate $\tilde{\bm{H}}^i$ using Algorithm \ref{OFDM_DSFC_generation}\;
			Generate $\tilde{\bm{y}}^i$ using (\ref{DFT_received_samples_3})\;
			Generate $\bm{x}_i$ using (\ref{training_input_i})\; 
			Generate $\bm{y}_i^{*}$ using (\ref{training_label_vector_1})\; 
	}
	Form $\mathcal{D}=\big\{(\bm{x}_i, \bm{y}_i^{*}) \big\}_{i=1}^{\breve{n}}$\;  	
	\For{$i\leftarrow \breve{n}+1$ \KwTo $\breve{n}+\check{n}$}{\nllabel{forins}	
		Generate $\tilde{\bm{H}}^i$ using Algorithm \ref{OFDM_DSFC_generation}\;
		Generate $\tilde{\bm{y}}^i$ using (\ref{DFT_received_samples_3})\;
		Generate $\check{\bm{x}}_i$ using (\ref{check_x_i})\;
		Generate $\check{\bm{y}}_i^{*}$ using (\ref{test_label_vector_1})\;
	}
Form $\mathcal{T}=\big\{ (\check{\bm{x}}_i, \check{\bm{y}}_i^{*})  \big\}_{i=\breve{n}+1}^{\breve{n}+\check{n}}$ 	
\end{algorithm}

\subsection{Generation of Training and Testing Sets}
\label{subsec: training_and_testing_data_generation}
Employing the generation parameters of Table \ref{table: sim_parameters} and Algorithm \ref{OFDM_DSFC_generation}, we generated the training and testing sets used by our computer experiments using Algorithm \ref{Training_testing_datasets_generation} (In Algorithm \ref{Training_testing_datasets_generation}'s generation, the SNR is defined -- without loss of generality -- as $\gamma_{snr}=\frac{\|\tilde{\bm{H}}^i\bm{d}^i\|_2^2}{\sigma^2}$). Following Algorithm \ref{Training_testing_datasets_generation}'s generation, we continue to the training of over-parameterized (deep) ReLU FNNs.         

\begin{table*}[!t]	
	\caption{FNN training and testing (hyper)parameters unless otherwise mentioned.}		
	\label{table: Hyperparameters}
	\centering
	\begin{tabular}{ | l | l | l | } 
		\hline      
		\textbf{(Hyper)parameters} & \textbf{Type/Value} & \textbf{Remark(s)}     \\  \hline  
		Learning rate  & 0.001    & Initial learning rate value  \\  \hline      
		Epoch size & 10      & Maximum epoch size. Having tried much larger epoch sizes, no      \\     
		&     &  difference in the training and testing performance was observed.           \\   \hline       
		Batch size & 50       & Very large and small batch sizes were tried. No visible difference        \\    
		&    &   in the training and testing outcomes were observed.  \\      \hline 
		Optimizer & \texttt{Adam} \cite{Adam_ICLR_15} & We also experimented with other optimizers. However, \texttt{Adam} produced               \\   
		&    & the smallest MSE and converged fast.      \\   \hline
		Activation function & ReLU     & This work's focus has been on over-parameterized ReLU FNNs.                     \\   
		&     & However, training was also conducted using other activation          \\   
		&     &  functions \cite{Comm_Act_Functions_2020,Rasamoelina_Activation_Function_20}. ReLU networks performed the best.              \\   \hline
		DNN depth ($K+2$) & $\{6,3\}$  & Different depth values have been tried. No significant difference      \\      
		&     &  in the training and testing outcomes were observed.         \\   \hline
		Over-parameterization & $m=50N$     & This is also the number of neurons in every hidden layer         \\  
		parameter ($m$)&     &             \\   \hline
		Training set & $\breve{n}=30 \hspace{1mm}000$    & Generated in MATLAB$^{\textregistered}$ using 30 000 OFDM symbols via             \\   
			&     &  Algorithm \ref{Training_testing_datasets_generation} for $N\in\{32, 64, 128\}$. Then, uploaded to \textsc{ENKI}.            \\   \hline
		Testing set &  $\check{n}=15 \hspace{1mm} 000$    & Generated in MATLAB$^{\textregistered}$ using 15 000 OFDM symbols via          \\   
			&     &  Algorithm \ref{Training_testing_datasets_generation} for $N\in\{32, 64, 128\}$. Then, uploaded to \textsc{ENKI}.         \\   \hline
		Training-validation split & 75 $\%$ to 25 $\%$    & 75 $\%$ of the training set was used for training and the remaining              \\   
		&     & 25 $\%$ were used for validation.           \\   \hline
		Training SNR values & randomly distributed        & In MATLAB$^{\textregistered}$, $n$ random numbers in the interval $(a,b)$ can be  
		             \\   
		& over $(-10, 40)$ dB     & generated as $r = a + (b-a).*\texttt{rand}(n,1)$.         \\   \hline
		Testing SNR ($\gamma_{snr}$) ranges & $[-10,40]$ dB     & $\gamma_{snr}\in \big\{-10, 0, 10, 20, 30, 40\big\}$ dB.            \\   \hline
		Initializer for  & \texttt{He normal} \cite{He_delving_deep'15} & Other initializers such as \texttt{LeCun normal} \cite{LeCun2012} were also tried.              \\   
		all layers &     &  However, no significant difference was observed.         \\   \hline
		Keras \textit{callbacks} \cite{Chollet_DL_with_Python'18} & Four callbacks  & \textit{Model checkpointing}, \textit{TensorBoard}, \textit{learning rate reduction}, and              \\   
		&     &   \textit{early stopping} callbacks \cite[Ch. 7]{Chollet_DL_with_Python'18} were used. Early stopping          \\   
		&     & and learning rate reduction were set to have patience over five     \\   
		&     &  epochs. Per five epochs rendering training stagnation, learning     \\   
		&     &  rate reduction was set to reduce the learning rate by 0.1.           \\   \hline
	\end{tabular}
\end{table*}

\subsection{Training of Over-Parameterized (Deep) ReLU FNNs}
\label{subsec: deep_ReLU_FNN_training}
Implemented on the NIST GPU cluster named \textsc{ENKI}, \textsc{Keras} \cite{Keras_web} with \textsc{TensorFlow} \cite{TensorFlow_web} as a backend is used for the offline training and online testing of a deep ReLU FNN per Algorithms \ref{Blind_DSCE_Alg} and \ref{Blind_DSCE_Alg_cont}, respectively. Algorithms \ref{Blind_DSCE_Alg} and \ref{Blind_DSCE_Alg_cont} are implemented in cascade as in Fig. \ref{proposed_blind_channel_estimator} to produce each subcarrier coupling matrix's estimates. The estimates' quality is quantified in terms of the testing MSE defined in (\ref{MSE_alt_defn}) while employing the training and testing (hyper)parameters -- unless otherwise mentioned -- of Table \ref{table: Hyperparameters}. As listed in Table \ref{table: Hyperparameters}, we have carried out our training by employing the \texttt{He normal} \cite{He_delving_deep'15} initialization. On the contrary, several the state-of-the-art DL-based channel estimation algorithms -- such as the algorithms of \cite{YYF_DL_DSFC_19,MSo_CommL_19,Liao_DL_TV_MIMO-OFDM'20,Gizzini_DL_for_IEEE_802.11p'20} -- employ initialization based on an LS estimation. Since an LS estimation relies on many pilot symbols, the existing DL-based techniques rely on the availability of several pilot symbols and they get more computationally complex due to the complexity of an LS estimation coupled with the offline training complexity of their DL networks. Employing such networks, our ReLU FNN-based estimation algorithm is the first blind estimation algorithm for a doubly selective fading channel. Toward this end, we hereinafter present our computer experimental results.        
   
\begin{figure}[t!]
	\centering
	\includegraphics{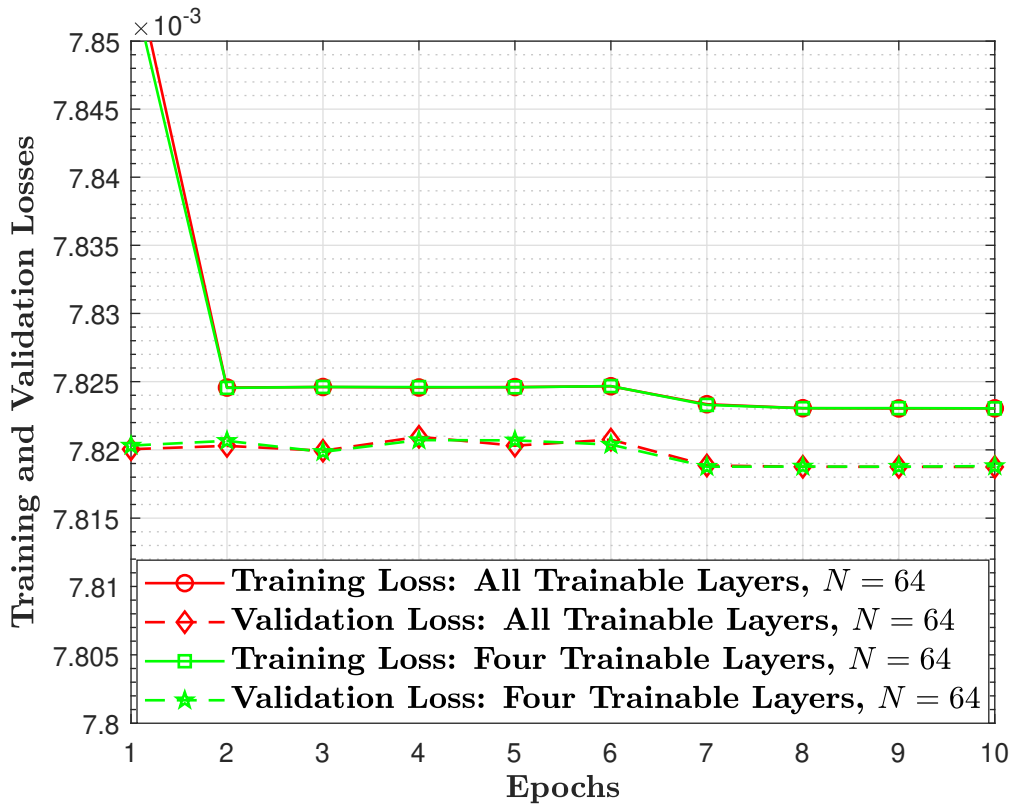}
	\caption{Training and validation losses versus epochs for the training scenarios of all trainable layers and four trainable layers: FNN depth ($K+2$)$=6$; the training set was generated using 100 000 OFDM symbols at random SNR; and an over-parameterization parameter $m=2N^2$.}
	\label{fig: ReLU_training_and_validation_Losses_Plot_All_Cases_K_4}
\end{figure}
\subsection{Results}
\label{subsec: results}
\begin{figure}[htb!]
	\centering
	\includegraphics{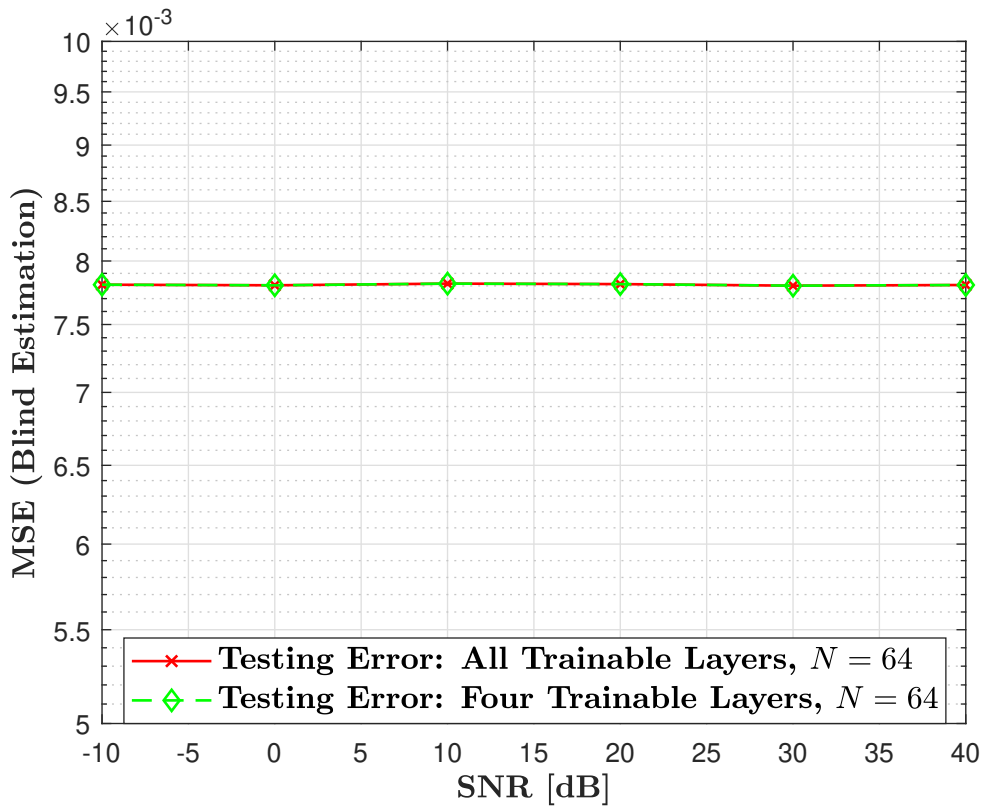}  \\ [4mm]
	\caption{Testing MSE versus SNR after training under the training scenarios of all trainable layers and four trainable layers: FNN depth ($K+2$)$=6$; training set (at random SNR) and testing set were generated using 100 000 and 40 000 OFDM symbols, respectively; and $m=2N^2$.}  \vspace{5mm}
	\label{fig: Deep_ReLU_FNN_Performance_Plot_All_Cases_K_4}
\end{figure}

\begin{figure}[htb!]
	\centering
	\includegraphics[width=\textwidth]{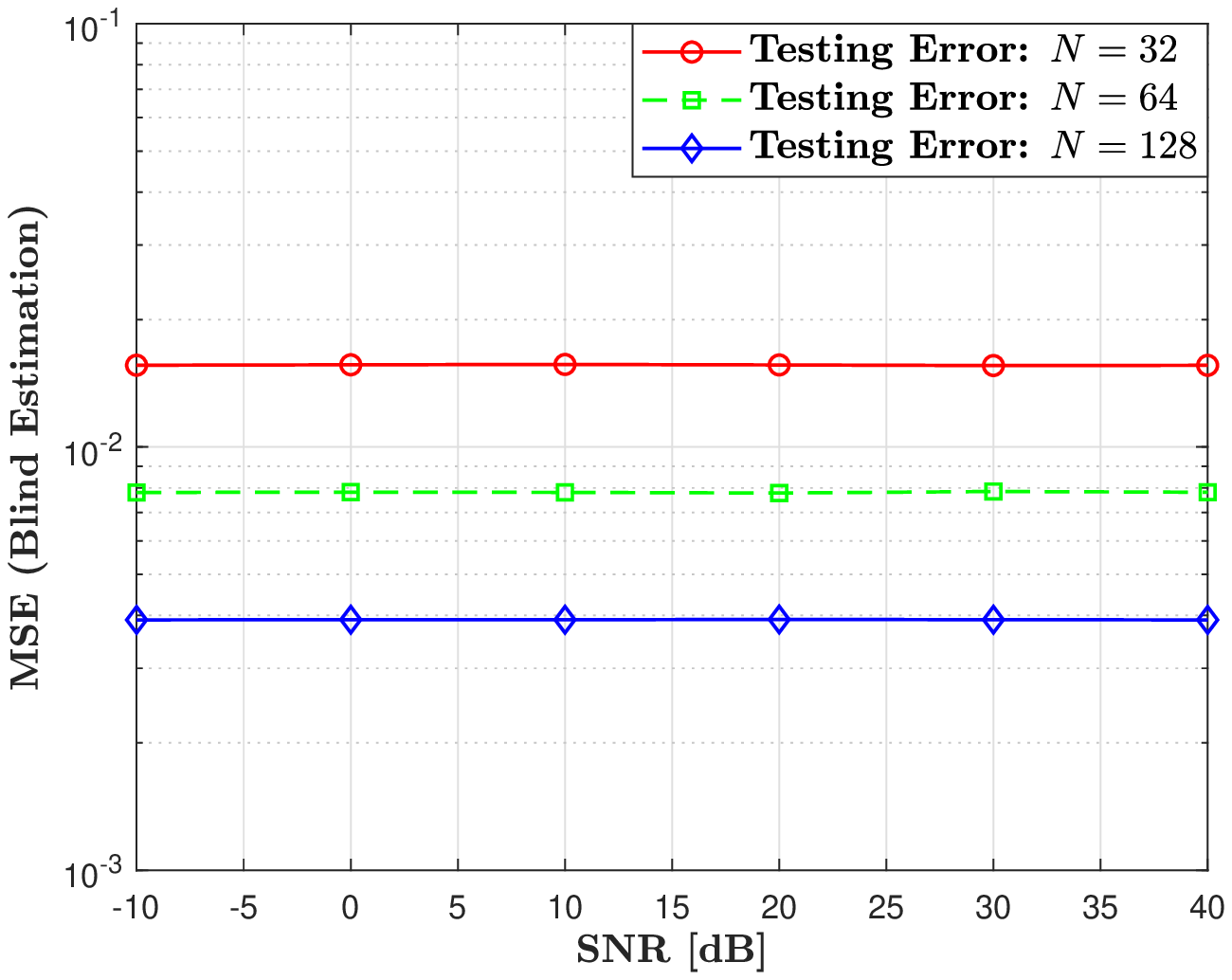}  \\ [5mm]
	\caption{Testing MSE versus SNR for $N\in\{32, 64, 128\}$ after training under the partial training scenario of Fig. \ref{ReLU_FNN_matrix_model}: FNN depth ($K+2$)$=3$; one trainable hidden layer.}  \vspace{5mm}
	\label{fig: ReLU_FNN_Performance_partial_trainable_layers_K_1}  
\end{figure}

\begin{figure}[htb!]
	\centering
	\includegraphics[width=\textwidth]{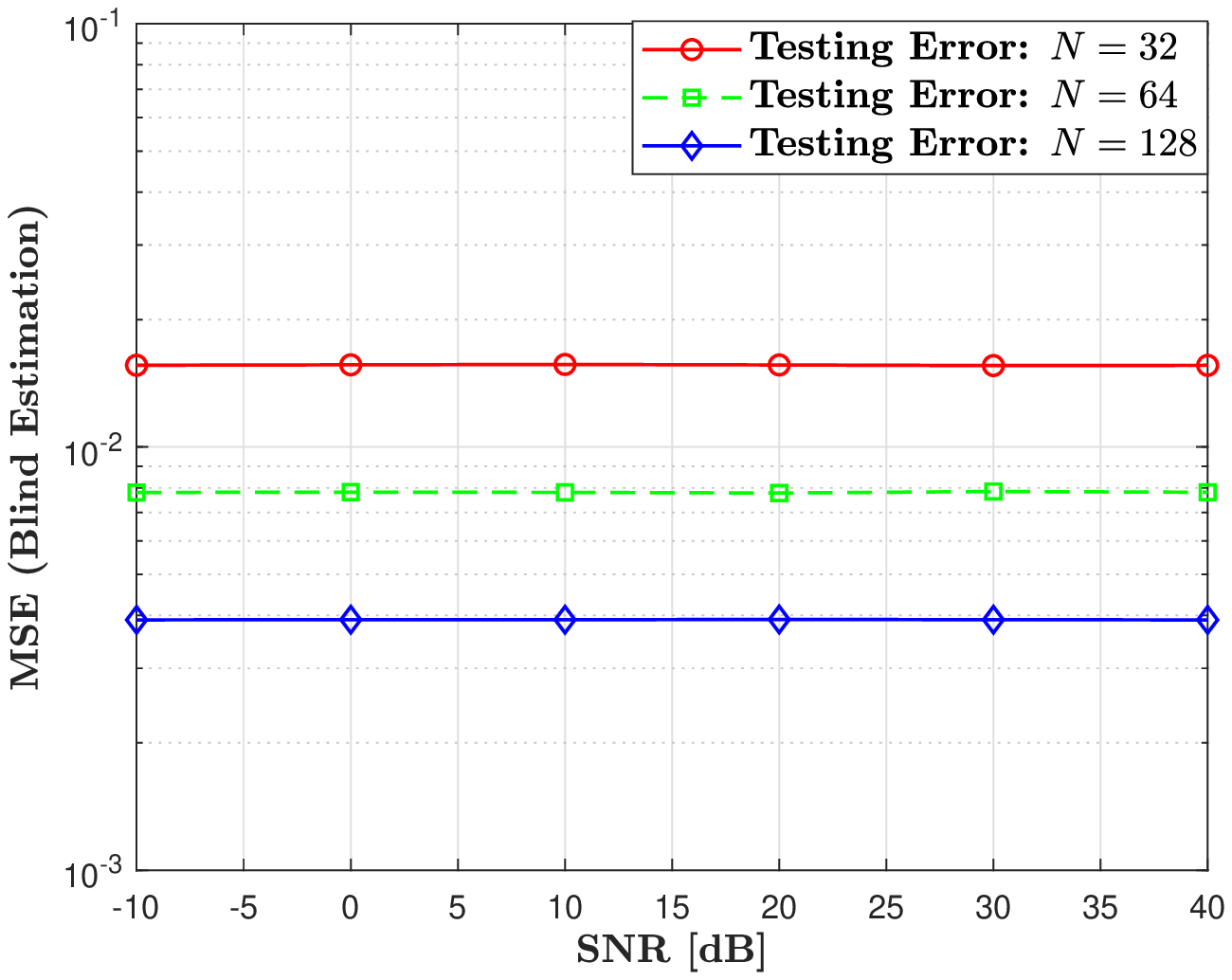}  \\ [5mm]
	\caption{Testing MSE versus SNR for $N\in\{32, 64, 128\}$ after training under the training scenario of all trainable layers: FNN depth ($K+2$)$=3$; all trainable layers.}   \vspace{5mm} 
	\label{fig: ReLU_FNN_Performance_all_trainable_layers_K_1}
\end{figure}

\begin{figure}[htb!]
	\centering
	\includegraphics[width=\textwidth]{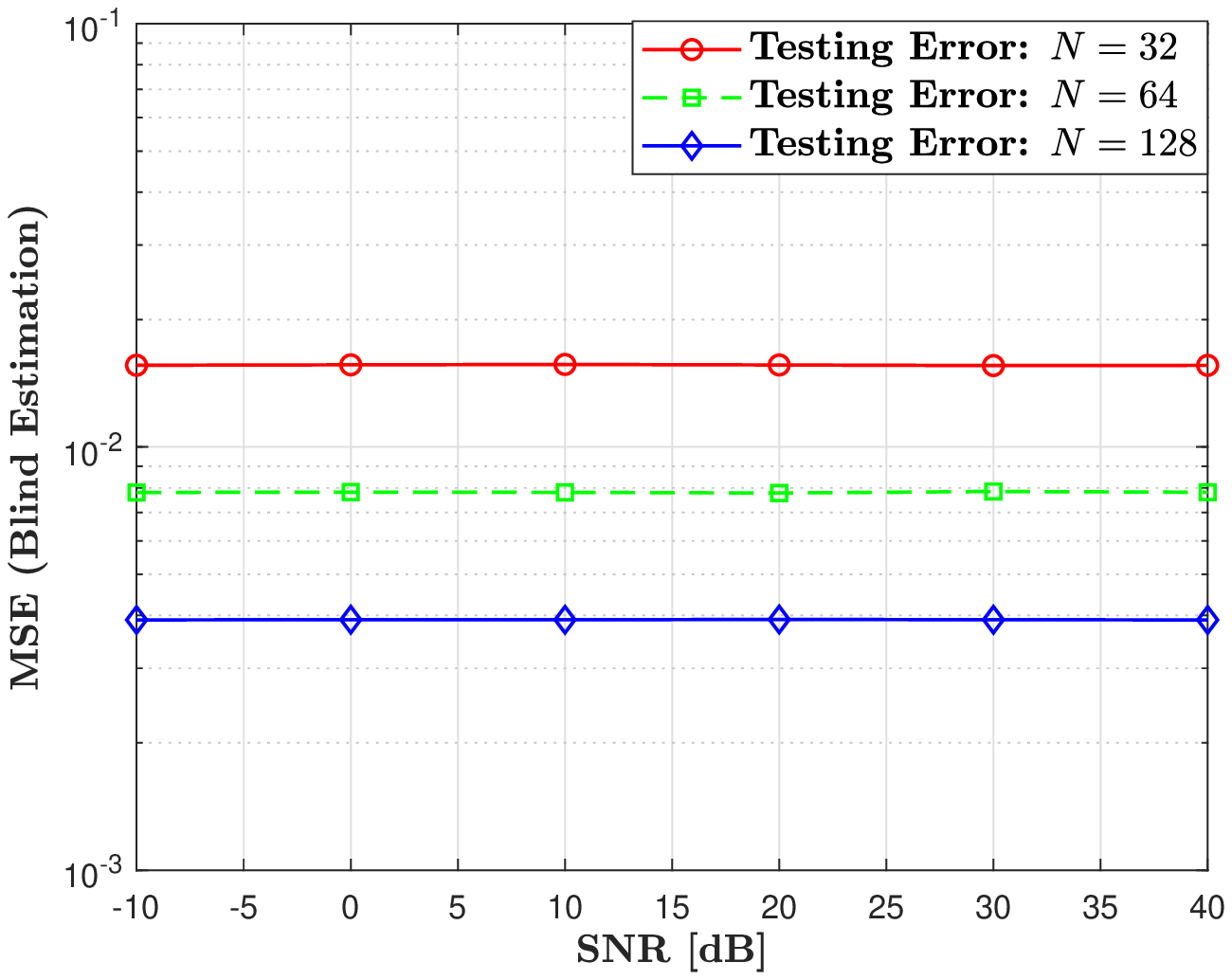}  \\ [5mm]
	\caption{Testing MSE versus SNR for $N\in\{32, 64, 128\}$ after training under the partial training scenario of Fig. \ref{ReLU_FNN_matrix_model}: FNN depth ($K+2$)$=6$; 4 trainable hidden layers.}   \vspace{5mm}
	\label{fig: ReLU_FNN_Performance_partial_trainable_layers}
\end{figure}

\begin{figure}[htb!]
	\centering
	\includegraphics[width=\textwidth]{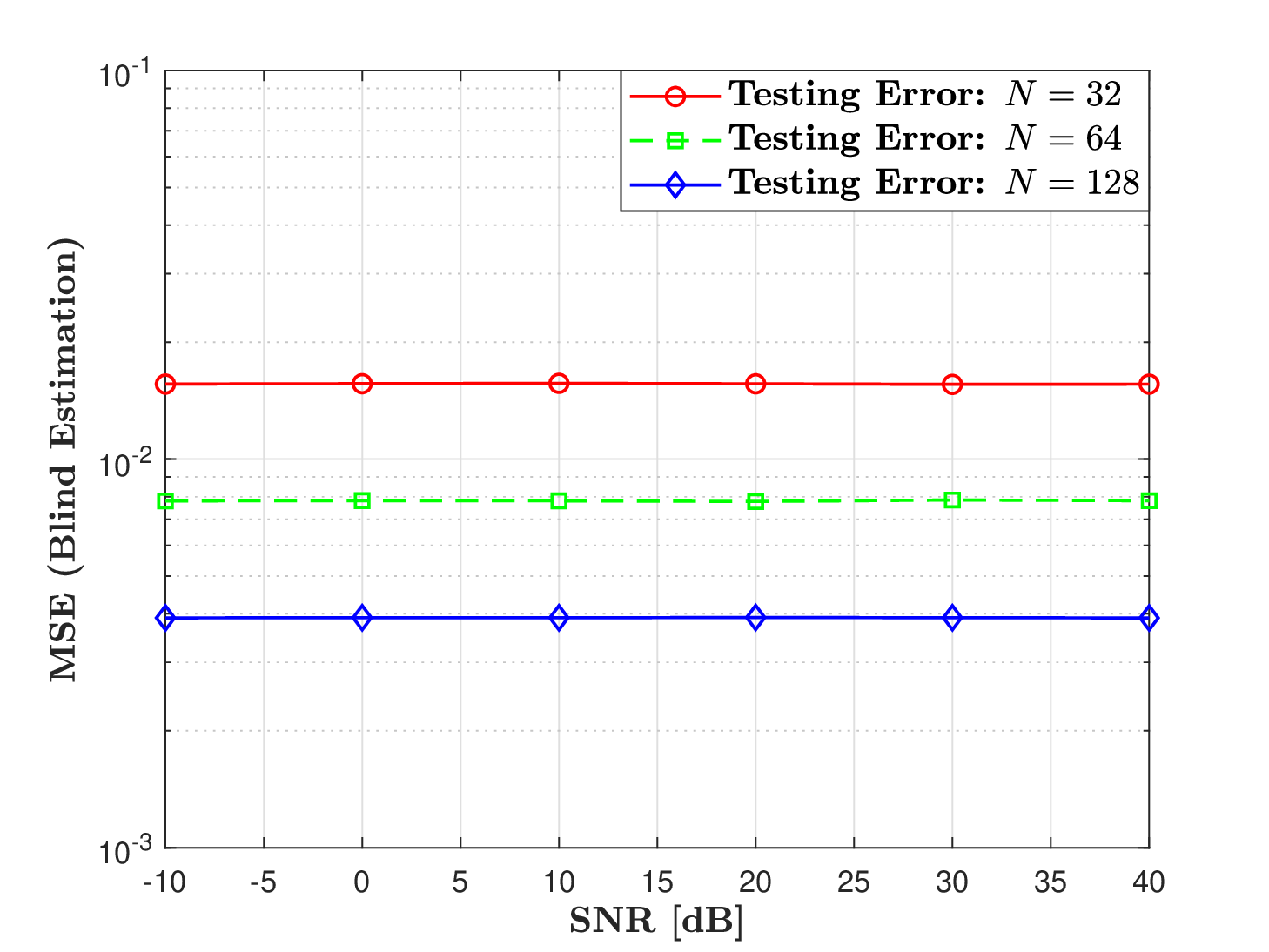}  \\ [4mm]
	\caption{Testing MSE versus SNR for $N\in\{32, 64, 128\}$ after training under the training scenario of all trainable layers: FNN depth ($K+2$)$=6$; all trainable layers.}   \vspace{5mm}
	\label{fig: ReLU_FNN_Performance_all_trainable_layers}
\end{figure}

The training performance and testing performance of our investigated blind channel estimator are displayed in Fig. \ref{fig: ReLU_training_and_validation_Losses_Plot_All_Cases_K_4} and Figs. \ref{fig: Deep_ReLU_FNN_Performance_Plot_All_Cases_K_4}-\ref{fig: ReLU_FNN_Performance_all_trainable_layers}, respectively. Figs. \ref{fig: ReLU_FNN_Performance_partial_trainable_layers_K_1}-\ref{fig: ReLU_FNN_Performance_all_trainable_layers} (also Tables \ref{table: ReLU_FNN_Performance_partial_trainable_layers_K_1}-\ref{table: ReLU_FNN_Performance_all_trainable_layers}) illustrate the blind channel estimator's testing MSE performance versus SNR for different $N$. Produced using the training generated per Algorithm \ref{Training_testing_datasets_generation} using $10^5$ OFDM symbols exhibiting random SNR, Fig. \ref{fig: ReLU_training_and_validation_Losses_Plot_All_Cases_K_4} displays a depth-6 ReLU FNN's training and validation losses over 10 epochs with $N=64$ and $m=2N^2$. Fig. \ref{fig: ReLU_training_and_validation_Losses_Plot_All_Cases_K_4} shows that the training and validation losses do not improve after the second epoch despite our exhaustive training endeavor over several seasons.\footnote{We have exhaustively tried all the known training tricks and techniques of deep learning. Meanwhile, since there are no blind doubly selective OFDM channel estimators proposed to date, we were not able to make a fair performance comparison between our proposed estimator and any other blind OFDM channel estimator. Toward the latter end, we hope that our DL-based OFDM channel estimator will serve as a performance baseline when future blind estimators are developed for an OFDM channel.} Besides, obtained using the training set (manifesting random SNR) and testing set generated per Algorithm \ref{Training_testing_datasets_generation}, respectively, using $10^5$ and $4\times 10^4$ OFDM symbols, Fig. \ref{fig: Deep_ReLU_FNN_Performance_Plot_All_Cases_K_4} shows a blind testing MSE performance of a depth-6 over-parameterized ReLU FNN. This over-parameterized ReLU FNN -- whose training and validation losses are shown in Fig. \ref{fig: ReLU_training_and_validation_Losses_Plot_All_Cases_K_4} -- has been tested for SNR values $\gamma_{snr}\in \big\{\minus 10, 0, 10, 20, 30, 40\big\}$ dB. At these SNR values, Fig. \ref{fig: Deep_ReLU_FNN_Performance_Plot_All_Cases_K_4} corroborates that the proposed doubly selective fading channel estimator exhibits similar blind estimation performance in terms of MSE. The MSE stagnation visible in Fig.~\ref{fig: ReLU_training_and_validation_Losses_Plot_All_Cases_K_4} implies an inference that the trained ReLU FNN attempted to learn a zero mean normal distribution which characterizes the subcarrier coupling matrix's non-zero elements modeled by the CE-BEM given by (\ref{BEM_equation}).

Figs. \ref{fig: ReLU_FNN_Performance_partial_trainable_layers_K_1}-\ref{fig: ReLU_FNN_Performance_all_trainable_layers} (also Tables \ref{table: ReLU_FNN_Performance_partial_trainable_layers_K_1}-\ref{table: ReLU_FNN_Performance_all_trainable_layers}) show the blind testing MSE performance of our proposed channel estimator for different $N$ and $K$ under the training scenarios of all trainable layers and partial trainable layers (of Fig. \ref{ReLU_FNN_matrix_model}). Per the partial trainable layers of Fig. \ref{ReLU_FNN_matrix_model}, Figs. \ref{fig: ReLU_FNN_Performance_partial_trainable_layers_K_1} and \ref{fig: ReLU_FNN_Performance_partial_trainable_layers} display the performance of our blind channel estimator for $K=1$ and $K=4$, respectively. For $K=1$ and $K=4$, Figs. \ref{fig: ReLU_FNN_Performance_partial_trainable_layers_K_1} and \ref{fig: ReLU_FNN_Performance_partial_trainable_layers} validate that the testing MSE performance of the ReLU FNN based blind channel estimator reduces by half as $N$ increases both from 32 to 64 and from 64 to 128. Such reductions corroborate the theoretical result predicted by Theorem \ref{Thm_lim_expectation_MSE}. In accord with Theorem \ref{Thm_lim_expectation_MSE}, Figs. \ref{fig: ReLU_FNN_Performance_all_trainable_layers_K_1} and ~\ref{fig: ReLU_FNN_Performance_all_trainable_layers} -- produced through the training scheme of all trainable hidden layers -- confirm that the testing MSE performance of the investigated blind channel estimator decreases by half when $N$ increases from 32 to 64 and from 64 to 128. Drawing lessons from these results and our developed theory, we eventually follow with our concluding remarks and research outlook, as stated in Section \ref{sec: DSFC_conc_rem_and_res_outlook}.

\begin{table*}[!htb]
	\centering
	\begin{tabular}{ | l | c | c | c | c | c | c | c | c | c | c | c | }
		\hline
		\multirow{2}{*}{Testing MSE under} & \multicolumn{6}{ c | }{SNR [dB]}   \\ \cline{2-7}
		 & -10 & 0 & 10 & 20 & 30 & 40         \\ \hline
		All trainable layers scenario & 0.00780965  & 0.00780495  & 0.00781821 & 0.00781436 & 0.00780191  & 0.00780724          \\ \hline
		Four trainable layers scenario &0.00780964  &0.00780488  &0.00781823  &0.00781432  &0.00780196  &0.00780713            \\ \hline
		
	\end{tabular}   \\ [5mm]
	\caption{Testing MSE versus SNR after training under the training scenarios of all trainable layers and four trainable layers: FNN depth ($K+2$)$=6$; training set (at random SNR) and testing set were generated using 100 000 and 40 000 OFDM symbols, respectively; and $m=2N^2$.}
	\label{table: Deep_ReLU_FNN_Performance_Plot_All_Cases_K_4}	
\end{table*}

\begin{table*}[!htb]
	\centering
	\begin{tabular}{ | l | c | c | c | c | c | c | c | c | c | c | c | }
		\hline
		\multirow{2}{*}{Testing MSE for} & \multicolumn{6}{ c | }{SNR [dB]}   \\ \cline{2-7}
		& -10 & 0 & 10 & 20 & 30 & 40        \\ \hline
		$N=32$ & 0.01559903  & 0.01563617  & 0.01566507  & 0.01562372  & 0.01557177  & 0.01558988         \\ \hline
		$N=64$ & 0.00780719  &0.00782051  &0.00781013  &0.00778281  &0.00784908  &0.00781623          \\ \hline
		$N=128$ & 0.00390365  &0.00390822  &0.00390656  &0.00391313  &0.00390769  &0.00390315         \\ \hline
		
	\end{tabular}   \\ [5mm]
	\caption{Testing MSE versus SNR for $N\in\{32, 64, 128\}$ after training under the partial training scenario of Fig. \ref{ReLU_FNN_matrix_model}: FNN depth ($K+2$)$=3$; one trainable hidden layer.}
	\label{table: ReLU_FNN_Performance_partial_trainable_layers_K_1}	
\end{table*}

\begin{table*}[!htb]
	\centering
	\begin{tabular}{ | l | c | c | c | c | c | c | c | c | c | c | c | }
		\hline
		\multirow{2}{*}{Testing MSE for} & \multicolumn{6}{ c | }{SNR [dB]}   \\ \cline{2-7}
		& -10 & 0 & 10 & 20 & 30 & 40        \\ \hline
		$N=32$ &0.01559796  &0.01563374  &0.01566395  &0.01562251  &0.01557054  &0.01558886            \\ \hline
		$N=64$ &0.0078073  &0.00782071  &0.00781014  &0.00778298  &0.00784912  &0.00781623            \\ \hline
		$N=128$ &0.00390379  &0.00390808  &0.00390675  &0.0039129  &0.00390762  &0.00390309            \\ \hline
		
	\end{tabular}   \\ [5mm]
	\caption{Testing MSE versus SNR for $N\in\{32, 64, 128\}$ after training under the training scenario of all trainable layers: FNN depth ($K+2$)$=3$; all trainable layers.}
	\label{table: ReLU_FNN_Performance_all_trainable_layers_K_1}	
\end{table*}

\begin{table*}[!htb]
	\centering
	\begin{tabular}{ | l | c | c | c | c | c | c | c | c | c | c | c | }
		\hline
		\multirow{2}{*}{Testing MSE for} & \multicolumn{6}{ c | }{SNR [dB]}   \\ \cline{2-7}
		& -10 & 0 & 10 & 20 & 30 & 40      \\ \hline
		$N=32$ &0.01559665  &0.01563301  &0.01566342  &0.0156216  &0.01556964  &0.01558796       \\ \hline
		$N=64$ &0.0078072  &0.0078205  &0.00781005  &0.00778276  &0.007849  &0.00781622       \\ \hline
		$N=128$ &0.00390333  &0.00390778  &0.00390632  &0.00391249  &0.00390729  & 0.00390274     \\ \hline
		
	\end{tabular}   \\ [5mm]
	\caption{Testing MSE versus SNR for $N\in\{32, 64, 128\}$ after training under the partial training scenario of Fig. \ref{ReLU_FNN_matrix_model}: FNN depth ($K+2$)$=6$; 4 trainable hidden layers.}
	\label{table: ReLU_FNN_Performance_partial_trainable_layers}	
\end{table*}

\begin{table*}[!htb]
	\centering
	\begin{tabular}{ | l | c | c | c | c | c | c | c | c | c | c | c | }
		\hline
		\multirow{2}{*}{Testing MSE for} & \multicolumn{6}{ c | }{SNR [dB]}   \\ \cline{2-7}
		& -10 & 0 & 10 & 20 & 30 & 40   \\ \hline
		$N=32$ &0.01559668  &0.01563304  &0.01566354  &0.0156215  &0.01556969  &0.01558812        \\ \hline
		$N=64$ &0.00780741  &0.00782064  &0.0078104  &0.00778296  &0.00784944  &0.00781625        \\ \hline
		$N=128$ &0.00390356  &0.00390786  &0.00390648  &0.0039127  &0.00390759  &0.00390293       \\ \hline
		
	\end{tabular}   \\ [5mm]
	\caption{Testing MSE versus SNR for $N\in\{32, 64, 128\}$ after training under the training scenario of all trainable layers: FNN depth ($K+2$)$=6$; all trainable layers.}
	\label{table: ReLU_FNN_Performance_all_trainable_layers}	
\end{table*}

\newpage

\section{Concluding Remarks and Research Outlook}
\label{sec: DSFC_conc_rem_and_res_outlook}
\subsection{Concluding Remarks}
\label{subsec: DSFC_conc_rem}
Inspired by the advancements of DL, we studied a DL-based blind channel estimator for an OFDM system operating over a doubly selective fading channel. For this channel filtering the transmitted OFDM symbols, we estimated the subcarrier coupling matrix by training and testing deep ReLU FNNs without any pilot symbols. Without any pilot symbols, we trained and tested a deep (and shallow) ReLU FNN-based OFDM channel estimator. In light of this estimator, we developed a theory -- at the junction of the theory of DL, high-dimensional probability, wireless communications, and signal processing -- on the testing MSE performance of a blind OFDM channel estimator based on over-parameterized ReLU FNNs. For over-parameterized ReLU FNN-based OFDM channel estimator, we quantified its asymptotic testing MSE performance w.r.t. infinite subcarriers. For subcarriers that are finite, we also derived a probabilistic lower bound for the non-asymptotic testing MSE manifested by a blind OFDM channel estimator based on over-parameterized ReLU FNNs. 
    
\subsection{Research Outlook}
\label{subsec: DSFC_res_outlook}
Apart from channel estimation and the respective testing MSE performance analysis restricted to OFDM systems, this paper inspires the testing MSE performance analysis of a generic DL-based regression algorithm and the development of a doubly selective channel modeling using DL. Employing DL for modeling and estimation further, this paper also points the way toward DL techniques for the modeling and estimation of a \textit{triply selective} MIMO fading channel \cite{C_Xiao_TS_MIMO_Ch_04}, which takes into account time-, frequency-, and angle-selectivity. Furthermore, the theoretical developments of this paper inspire fundamental research toward a unified theory of DL as applied to wireless communications, signal processing, and networking (e.g., the efficacy of DL-based algorithms for low SNR regimes \cite{fan2019selective}.).


\appendices

\newpage
\section{Proof of Theorem \ref{Thm_lim_expectation_MSE}}
\label{proof_Thm_lim_expectation_MSE}
We first underscore that the over-parameterized ReLU FNN-based blind OFDM channel estimator has been trained using \texttt{SGD} and its convergence can be characterized using Theorem \ref{SGD_over_parameterized_net_convergence} and Corollary \ref{SGD_convergence_prop}. Thus, it follows through Corollary \ref{SGD_convergence_prop} that     
\begin{equation}
\label{Conv_constraint_at_App_1}
\big\|\check{\bm{W}}-\bm{W} \big\|_{F}\leq\omega, 
\end{equation}	
where -- per Corollary \ref{SGD_convergence_prop} -- $\omega= O\bigg(\frac{\breve{n}^{3.5}\sqrt{d_y}}{\delta \sqrt{bm}}\log m \bigg)$, $\check{\bm{W}}\eqdef\big[\check{\bm{W}}_1 \hspace{1mm} \check{\bm{W}}_2 \ldots \check{\bm{W}}_K\big]\in\mathbb{R}^{m\times mK}$, and $\bm{W}\eqdef[\bm{W}_1 \hspace{1mm} \bm{W}_2 \ldots \hspace{1mm} \bm{W}_K ]\in\mathbb{R}^{m\times mK}$ for $\bm{W}_k\sim \mathcal{N}\big(\bm{0}, \frac{2}{m}\bm{I}_m\big)$ (per Definition \ref{Def_random_initialization}). With this in mind, we begin by applying expectation -- a linear operation -- to (\ref{MSE_cost_2}) as   
\begin{equation}
\label{Expect_MSE_cost_1}
\mathbb{E}\{ F(\check{\bm{W}}) \}=\frac{1}{d_y\check{n}} \sum_{i=\breve{n}+1}^ {\breve{n}+\check{n}}  \mathbb{E}\{F_i(\check{\bm{W}})\}, 
\end{equation}
where $d_y=2N^2$ and    
\begin{equation}
\label{MSE_definition_1}
F_i(\check{\bm{W}})=\| \bm{B}\check{\bm{x}}_{i,K}-\check{\bm{y}}_i^{*} \|^2.
\end{equation}
For $\bm{x}$, $\bm{y}$, $\bm{z}\in\mathbb{R}^{d_y}$, recall that $\|\bm{x}\|^2=\langle\bm{x},\bm{x}\rangle$, $\langle\bm{x}+\bm{y},\bm{z}\rangle=\langle\bm{x},\bm{z}\rangle+\langle\bm{y},\bm{z}\rangle$, and $\langle\bm{x},\bm{y}\rangle=\langle\bm{y},\bm{x}\rangle$ \cite[p. 15]{James_matrix_algebra_07}; and $\langle\bm{x},\bm{y}+\bm{z}\rangle=\langle\bm{x},\bm{y}\rangle+\langle\bm{x},\bm{z}\rangle$ \cite[p. 315-316]{RHCR13}. As a result,    
\begin{subequations}
\begin{align}    
\label{inner_product_1} 
F_i(\check{\bm{W}})& =\| \bm{B}\check{\bm{x}}_{i,K}-\check{\bm{y}}_i^{*} \|^2= \langle\bm{B}\check{\bm{x}}_{i,K}-\check{\bm{y}}_i^{*},\bm{B}\check{\bm{x}}_{i,K}-\check{\bm{y}}_i^{*}\rangle   \\
\label{inner_product_2}
&=\langle\bm{B}\check{\bm{x}}_{i,K}-\check{\bm{y}}_i^{*},\bm{B}\check{\bm{x}}_{i,K}\rangle+\langle\bm{B}\check{\bm{x}}_{i,K}-\check{\bm{y}}_i^{*},-\check{\bm{y}}_i^{*}\rangle    \\
\label{inner_product_3}
&=\langle\bm{B}\check{\bm{x}}_{i,K},\bm{B}\check{\bm{x}}_{i,K}-\check{\bm{y}}_i^{*}\rangle+\langle\bm{B}\check{\bm{x}}_{i,K}-\check{\bm{y}}_i^{*},-\check{\bm{y}}_i^{*}\rangle \\
\label{inner_product_4} 
&=\|\bm{B}\check{\bm{x}}_{i,K}\|^2-2\langle\bm{B}\check{\bm{x}}_{i,K},\check{\bm{y}}_i^{*}\rangle + \|\check{\bm{y}}_i^{*}\|^2    \\ 
\label{inner_product_5} 
&\leq \|\bm{B}\check{\bm{x}}_{i,K}\|^2 + \|\check{\bm{y}}_i^{*}\|^2.
\end{align}
\end{subequations}
Applying expectation to both sides of (\ref{inner_product_5}) gives 
\begin{equation}
\label{expect_inner_product_1} 
0\stackrel{(a)}{\leq}  \mathbb{E}\big\{F_i(\check{\bm{W}})\big\}\leq \mathbb{E}\big\{\|\bm{B}\check{\bm{x}}_{i,K}\|^2 \big\}+ \mathbb{E}\big\{\|\check{\bm{y}}_i^{*}\|^2\big\}, 
\end{equation}
where $(a)$ follows from (\ref{MSE_definition_1}) that $F_i(\check{\bm{W}})\geq 0$. To bound the RHS of (\ref{expect_inner_product_1}) and then the RHS of (\ref{Expect_MSE_cost_1}), we need to bound $\mathbb{E}\big\{\|\bm{B}\check{\bm{x}}_{i,K}\|^2 \big\}$ and simplify $\mathbb{E}\big\{\|\check{\bm{y}}_i^{*}\|^2\big\}$. To proceed, we prove the following fact for completeness.       
 
\begin{fact}
\label{fact_matrix_vector_product}
Let $r$ be the rank of $\tilde{\bm{A}}\in\mathbb{R}^{m\times n}$ and $\bm{x}\in\mathbb{R}^{n}$. If $\sigma_i$ is the $i$-th largest singular value of $\tilde{\bm{A}}$, then     
\begin{subequations}
\begin{align}
\label{Fact_1_1}
\sigma_i&\leq \frac{1}{\sqrt{i}} \|\tilde{\bm{A}}\|_F  \hspace{1mm}\text{\cite[Exercise 4.1.2]{vershynin_2018}}   \\
\label{Fact_1_2}
\|\tilde{\bm{A}}\bm{x}\|&\leq\sigma_1 \|\bm{x}\|   \leq \|\tilde{\bm{A}}\|_F \|\bm{x}\|   \\
\label{Fact_1_3}
\|\tilde{\bm{A}}\|_2 & \leq \|\tilde{\bm{A}}\|_F\leq \sqrt{r}\|\tilde{\bm{A}}\|_2\leq\sqrt{\textnormal{max}\{m,n\}}\|\tilde{\bm{A}}\|_2.
\end{align}
\end{subequations}

\proof From the compact SVD of $\tilde{\bm{A}}$, $\tilde{\bm{A}}=\bm{U}\bm{\Sigma}\bm{V}^T$, where $\bm{U}\in\mathbb{R}^{m\times r}$, $\bm{V}\in\mathbb{R}^{n\times r}$, $\bm{U}^T\bm{U}=\bm{V}^T\bm{V}=\bm{I}_r$, and $\bm{\Sigma}=\textnormal{diag}(\sigma_1, \sigma_2, \ldots,  \sigma_r)\in\mathbb{R}^{r\times r}$ for $r\leq\min\{m,n\}$. Following the definition of Frobenius norm (see Definition \ref{Frob_and_spec_norma}) \cite[p. 72]{vershynin_2018},  
\begin{equation}
\label{Frob_norm_1}
\|\tilde{\bm{A}}\|_F^2=\textnormal{tr}(\tilde{\bm{A}}^T\tilde{\bm{A}})=\textnormal{tr}\big(\bm{V}\bm{\Sigma}^T\bm{U}^T\bm{U}\bm{\Sigma}\bm{V}^T\big)=\textnormal{tr}\big(\bm{V}\bm{\Sigma}^2\bm{V}^T\big), 
\end{equation}
where $\bm{\Sigma}^2=\bm{\Sigma}^T\bm{\Sigma}$. Employing the trace property $\textnormal{tr}(\tilde{\bm{A}}\tilde{\bm{B}})=\textnormal{tr}(\tilde{\bm{B}}\tilde{\bm{A}})$ \cite{Magnus_07} in (\ref{Frob_norm_1}) simplifies the RHS of (\ref{Frob_norm_1}) to   
\begin{equation}
\label{Frob_norm_2}
\|\tilde{\bm{A}}\|_F^2=\textnormal{tr}\big(\bm{V}^T\bm{V}\bm{\Sigma}^2\big)=\textnormal{tr}\big(\bm{\Sigma}^2\big)=\sum_{i=1}^r \sigma_i^2 \stackrel{(a)}{\geq} i\sigma_i^2, \hspace{0.5mm} \forall i\in[r],  
\end{equation}
where $(a)$ follows from the fact that $\sigma_1\geq \sigma_2 \geq \ldots \geq \sigma_r$. Hence, $i\sigma_i^2\leq \|\tilde{\bm{A}}\|_F^2\Leftrightarrow \sigma_i\leq\frac{1}{\sqrt{i}}\|\tilde{\bm{A}}\|_F$. This completes the proof of (\ref{Fact_1_1}). 

To continue to the proof of (\ref{Fact_1_2}), we first recall that $\|\bm{x}\|=\sqrt{\langle\bm{x},\bm{x}\rangle}$. Using this identity,    
\begin{equation}
\label{Ax_bound_1}
\|\tilde{\bm{A}}\bm{x}\|=\sqrt{\langle\tilde{\bm{A}}\bm{x},\tilde{\bm{A}}\bm{x}\rangle}=\sqrt{\bm{x}^T\tilde{\bm{A}}^T\tilde{\bm{A}}\bm{x} }\stackrel{(a)}{=}\sqrt{\bm{x}^T\bm{V}\bm{\Sigma}^2\bm{V}^T\bm{x}}, 
\end{equation}
where $(a)$ follows from the compact SVD of $\tilde{\bm{A}}$ (i.e., $\tilde{\bm{A}}=\bm{U}\bm{\Sigma}\bm{V}^T$) and (\ref{Frob_norm_1}). Since $\sigma_1^2\bm{I}_r-\bm{\Sigma}^2\succeq \bm{0}$, $\bm{x}^T\bm{V}(\sigma_1^2\bm{I}_r-\bm{\Sigma}^2 ) \bm{V}^T\bm{x} \geq 0$. Thus, $\sigma_1^2\bm{x}^T\bm{V}\bm{V}^T\bm{x}-\bm{x}^T\bm{V}\bm{\Sigma}^2\bm{V}^T\bm{x}\geq 0$ and   
\begin{subequations}
	\begin{align}
		\label{Inequality_eq_1}
		\sigma_1^2\bm{x}^T\bm{V}\bm{V}^T\bm{x} & \geq \bm{x}^T\bm{V}\bm{\Sigma}^2\bm{V}^T\bm{x}   \\
		\label{Inequality_eq_2}
		\bm{x}^T\bm{V}\bm{\Sigma}^2\bm{V}^T\bm{x} &	\leq \sigma_1^2\bm{x}^T\bm{V}\bm{V}^T\bm{x}.	
	\end{align}
\end{subequations}  
Employing (\ref{Inequality_eq_2}) in the RHS of (\ref{Ax_bound_1}) leads to the inequality  
\begin{equation}
\label{Ax_bound_2}
\|\tilde{\bm{A}}\bm{x}\|\leq \sigma_1\sqrt{\bm{x}^T\bm{V}\bm{V}^T\bm{x}}\stackrel{(a)}{=}\sigma_1\sqrt{\bm{x}^T\bm{x}}=\sigma_1\sqrt{\langle\bm{x},\bm{x}\rangle} \stackrel{(b)}{=} \sigma_1\|\bm{x}\|, 
\end{equation}
where $(a)$ is due to the fact that $\bm{V}\bm{V}^T=\bm{I}_n$ which directly follows from $\bm{V}^T\bm{V}=\bm{I}_r$ for $\bm{V}\in\mathbb{R}^{n\times r}$ and $(b)$ is because of the fact that $\langle\bm{x},\bm{x}\rangle=\|\bm{x}\|^2$. Moreover, since $\sigma_1\leq \sqrt{\sum_{i=1}^r \sigma_i^2}$, it directly follows through (\ref{Ax_bound_2}) and (\ref{Frob_norm_2}) that
\begin{equation}
	\label{Ax_bound_3}
	\|\tilde{\bm{A}}\bm{x}\| \leq  \sigma_1 \|\bm{x}\| \leq \|\tilde{\bm{A}}\|_F \|\bm{x}\|.   
\end{equation} 
This finishes the proof of (\ref{Fact_1_2}).      

Finally, to prove (\ref{Fact_1_3}), we exploit the compact SVD (economic SVD) of $\tilde{\bm{A}}$, as stated above, with rank $r\leq\textnormal{min}\{m,n\}$: following the simplifications that led to (\ref{Frob_norm_2}), $\sigma_1^2 \leq \|\tilde{\bm{A}}\|_F^2=\sum_{i=1}^r \sigma_i^2\leq r \sigma_1^2$ and hence       
\begin{equation}
\label{Frob_norm_3}
\sigma_1\leq \|\tilde{\bm{A}}\|_F \leq \sqrt{r}\sigma_1.
\end{equation}
Recalling that $\|\tilde{\bm{A}}\|_2=\sigma_1$ \cite[eq. (5.2.7), p. 281]{CDM00} and the rank of $\tilde{\bm{A}}$ is bounded as $r\leq\textnormal{min}\{m,n\}\leq\textnormal{max}\{m,n\}$, (\ref{Frob_norm_3}) leads to   
\begin{equation}
\label{Frob_norm_4}
\|\tilde{\bm{A}}\|_2\leq \|\tilde{\bm{A}}\|_F \leq \sqrt{r}\|\tilde{\bm{A}}\|_2\leq\sqrt{\textnormal{max}\{m,n\}}\|\tilde{\bm{A}}\|_2.
\end{equation}
This completes the proof of (\ref{Fact_1_3}). This also ends the proof of Fact \ref{fact_matrix_vector_product}.    \QED
\end{fact}

Deploying Fact \ref{fact_matrix_vector_product} -- via (\ref{Fact_1_2}) -- and denoting that $\tilde{\sigma}_1$ is the maximum singular value of $\bm{B}$,     
\begin{equation}
\label{B_h_i_K}
\|\bm{B}\check{\bm{x}}_{i,K}\|^2\leq\tilde{\sigma}_1^2\|\check{\bm{x}}_{i,K}\|^2\leq\|\bm{B}\|_F^2\|\check{\bm{x}}_{i,K}\|^2. 
\end{equation}
Applying expectation to both sides of (\ref{B_h_i_K}),  
\begin{subequations}
\begin{align}
\label{Expec_B_h_i_K_1}
\mathbb{E}\big\{  \|\bm{B}\check{\bm{x}}_{i,K}\|^2  \big\}\leq &
\mathbb{E}\big\{\|\bm{B}\|_F^2\|\check{\bm{x}}_{i,K}\|^2\big\}    \\
\label{Expec_B_h_i_K_2}
\stackrel{(a)}{\leq} &
\mathbb{E}\big\{\|\bm{B}\|_F^2\|\big\} \mathbb{E}\big\{\|\check{\bm{x}}_{i,K}\|^2\big\}, 
\end{align}
\end{subequations}
where $(a)$ follows from the independence of the entries of $\bm{B}$ (a fixed Gaussian matrix which is not trainable) and $\check{\bm{x}}_{i,K}$ -- defined via (\ref{bm_x_i_K_definition}). From (\ref{bm_x_i_K_definition}), it follows that 
\begin{equation}
\label{ReLU_net_output}
\check{\bm{x}}_{i,K}=\Big(\prod_{k=0}^{K-1}\check{\bm{\Sigma}}_{i,K-k}\check{\bm{W}}_{K-k}\Big) \check{\bm{\Sigma}}_{i,0}\bm{A}\check{\bm{x}}_i.
\end{equation}
Employing Fact \ref{fact_matrix_vector_product} -- via (\ref{Fact_1_2}) -- in (\ref{ReLU_net_output}),    
\begin{subequations}
\begin{align}
\label{ReLU_net_output_1_1}
\|\check{\bm{x}}_{i,K}\|&\leq\big\|\big( \prod_{k=0}^{K-1} \check{\bm{\Sigma}}_{i,K-k} \check{\bm{W}}_{K-k} \big) \check{\bm{\Sigma}}_{i,0}\bm{A}\big\|_F \|\check{\bm{x}}_i\|       \\
\label{ReLU_net_output_1}
&\stackrel{(a)}{=}\big\|\big(\prod_{k=0}^{K-1} \check{\bm{\Sigma}}_{i,K-k} \check{\bm{W}}_{K-k} \big) \check{\bm{\Sigma}}_{i,0}\bm{A}\big\|_F,
\end{align}		
\end{subequations}	
where $(a)$ is due to Assumption \ref{normalization_assumption} that $\|\check{\bm{x}}_i\|=1$. Since $\|\bm{AB}\|_F\leq \|\bm{A}\|_F\|\bm{B}\|_F$ \cite[p. 279]{CDM00}, (\ref{ReLU_net_output_1}) leads to    
\begin{subequations}
\begin{align}
\label{ReLU_net_output_2_1}
\|\check{\bm{x}}_{i,K}\|&\leq \big\| \prod_{k=0}^{K-1} \check{\bm{\Sigma}}_{i,K-k} \check{\bm{W}}_{K-k} \big\|_F \big\|\check{\bm{\Sigma}}_{i,0}\big\|_F \|\bm{A}\|_F     \\
\label{ReLU_net_output_2}
&\stackrel{(a)}{\leq} \sqrt{m} \big\| \prod_{k=0}^{K-1} \check{\bm{\Sigma}}_{i,K-k} \check{\bm{W}}_{K-k} \big\|_F \|\bm{A}\|_F  \\
\label{ReLU_net_output_2_2}
&\stackrel{(b)}{\leq} \sqrt{m}  \prod_{k=0}^{K-1} \underbrace{ \big\| \check{\bm{\Sigma}}_{i,K-k}\check{\bm{W}}_{K-k}  \big\|_F }_{=V_k} \|\bm{A}\|_F, 
\end{align}
\end{subequations}
where $(a)$ is due to the fact $\|\check{\bm{\Sigma}}_{i,0}\|_F\leq\|\bm{I}_m\|_F=\sqrt{m}$ for $\check{\bm{\Sigma}}_{i,0}=\textnormal{diag}\big( \{0,1\}^{1 \times m} \big)$ and $(b)$ follows from the repeated application of the identity $\|\bm{AB}\|_F\leq \|\bm{A}\|_F\|\bm{B}\|_F$ \cite[p. 279]{CDM00}.         

To proceed further, we simplify $V_k$ as defined in (\ref{ReLU_net_output_2_2}). Accordingly,      
\begin{subequations}
\begin{align}
\label{V_k_expression_1}
V_k=\| \check{\bm{\Sigma}}_{i,K-k}\check{\bm{W}}_{K-k} \|_F &\stackrel{(a)}{=} \|(\check{\bm{\Sigma}}_{i,K-k}\check{\bm{W}}_{K-k}-\check{\bm{\Sigma}}_{i,K-k}\bm{W}_{K-k})+\check{\bm{\Sigma}}_{i,K-k}\bm{W}_{K-k}\|_F   \\
\label{V_k_expression_2}
&\stackrel{(b)}{\leq} \|\underbrace{\check{\bm{\Sigma}}_{i,K-k}\check{\bm{W}}_{K-k}-\check{\bm{\Sigma}}_{i,K-k}\bm{W}_{K-k}}_{=\check{\bm{X}}_k}\|_F+ \|\underbrace{\check{\bm{\Sigma}}_{i,K-k}\bm{W}_{K-k}}_{=\check{\bm{Y}}_k}\|_F   \\
\label{V_k_expression_3}
&=\|\check{\bm{X}}_k\|_F+ \|\check{\bm{Y}}_k\|_F, 
\end{align}
\end{subequations}
where $(a)$ is due to adding $\bm{0}=\check{\bm{\Sigma}}_{i,K-k}\bm{W}_{K-k}-\check{\bm{\Sigma}}_{i,K-k}\bm{W}_{K-k}$ for $\bm{W}_k\sim \mathcal{N}\big(\bm{0}, \frac{2}{m}\bm{I}_m\big)$ and $(b)$ follows from the fact that Frobenius norm satisfies the definition of matrix norm \cite[p. 280]{CDM00}. To bound (\ref{V_k_expression_3}), we are going to bound $\|\check{\bm{X}}_k\|_F$ and $\|\check{\bm{Y}}_k\|_F$ -- as defined in (\ref{V_k_expression_2}). To begin with $\|\check{\bm{Y}}_k\|_F$,     
\begin{equation}
\label{Y_k_check_expression_1}
\| \check{\bm{Y}}_k \|_F =\|  \check{\bm{\Sigma}}_{i,K-k}\bm{W}_{K-k} \|_F \stackrel{(a)}{\leq} \| \check{\bm{\Sigma}}_{i,K-k} \|_F \| \bm{W}_{K-k} \|_F \stackrel{(b)}{\leq} \sqrt{m} \|\bm{W}_{K-k}\|_F, 
\end{equation}      
where $(a)$ follows from the identity $\|\bm{AB}\|_F\leq \|\bm{A}\|_F\|\bm{B}\|_F$ \cite[p. 279]{CDM00} and $(b)$ is due to the fact $\|\bm{\Sigma}_{i,K-k}\|_F\leq\|\bm{I}_m\|_F=\sqrt{m}$ for $\check{\bm{\Sigma}}_{i,K-k} = \textnormal{diag}\big( \{0,1\}^{1 \times m} \big)$. To continue, we are going to prove the following lemma.                 
\begin{lemma}
\label{Lem_Frob_norm_Y_k_check}
Let $\bar{C}$ be an absolute constant and $m<< \bar{C} <\infty$. Let $\bm{W}_{K-k}\sim \mathcal{N}\big(\bm{0}, \frac{2}{m}\bm{I}_m\big)$. Then, with a high probability,   
\begin{equation}
\label{Lem_Frob_norm_Y_k_check_bound}
\sqrt{m} \| \bm{W}_{K-k} \|_F \leq \bar{C}. 
\end{equation}  
\proof We are going to derive the upper bound of $\mathbb{P}(\sqrt{m} \| \bm{W}_{K-k} \|_F > \bar{C})$. From the to be derived upper bound, we will then deduce (\ref{Lem_Frob_norm_Y_k_check_bound}). To begin, assume an absolute constant $C>0$. W.r.t. $C>0$, it follows that      
\begin{subequations}
\begin{align}
\label{Lem_Frob_norm_Y_k_check_bound_1}
\mathbb{P}\big(\sqrt{m} \|\bm{W}_{K-k}\|_F > C\big)& \stackrel{(a)}{\leq} \mathbb{P}\big(\sqrt{m} \|\bm{W}_{K-k}\|_F \geq C\big)     \\
\label{Lem_Frob_norm_Y_k_check_bound_2}
&=\mathbb{P}\big(\| \bm{W}_{K-k} \|_F \geq C/\sqrt{m}\big)       \\
\label{Lem_Frob_norm_Y_k_check_bound_3}
&\stackrel{(b)}{\leq} \frac{\sqrt{m}}{C} \mathbb{E}\big\{ \|\bm{W}_{K-k} \|_F \big\}   \\
\label{Lem_Frob_norm_Y_k_check_bound_4}  
&= \frac{\sqrt{m}}{C} \mathbb{E}\big\{ \sqrt{ \|\bm{W}_{K-k} \|_F^2} \big\}         \\
\label{Lem_Frob_norm_Y_k_check_bound_5}
&\stackrel{(c)}{\leq}  \frac{\sqrt{m}}{C}  \sqrt{ \mathbb{E}\big\{ \|\bm{W}_{K-k} \|_F^2 \big\}}, 
\end{align}
\end{subequations}
where $(a)$ is because of the fact that $\mathbb{P}\big(\sqrt{m} \|\bm{W}_{K-k} \|_F \geq C\big)=\mathbb{P}\big(\sqrt{m} \|\bm{W}_{K-k} \|_F > C\big)+\mathbb{P}\big(\sqrt{m} \| \bm{W}_{K-k} \|_F = C\big)\geq \mathbb{P}\big(\sqrt{m} \|\bm{W}_{K-k} \|_F > C\big)$; $(b)$ is because of Markov's inequality (see Proposition \ref{Prop_Markov_inequality}); and $(c)$ is due to Jensen's inequality (see Proposition \ref{Prop_Jensen_inequality}) w.r.t. the concavity of the square-root function. Employing the definition of Frobenius norm (see Definition \ref{Frob_and_spec_norma}) w.r.t. $\bm{W}_{K-k}\in\mathbb{R}^{m\times m}$ and $\big(\bm{W}_{K-k}\big)_{i,j}\sim \mathcal{N}(0, 2/m)$ in regards to Definition \ref{Def_random_initialization},              
\begin{subequations}
\begin{align}
\label{Lem_Frob_norm_Y_k_check_bound_6}
\sqrt{ \mathbb{E}\big\{ \|\bm{W}_{K-k} \|_F^2 \big\}} &=\sqrt{ \sum_{i=1}^m \sum_{j=1}^m \mathbb{E}\big\{ \big(\bm{W}_{K-k}\big)_{i,j}^2 \big\} }   \\
\label{Lem_Frob_norm_Y_k_check_bound_7}
&=\sqrt{m^2 \times \frac{2}{m}}=\sqrt{2m}, 
\end{align}
\end{subequations}
where (\ref{Lem_Frob_norm_Y_k_check_bound_7}) is valid for all $k\in[K]$. Substituting (\ref{Lem_Frob_norm_Y_k_check_bound_7}) into the RHS of (\ref{Lem_Frob_norm_Y_k_check_bound_5}),      
\begin{equation}
\label{Lem_Frob_norm_Y_k_check_bound_8}
\mathbb{P}\big(\sqrt{m} \| \bm{W}_{K-k} \|_F > C\big) \leq \sqrt{2} \frac{m}{C}.
\end{equation}
If we now choose $C=\bar{C}>>m$ and $C=\bar{C}<\infty$, the axiomatic definition of probability and (\ref{Lem_Frob_norm_Y_k_check_bound_8}) give $0\leq \mathbb{P}\big(\sqrt{m} \|\bm{W}_{K-k} \|_F > \bar{C}\big) \leq 0$. As a result, when $m<< C=\bar{C}<\infty$,      
\begin{equation}
\label{Lem_Frob_norm_Y_k_check_bound_9}
\mathbb{P}\big(\sqrt{m} \| \bm{W}_{K-k} \|_F \leq \bar{C}\big)=1-\mathbb{P}\big(\sqrt{m} \|\bm{W}_{K-k} \|_F > \bar{C}\big)=1.
\end{equation}
Therefore, as long as $m<<\bar{C}<\infty$, (\ref{Lem_Frob_norm_Y_k_check_bound_9}) corroborates (\ref{Lem_Frob_norm_Y_k_check_bound}) and hence Lemma \ref{Lem_Frob_norm_Y_k_check}. This concludes the proof of Lemma \ref{Lem_Frob_norm_Y_k_check}.           \QED       
\end{lemma}       

For $m<<\bar{C}<\infty$, thus, plugging (\ref{Lem_Frob_norm_Y_k_check_bound}) into the RHS of (\ref{Y_k_check_expression_1}) gives the bound   
\begin{equation}
\label{Y_k_check_expression_2}
\| \check{\bm{Y}}_k \|_F \leq \bar{C}.  
\end{equation}
Meanwhile, employing (\ref{Y_k_check_expression_2}) in the RHS of (\ref{V_k_expression_3}),
\begin{equation}
\label{V_k_expression_4}
V_k\leq \| \check{\bm{X}}_k \|_F+ \bar{C}, 
\end{equation} 
where $m<<\bar{C}<\infty$. Substituting (\ref{V_k_expression_4}) into (\ref{ReLU_net_output_2_2}),    
\begin{subequations}
\begin{align}  
\label{ReLU_net_output_2_2_1}
\|\check{\bm{x}}_{i,K}\| &\leq \sqrt{m} \|\bm{A}\|_F \prod_{k=0}^{K-1}   \big(\|\check{\bm{X}}_k \|_F+ \bar{C}\big)        \\
\label{ReLU_net_output_2_2_2}
&\stackrel{(a)}{\leq} \sqrt{m} \|\bm{A}\|_F \prod_{k=0}^{K-1}   \exp\big(\|\check{\bm{X}}_k\|_F+ \bar{C}\big)      \\
\label{ReLU_net_output_2_2_3}
&= \sqrt{m}\exp\big(\bar{C}K\big) \|\bm{A}\|_F \prod_{k=0}^{K-1}  \exp\big(\|\check{\bm{X}}_k\|_F\big)    \\
\label{ReLU_net_output_2_2_4}
&\stackrel{(b)}{\leq} \sqrt{m}\exp\big(\bar{C}K\big) \|\bm{A}\|_F \prod_{k=0}^{K-1}   \exp\big(\|\check{\bm{X}}_k\|_F^2\big),
\end{align}
\end{subequations}
where $(a)$ follows from the fact that $x\leq\exp(x)$ and for $m<<\bar{C}<\infty$; $(b)$ is due to the inequality $\exp(x)\leq\exp(x^2)$ for $x\geq 0$. To proceed, we require the following lemma.      
\begin{lemma}
\label{Lem_Frob_norm_X_k_check}
For $\check{\bm{X}}_k$ defined in (\ref{V_k_expression_2}) and $\bm{W}_{K-k}\sim \mathcal{N}\big(\bm{0}, \frac{2}{m}\bm{I}_m\big)$,    
\begin{equation}
\label{Lem_Frob_norm_X_k_check_bound_1}
\prod_{k=0}^{K-1}  \exp\big(\| \check{\bm{X}}_k \|_F^2\big)\leq \exp(m\omega^2K).
\end{equation} 

\proof Substituting $\check{\bm{X}}_k=\check{\bm{\Sigma}}_{i,K-k}\check{\bm{W}}_{K-k}-\check{\bm{\Sigma}}_{i,K-k}\bm{W}_{K-k}$ into the LHS of (\ref{Lem_Frob_norm_X_k_check_bound_1}),    
\begin{equation}
\label{Lem_Frob_norm_X_k_check_bound_2}
\prod_{k=0}^{K-1} \exp\big(\|\check{\bm{X}}_k \|_F^2\big)=\prod_{k=0}^{K-1}   \exp\big(\|\check{\bm{\Sigma}}_{i,K-k}\check{\bm{W}}_{K-k}-\check{\bm{\Sigma}}_{i,K-k}\bm{W}_{K-k} \|_F^2\big).
\end{equation} 
From the convergence constraint given by (\ref{Conv_constraint_at_App_1}), it follows that   
\begin{subequations}
\begin{align}
\label{Lem_Frob_norm_X_k_check_bound_3}
\|\check{\bm{W}}-\bm{W} \|_{F}^2=\sum_{k=0}^{K-1} \|\check{\bm{W}}_{K-k}-\bm{W}_{K-k} \|_{F}^2 & \leq\omega^2   \\
\label{Lem_Frob_norm_X_k_check_bound_4}
\underbrace{\sum_{k'=0}^{K-1} \| \check{\bm{\Sigma}}_{i,K-k'} \|_F^2\sum_{k=0}^{K-1} \|\check{\bm{W}}_{K-k}-\bm{W}_{K-k} \|_{F}^2 }_{=f(x)} & \stackrel{(a)}{\leq} \omega^2 \sum_{k'=0}^{K-1} \| \check{\bm{\Sigma}}_{i,K-k'} \|_F^2   \\
\label{Lem_Frob_norm_X_k_check_bound_5}
f(x) & \stackrel{(b)}{\leq} m\omega^2K,   
\end{align}
\end{subequations}
where $(a)$ follows from multiplying both sides of (\ref{Lem_Frob_norm_X_k_check_bound_3}) by $\sum_{k'=0}^{K-1} \|\check{\bm{\Sigma}}_{i,K-k'}\|_F^2$ and $(b)$ follows from the bound $\sum_{k'=0}^{K-1} \| \check{\bm{\Sigma}}_{i,K-k'} \|_F^2\leq mK$ in respect to $\check{\bm{\Sigma}}_{i,K-k'}=\textnormal{diag}\big( \{0,1\}^{1 \times m} \big)$ which is valid for all $k'$ and letting  
\begin{subequations}
	\begin{align}
	\label{f_x_def_1}
	\hspace{-2mm}f(x)= &\sum_{k'=0}^{K-1} \sum_{k=0}^{K-1} \| \check{\bm{\Sigma}}_{i,K-k'} \|_F^2 \|\check{\bm{W}}_{K-k}-\bm{W}_{K-k} \|_{F}^2    \\
	\label{f_x_def_2}
	=& \sum_{k=0}^{K-1} \| \check{\bm{\Sigma}}_{i,K-k} \|_F^2 \|\check{\bm{W}}_{K-k}-\bm{W}_{K-k} \|_{F}^2 + \sum_{k'=0}^{K-1} \sum_{k=0, k\neq k'}^{K-1} \| \check{\bm{\Sigma}}_{i,K-k'} \|_F^2 \|\check{\bm{W}}_{K-k}-\bm{W}_{K-k} \|_{F}^2   \\
	\label{f_x_def_3}
	\geq &  \sum_{k=0}^{K-1} \| \check{\bm{\Sigma}}_{i,K-k} \|_F^2 \|\check{\bm{W}}_{K-k}-\bm{W}_{K-k} \|_{F}^2. 
	\end{align}
\end{subequations} 

It then follows directly from (\ref{f_x_def_3}) and (\ref{Lem_Frob_norm_X_k_check_bound_5}) that  
\begin{subequations}
\begin{align}
\label{Lem_Frob_norm_X_k_check_bound_6}
\sum_{k=0}^{K-1} \| \check{\bm{\Sigma}}_{i,K-k} \|_F^2 \|\check{\bm{W}}_{K-k}-\bm{W}_{K-k} \|_{F}^2 &\leq m\omega^2K         \\  
\label{Lem_Frob_norm_X_k_check_bound_7}
\sum_{k=0}^{K-1} \|\check{\bm{\Sigma}}_{i,K-k}\check{\bm{W}}_{K-k}-\check{\bm{\Sigma}}_{i,K-k}\bm{W}_{K-k}\|_{F}^2 &\stackrel{(a)}{\leq} m\omega^2K                  \\ 
\label{Lem_Frob_norm_X_k_check_bound_8}
\prod_{k=0}^{K-1} \exp\big(\|\check{\bm{\Sigma}}_{i,K-k}\check{\bm{W}}_{K-k}-\check{\bm{\Sigma}}_{i,K-k}\bm{W}_{K-k} \|_{F}^2\big) &\stackrel{(b)}{\leq} \exp\big(m\omega^2K\big), 
\end{align}
\end{subequations}
where $(a)$ is due to the identity $\|\bm{AB}\|_F\leq \|\bm{A}\|_F\|\bm{B}\|_F$ \cite[p. 279]{CDM00} and $(b)$ follows by applying the exponential function to both sides of (\ref{Lem_Frob_norm_X_k_check_bound_7}). 

Therefore, exploiting (\ref{Lem_Frob_norm_X_k_check_bound_8}) in the RHS of (\ref{Lem_Frob_norm_X_k_check_bound_2}),           
\begin{equation}
\label{Lem_Frob_norm_X_k_check_bound_9}
\prod_{k=0}^{K-1}   \exp\big(\big\|\check{\bm{X}}_k\big\|_F^2\big)\leq\exp\big(m\omega^2K\big).
\end{equation} 
This is exactly (\ref{Lem_Frob_norm_X_k_check_bound_1}) and the end of Lemma \ref{Lem_Frob_norm_X_k_check}'s proof. \QED
\end{lemma}

Employing (\ref{Lem_Frob_norm_X_k_check_bound_9}) in the RHS of (\ref{ReLU_net_output_2_2_4}),     
\begin{subequations}
\begin{align}  
\label{ReLU_net_output_2_2_5}
\|\check{\bm{x}}_{i,K}\| & \leq \sqrt{m}\exp\big(\bar{C}K+m\omega^2K\big) \|\bm{A}\|_F   \\
\label{ReLU_net_output_2_2_6}
\|\check{\bm{x}}_{i,K}\|^2 & \leq m\exp\big(2\bar{C}K+2m\omega^2K\big) \|\bm{A}\|_F^2,       
	\end{align}
\end{subequations}
where $m<< \bar{C}< \infty$. Applying expectation to (\ref{ReLU_net_output_2_2_6}),    
\begin{equation}
\label{Expect_ReLU_net_output_1}
\mathbb{E}\big\{\|\check{\bm{x}}_{i,K}\|^2\big\}\leq m\exp\big(2\bar{C}K+2m\omega^2K\big)\mathbb{E}\big\{\|\bm{A}\|_F^2\big\}.
\end{equation}
Meanwhile, deploying (\ref{Expect_ReLU_net_output_1}) in the RHS of (\ref{Expec_B_h_i_K_2}),     
\begin{equation}
\label{Expec_B_h_i_K_3}
\mathbb{E}\big\{ \|\bm{B}\check{\bm{x}}_{i,K}\|^2  \big\} \leq 
me^{(2\bar{C}K+2m\omega^2K)}\mathbb{E}\big\{\|\bm{B}\|_F^2\|\big\}\mathbb{E}\big\{\|\bm{A}\|_F^2\big\}.
\end{equation}
To simplify the RHS of (\ref{Expec_B_h_i_K_3}), we require the following lemma.    
\begin{lemma}
\label{expec_square_simp}
With regard to the random initialization of Definition \ref{Def_random_initialization} and our channel estimation problem setting,    
\begin{equation*}
\label{Expec_bm_A_B_F_sq_lemma}
\mathbb{E}\big\{\|\bm{A}\|_F^2\big\}=2(2N+1) \hspace{2mm} \textnormal{and} \hspace{2mm} \mathbb{E}\big\{\|\bm{B}\|_F^2\big\}=m.
\end{equation*} 

\proof Deploying Definition \ref{Def_random_initialization} and the linearity of expectation,      
\begin{subequations}
\begin{align}
\label{Expec_bm_A_F_sq_1}
\mathbb{E}\big\{\|\bm{A}\|_F^2\big\}=&\sum_{i=1}^{m}\sum_{j=1}^{d_x} \mathbb{E}\big\{(\bm{A})_{i,j}^2\big\}\stackrel{(a)}{=}\sum_{i=1}^{m}\sum_{j=1}^{2N+1} \mathbb{E}\big\{(\bm{A})_{i,j}^2\big\}   \\
\label{Expec_bm_B_F_sq_1}
\mathbb{E}\big\{\|\bm{B}\|_F^2\big\}=&\sum_{i=1}^{d_y}\sum_{j=1}^{m} \mathbb{E}\big\{(\bm{B})_{i,j}^2\big\}\stackrel{(b)}{=}\sum_{i=1}^{2N^2}\sum_{j=1}^{m} \mathbb{E}\big\{(\bm{B})_{i,j}^2\big\}, 
\end{align}
\end{subequations} 
where $(a)$ and $(b)$ follow from our problem setup that $d_x=2N+1$ and $d_y=2N^2$, respectively. Recalling the settings of Definition \ref{Def_random_initialization}, $\mathbb{E}\big\{(\bm{A})_{i,j}^2\big\}=2/m$ and $\mathbb{E}\big\{(\bm{B})_{i,j}^2\big\}=1/2N^2$. As a result,       
\begin{equation}
\label{Expec_bm_A_B_F_sq_1}
\mathbb{E}\big\{\|\bm{A}\|_F^2\big\}=2(2N+1) \hspace{2mm} \textnormal{and} \hspace{2mm} \mathbb{E}\big\{\|\bm{B}\|_F^2\big\}=m.      
\end{equation}
This completes the proof of Lemma \ref{expec_square_simp}.   \QED
\end{lemma}

Substituting (\ref{Expec_bm_A_B_F_sq_1}) into (\ref{Expec_B_h_i_K_3}) then results in the bound       
\begin{equation}
\label{Expec_B_h_i_K_4}
\mathbb{E}\big\{  \|\bm{B}\check{\bm{x}}_{i,K}\|^2  \big\} \leq 2(2N+1)m^2e^{(2\bar{C}K+2m\omega^2K)}.
\end{equation}

With regard to our progress of bounding (\ref{expect_inner_product_1}), we are now left with simplifying $\mathbb{E}\big\{\|\check{\bm{y}}_i^{*}\|^2\big\}$. Toward this end, we continue to the simplification of $\|\check{\bm{y}}_i^{*}\|^2$ as pursued in the subsequent simplifications that led to the following lemma.          
\begin{lemma}
\label{On_channel_energy}
Concerning the system model of Sec. \ref{sec: system_model}, the CE-BEM model of Sec. \ref{subsec: DSC_prelims}, and the doubly selective fading channel generation technique of Sec. \ref{subsec: Gen_DSF_channel},        	
\begin{equation*}
\label{Expect_norm_check_y_i_lemma}
\mathbb{E}\big\{\|\check{\bm{y}}_i^{*}\|^2\big\}=N.
\end{equation*}

\proof Recalling that the training labels are the real and the imaginary parts of the subcarrier coupling matrix, it follows through (\ref{test_label_vector_1}) that    
\begin{equation}
\label{norm_check_y_i_2_1}
\|\check{\bm{y}}_i^{*}\|^2=\big\|\tilde{\bm{H}}^i\big\|_F^2=\textnormal{tr}\big((\tilde{\bm{H}}^i)^H\tilde{\bm{H}}^i\big).
\end{equation}
Substituting (\ref{H_sc_matrix}) into (\ref{norm_check_y_i_2_1}) and recalling that the DFT matrix $\bm{F}_N$ satisfies $\bm{F}_N^H\bm{F}_N=\bm{F}_N\bm{F}_N^H=\bm{I}_N$ leads to  
\begin{subequations}
\begin{align}
\label{norm_check_y_i_2_2}
\|\check{\bm{y}}_i^{*}\|^2&=\textnormal{tr}\big(\big(\bm{F}_N\bm{H}^{i}\bm{F}_N^H\big)^H\bm{F}_N\bm{H}^{i}\bm{F}_N^H\big)   \\
\label{norm_check_y_i_2_3}
&=\textnormal{tr}\big(\bm{F}_N\big(\bm{H}^{i}\big)^H\bm{F}_N^H\bm{F}_N\bm{H}^{i}\bm{F}_N^H\big)   \\
\label{norm_check_y_i_2_4}
&\stackrel{(a)}{=}\textnormal{tr}\big(\bm{F}_N\big(\bm{H}^{i}\big)^H\bm{H}^{i}\bm{F}_N^H\big)  \\
\label{norm_check_y_i_2_4_1}
&\stackrel{(b)}{=}\textnormal{tr}\big(\bm{F}_N^H\bm{F}_N\big(\bm{H}^{i}\big)^H\bm{H}^{i}\big)  \\
\label{norm_check_y_i_2_4_1_1}
&\stackrel{(c)}{=}\textnormal{tr}\big(\big(\bm{H}^{i}\big)^H\bm{H}^{i}\big)     \\
\label{norm_check_y_i_2_5}
&=\big\|\bm{H}^i\big\|_F^2=\sum_{n=0}^{N-1}\sum_{l=0}^{N-1} \big|\big(\bm{H}^{i}\big)_{n,l}\big|^2,  
\end{align}
\end{subequations}
where $(a)$ is due to the unitary DFT matrix such that $\bm{F}_N^H\bm{F}_N=\bm{F}_N\bm{F}_N^H=\bm{I}_N$ \cite{Giannakis_ST_BBWC_book_03}, $(b)$ is because of the identity $\textnormal{tr}(\bm{A}\bm{B})=\textnormal{tr}(\bm{B}\bm{A})$ \cite{Magnus_07}, and $(c)$ is due to the identity $\bm{F}_N^H\bm{F}_N=\bm{I}_N$.     

\begin{figure*}[!t]
	\normalsize
	\begin{equation}
	\label{H_i_matrix_expr_1}
	\bm{H}^i=
	\begin{bmatrix}
	h^i[0;0] & 0 & \ldots & 0 & h^i[0;L-1] & \ldots & h^i[0;1] \\
	h^i[1;1] & h^i[1;0] & 0 & \ldots & 0 & \ddots  & \vdots    \\
	\vdots & \ddots & \ddots & \ddots & \ddots & \ddots & h^i[L-1;L-1]  \\ 
	h^i[L-1;L-1]  & \ddots & \ddots & \ddots & \ddots & \ddots & 0   \\ 
	0 & \ddots & \ddots & \ddots & \ddots & \ddots & \vdots    \\ 
	\vdots & \ddots & \ddots & \ddots & \ddots & \ddots & 0   \\ 
	0 & \ldots & 0 & h^i[N-1;L-1] & \ldots & h^i[N-1;1] & h^i[N-1;0]      
	\end{bmatrix}.
	\end{equation} 
\end{figure*}

Based on the definition expressed by (\ref{H_T_time_dom}), the consideration $h^{i}[n;l]=0$ for $l\not\in[0, L-1]$, and \cite[eq. (3)]{Yu_DSFC_05}, $\bm{H}^{i}$ is given by (\ref{H_i_matrix_expr_1}), as shown at the top of this page. Employing (\ref{H_i_matrix_expr_1}) in (\ref{norm_check_y_i_2_5}) leads to  
\begin{subequations}
\begin{align}
\label{norm_check_y_i_2_6}
\|\check{\bm{y}}_i^{*}\|^2&=\sum_{n=0}^{N-1}\sum_{l=0}^{L-1} \big| h^{i}[n; l] \big|^2      \\
\label{norm_check_y_i_2_7}
&\stackrel{(a)}{=}\sum_{n=0}^{N-1}\sum_{l=0}^{L-1} h^H[i\tilde{N}+n; l] h[i\tilde{N}+n; l],         
\end{align}	
\end{subequations}
where $(a)$ follows from (\ref{DT_CIR}). Following the CE-BEM model given by (\ref{BEM_equation}),        
\begin{equation}
\label{BEM_equation_simpli_1}
h[n;l]\eqdef \bm{h}_{n,l}\bm{v}_n, 
\end{equation}
where 
\begin{subequations} 
\begin{align}
\label{BEM_equation_simpli_1_1}
\bm{h}_{n,l}&=\big[ h_0[ \lfloor{n/\tilde{N}\rfloor}; l], h_1[ \lfloor{n/\tilde{N}\rfloor}; l], \ldots, h_Q[ \lfloor{n/\tilde{N}\rfloor}; l] \big]       \\
\label{BEM_equation_simpli_1_2}
\bm{v}_n&=\big[e^{\jmath \frac{2\pi (0-Q/2)n}{\tilde{N}}}, e^{\jmath \frac{2\pi (1-Q/2)n}{\tilde{N}}}, \ldots, e^{\jmath \frac{2\pi (Q/2)n}{\tilde{N}}}\big]^T.
\end{align}
\end{subequations}
From (\ref{BEM_equation_simpli_1}),    
\begin{equation}
\label{BEM_equation_simpli_2}
h[i\tilde{N}+n;l] \eqdef \bm{h}_{i\tilde{N}+n,l}\bm{v}_{i\tilde{N}+n}, 
\end{equation}   
where it is deduced from (\ref{BEM_equation_simpli_1_1}) and (\ref{BEM_equation_simpli_1_2}) that                
\begin{equation}
\label{BEM_coefficient_relation_3}
\bm{v}_{i\tilde{N}+n}=\big[e^{\jmath \frac{2\pi (0-Q/2)(i\tilde{N}+n)}{\tilde{N}}}, \ldots, e^{\jmath \frac{2\pi (Q/2)(i\tilde{N}+n)}{\tilde{N}}}\big]^T
\end{equation}
and 
\begin{subequations} 
\begin{align}
\label{BEM_coefficient_relation_1}
\bm{h}_{i\tilde{N}+n,l}=&\big[ h_0\big[ \lfloor{(i\tilde{N}+n)/\tilde{N}\rfloor}; l\big], \ldots, h_Q\big[ \lfloor{(i\tilde{N}+n)/\tilde{N}\rfloor}; l\big] \big]  \\
\label{BEM_coefficient_relation_2}
\stackrel{(a)}{=}& \big[ h_0\big[i+ \lfloor{n/\tilde{N}\rfloor}; l\big], \ldots, h_Q\big[i+ \lfloor{n/\tilde{N}\rfloor}; l\big] \big]      \\
\label{BEM_coefficient_relation_2_1}
\stackrel{(b)}{=}& \big[ h_0[i; l], \ldots, h_Q[i; l] \big], 
\end{align}
\end{subequations}
where $(a)$ is due to the identity $\lfloor{x+n\rfloor}=\lfloor{x\rfloor}+n$ for $n\in\mathbb{Z}$ and $(b)$ follows from the relation $n<N<\tilde{N}$, per (\ref{DT_CIR}). The relation in (\ref{BEM_coefficient_relation_2_1}) is harmonious with the fact that there are only $Q+1$ BEM coefficients -- $h_q[ i; l]\sim\mathcal{CN}(0, \sigma_{q,l}^2)$ -- per every $i$-th OFDM symbol and the BEM coefficients remain constant during the $i$-th OFDM symbol. Meanwhile, it is to be noted that the variances $\sigma_{q,l}^2$ follow the doubly selective fading channel's generation constraints given by (\ref{sigma_q_l_def}) and (\ref{gamma_def}) which, in turn, validate Assumption \ref{CE_BEM_coeffcients_normalization}.               

Per Assumption \ref{CE_BEM_coefficients}, $h_q[ i; l]\sim\mathcal{CN}(0, \sigma_{q,l}^2)$ and is independently generated for all $q\in\{0, 1, \ldots, Q\}$ and all $l\in\{0, 1, \ldots, L-1\}$. As a result, $h[i\tilde{N}+n;l]$ -- given by (\ref{BEM_equation_simpli_2}) -- will exhibit a complex normal distribution with zero mean and variance $\sigma_{i\tilde{N}+n,l}^2$ which is computed as       
\begin{subequations}
\begin{align}
\label{var_calc_1}
\sigma_{i\tilde{N}+n,l}^2&=\mathbb{E}\big\{ h^H[i\tilde{N}+n;l]h[i\tilde{N}+n;l] \big\}	   \\
\label{var_calc_1_2}
&\stackrel{(a)}{=}\mathbb{E}\big\{\bm{v}_{i\tilde{N}+n}^H \bm{h}_{i\tilde{N}+n,l}^H \bm{h}_{i\tilde{N}+n,l} \bm{v}_{i\tilde{N}+n} \big\}      \\
\label{var_calc_2}
&\stackrel{(b)}{=}\bm{v}_{i\tilde{N}+n}^H \mathbb{E}\big\{\bm{h}_{i\tilde{N}+n,l}^H \bm{h}_{i\tilde{N}+n,l}\big\} \bm{v}_{i\tilde{N}+n}, 
\end{align}
\end{subequations}
where $(a)$ follows from (\ref{BEM_equation_simpli_2}) and $(b)$ is due to the fact that $\bm{v}_{i\tilde{N}+n}$ is not a random vector. As each element of $\bm{h}_{i\tilde{N}+n,l}$ is a zero mean Gaussian distributed RV which is independently generated, it directly follows through (\ref{BEM_coefficient_relation_2_1}) that $\mathbb{E}\big\{\bm{h}_{i\tilde{N}+n,l}^H \bm{h}_{i\tilde{N}+n,l}\big\}=\textnormal{diag}\big(\sigma_{0,l}^2, \ldots, \sigma_{Q,l}^2 \big)$. Substituting this expression into (\ref{var_calc_2}) then leads to   
\begin{subequations}
	\begin{align}
	\label{var_calc_3_1}
	\sigma_{i\tilde{N}+n,l}^2&=\bm{v}_{i\tilde{N}+n}^H \textnormal{diag}\big(\sigma_{0,l}^2, \ldots, \sigma_{Q,l}^2 \big) \bm{v}_{i\tilde{N}+n}   \\
	\label{var_calc_3_1_1}
	&=\sum_{q=0}^Q \sigma_{q,l}^2 \overline{(\bm{v}_{i\tilde{N}+n})_q}(\bm{v}_{i\tilde{N}+n})_q  \\
	\label{var_calc_3}
	&\stackrel{(a)}{=}\sum_{q=0}^Q \sigma_{q,l}^2,  	
	\end{align}	
\end{subequations}
where $(a)$ is due to (\ref{BEM_coefficient_relation_3}) that $\overline{(\bm{v}_{i\tilde{N}+n})_q}(\bm{v}_{i\tilde{N}+n})_q=1$ for all $q\in\{0, 1, \ldots, Q\}$. Therefore, employing (\ref{var_calc_3}) in (\ref{var_calc_1}) gives       
\begin{equation}
\label{expect_relationship}
\mathbb{E}\big\{ h^H[i\tilde{N}+n;l]h[i\tilde{N}+n;l]\big\}=\sum_{q=0}^Q \sigma_{q,l}^2, 
\end{equation}
where $h[i\tilde{N}+n;l]\sim \mathcal{CN}\big(0, \sum_{q=0}^Q \sigma_{q,l}^2 \big)$. 

Moreover, applying expectation to both sides of (\ref{norm_check_y_i_2_7}),  
\begin{subequations}
	\begin{align}
	\label{Expect_norm_check_y_i_1}
	\mathbb{E}\big\{\|\check{\bm{y}}_i^{*}\|^2\big\}&=\sum_{n=0}^{N-1}\sum_{l=0}^{L-1} \mathbb{E}\big\{h^H[i\tilde{N}+n; l] h[i\tilde{N}+n; l]\big\}      \\
	\label{Expect_norm_check_y_i_2}
	&\stackrel{(a)}{=}\sum_{n=0}^{N-1}\sum_{l=0}^{L-1} \sum_{q=0}^Q \sigma_{q,l}^2,         
	\end{align}	
\end{subequations}
where $(a)$ follows from (\ref{expect_relationship}). Per Assumption \ref{CE_BEM_coeffcients_normalization} (also concerning the variances constrained per (\ref{sigma_q_l_def}) and (\ref{gamma_def})), $\sum_{l=0}^{L-1} \sum_{q=0}^Q \sigma_{q,l}^2=1$. As a result, (\ref{Expect_norm_check_y_i_2}) simplifies to    
\begin{equation}
\label{Expect_norm_check_y_i_3}
\mathbb{E}\big\{\|\check{\bm{y}}_i^{*}\|^2\big\}=\sum_{n=0}^{N-1} 1=N.
\end{equation}
This completes the proof of Lemma \ref{On_channel_energy}.     \QED
\end{lemma}
\begin{remark}
\label{On_Parseval's theorem}
Defining the energy of the subcarrier coupling matrix as $\|\tilde{\bm{H}}^i\|_F^2$, (\ref{norm_check_y_i_2_1}), (\ref{norm_check_y_i_2_5}), and (\ref{Expect_norm_check_y_i_3}) assert that the frequency-domain channel energy and time-domain channel energy are the same. Thus, Lemma \ref{On_channel_energy} corroborates Parseval's theorem \cite[p. 18]{Proakis_5th_ed_08}, \cite[p. 60]{AORS10}.
\end{remark} 

Meanwhile, employing (\ref{Expec_B_h_i_K_4}) and (\ref{Expect_norm_check_y_i_3}) in the RHS of   (\ref{expect_inner_product_1}),       
\begin{equation}
\label{expect_inner_product_2} 
0 \leq \mathbb{E}\big\{F_i(\check{\bm{W}})\big\}\leq 2(2N+1)m^2e^{(2\bar{C}K+2m\omega^2K)}+ N,   
\end{equation}
where $m<< \bar{C} <\infty$. Deploying (\ref{expect_inner_product_2}) in (\ref{Expect_MSE_cost_1}) and recalling that $d_y=2N^2$,     
\begin{equation}
\label{Expect_MSE_cost_2}
0 \leq \mathbb{E}\big\{F(\check{\bm{W}})\big\}\leq \frac{2(2N+1)m^2e^{(2\bar{C}K+2m\omega^2K)}+N}{2N^2}. 
\end{equation}
Because $m$ and $K$ are fixed; $m, K<\infty$; and $m<< \bar{C} <\infty$, applying limit -- as $N\to\infty$ -- to (\ref{Expect_MSE_cost_2}) gives  
\begin{equation}
\label{Expect_MSE_cost_4}
0 \leq \lim_{N\to\infty} \mathbb{E}\{ F(\check{\bm{W}}) \} \leq 0. 
\end{equation}
This leads to (\ref{lim_expectation_MSE}), which is the desired result. This concludes the proof of Theorem \ref{Thm_lim_expectation_MSE}.  \QEDclosed

\section{Proof of Lemma \ref{lem_TD_FD_MSE_relation}}
\label{proof_lem_TD_FD_MSE_relation}
Because $\tilde{\bm{H}}^{i}$ and $\bm{H}^{i}$ are related via (\ref{H_sc_matrix}), their estimated counterparts are also related as          
\begin{equation}
\label{estimated_SC_matrix}
\hat{\tilde{\bm{H}}}^{i}=\bm{F}_N\hat{\bm{H}}^{i}\bm{F}_N^H.
\end{equation}
Substituting (\ref{H_sc_matrix}) and (\ref{estimated_SC_matrix}) into (\ref{MSE_alt_defn}), 
\begin{subequations}
	\begin{align}
	\label{MSE_alt_defn_1}
	F(\check{\bm{W}}) &=\frac{1}{2\check{n}N^2}\sum_{i=\breve{n}+1}^ {\breve{n}+\check{n}}  \big\| \bm{F}_N\hat{\bm{H}}^{i}\bm{F}_N^H-\bm{F}_N\bm{H}^{i}\bm{F}_N^H \big\|_{F}^2   \\
	\label{MSE_alt_defn_2}
	&=\frac{1}{2\check{n}N^2}\sum_{i=\breve{n}+1}^ {\breve{n}+\check{n}}  \underbrace{\big\| \bm{F}_N\big(\hat{\bm{H}}^{i}-\bm{H}^{i}\big)\bm{F}_N^H\big\|_{F}^2}_{=F_i(\check{\bm{W}})}.  
	\end{align}
\end{subequations}
Following the definition of Frobenius norm (see Definition \ref{Frob_and_spec_norma}) and \cite[p. 72]{vershynin_2018},  
\begin{equation}
\label{Frob_norm_comp_def}
\|\bm{A}\|_F^2=\textnormal{tr}(\bm{A}^H\bm{A}).
\end{equation}
Employing (\ref{Frob_norm_comp_def}) in the RHS of (\ref{MSE_alt_defn_2}),  
\begin{subequations}
	\begin{align}  
	\label{MSE_alt_defn_3} 
	F_i(\check{\bm{W}}) &=\textnormal{tr}\big( \big[\bm{F}_N\big(\hat{\bm{H}}^{i}-\bm{H}^{i}\big)\bm{F}_N^H\big]^H\bm{F}_N\big(\hat{\bm{H}}^{i}-\bm{H}^{i}\big)\bm{F}_N^H \big)       \\
	\label{MSE_alt_defn_4}
	&=\textnormal{tr}\big(\bm{F}_N\big(\hat{\bm{H}}^{i}-\bm{H}^{i}\big)^H\bm{F}_N^H\bm{F}_N\big(\hat{\bm{H}}^{i}-\bm{H}^{i}\big)\bm{F}_N^H \big)     \\
	\label{MSE_alt_defn_5}
	&\stackrel{(a)}{=} \textnormal{tr}\big(\bm{F}_N\big(\hat{\bm{H}}^{i}-\bm{H}^{i}\big)^H\big(\hat{\bm{H}}^{i}-\bm{H}^{i}\big)\bm{F}_N^H \big)  \\
	\label{MSE_alt_defn_6}
	&\stackrel{(b)}{=} \textnormal{tr}\big(\bm{F}_N^H\bm{F}_N\big(\hat{\bm{H}}^{i}-\bm{H}^{i}\big)^H\big(\hat{\bm{H}}^{i}-\bm{H}^{i}\big)\big)    \\
	\label{MSE_alt_defn_7}
	&\stackrel{(c)}{=} \textnormal{tr}\big(\big(\hat{\bm{H}}^{i}-\bm{H}^{i}\big)^H\big(\hat{\bm{H}}^{i}-\bm{H}^{i}\big)\big)    \\
	\label{MSE_alt_defn_8}
	&\stackrel{(d)}{=} \big\|\hat{\bm{H}}^{i}-\bm{H}^{i}\big\|_F^2, 
	\end{align}
\end{subequations}
where $(a)$ is due to the identity $\bm{F}_N^H\bm{F}_N=\bm{I}_N$; $(b)$ follows from the identity $\textnormal{tr}(\bm{A}\bm{B})=\textnormal{tr}(\bm{B}\bm{A})$ \cite{Magnus_07}; $(c)$ is also due to the identity $\bm{F}_N^H\bm{F}_N=\bm{I}_N$; and $(d)$ follows through (\ref{Frob_norm_comp_def}).     

Finally, substituting (\ref{MSE_alt_defn_8}) into the RHS of (\ref{MSE_alt_defn_2}),        
\begin{equation}
\label{MSE_alt_defn_9}
F(\check{\bm{W}})=\frac{1}{2\check{n}N^2}\sum_{i=\breve{n}+1}^ {\breve{n}+\check{n}}  \big\|\hat{\bm{H}}^{i}-\bm{H}^{i}\big\|_F^2.  
\end{equation}
This is exactly (\ref{MSE_eq_defn}) and Lemma \ref{lem_TD_FD_MSE_relation}'s proof is ended.   \QEDclosed

\section{Proof of Theorem \ref{Thm_probabilistic_MSE_lower_bound}}
\label{proof_Thm_probabilistic_MSE_lower_bound}
Employing (\ref{MSE_cost_2}) and the dimension $d_y=2N^2$,    
\begin{subequations}
	\begin{align}
\label{Instant_MSE_bound_1}
\underbrace{\mathbb{P}\bigg( F(\check{\bm{W}}) \geq \frac{t}{2N^2} \bigg)}_{=p}=&\mathbb{P}\bigg( \frac{1}{d_y\check{n}} \sum_{i=\breve{n}+1}^ {\breve{n}+\check{n}}   \|\bm{B}\check{\bm{x}}_{i,K}-\check{\bm{y}}_i^{*} \|^2 \geq \frac{t}{2N^2}  \bigg)     \\
\label{Instant_MSE_bound_2}
=& \mathbb{P}\bigg(\frac{1}{\check{n}} \sum_{i=\breve{n}+1}^ {\breve{n}+\check{n}}   \|\bm{B}\check{\bm{x}}_{i,K}-\check{\bm{y}}_i^{*} \|^2 \geq t  \bigg)  \\
\label{Instant_MSE_bound_3}
=& \mathbb{P}\Big(\sum_{i=\breve{n}+1}^{\breve{n}+\check{n}} U_i \geq \check{n}t  \Big),
	\end{align}
\end{subequations}
where $U_i\geq 0$ and it is defined for all $i\in \{\breve{n}+1, \breve{n}+2, \ldots, \breve{n}+\check{n}\}$ as     
\begin{equation}
\label{RV_U_i_def}
U_i \eqdef \|\bm{B}\check{\bm{x}}_{i,K}-\check{\bm{y}}_i^{*} \|^2. 
\end{equation}      
To simplify the RHS of (\ref{Instant_MSE_bound_3}) by bounding it from below, we define events $\mathcal{E}_i$ for all $i\in \{\breve{n}+1, \breve{n}+2, \ldots, \breve{n}+\check{n}\}$ as   
\begin{equation}
\label{Event_i_def}
\mathcal{E}_i\eqdef \big\{ \|\bm{B}\check{\bm{x}}_{i,K}-\check{\bm{y}}_i^{*} \|^2 \geq t \big\}.
\end{equation}
Regarding the events in (\ref{Event_i_def}), their respective complements are denoted by $\mathcal{E}_i^c$ and defined for all $i\in \{\breve{n}+1, \breve{n}+2, \ldots, \breve{n}+\check{n}\}$ as  
\begin{equation}
\label{Event_i_c_def}
\mathcal{E}_i^c\eqdef \{ \|\bm{B}\check{\bm{x}}_{i,K}-\check{\bm{y}}_i^{*} \|^2 < t \}.
\end{equation} 
    
By employing total probability theorem (see Theorem \ref{Total_Prob_Thm}), we aim to bound the probability in the RHS of (\ref{Instant_MSE_bound_3}) from below by conditioning successively on the events in (\ref{Event_i_def}) or their complements in (\ref{Event_i_c_def}) considering that they would occur one by one. The consideration that the events in (\ref{Event_i_def}) or their complements in (\ref{Event_i_c_def}) would happen successively is a natural phenomenon pursuant to the fact that FNNs produce their outputs successively. Toward this end, applying total probability theorem w.r.t. $\mathcal{E}_{\breve{n}+1}$ and $\mathcal{E}_{\breve{n}+1}^c$ to the RHS of (\ref{Instant_MSE_bound_3}),      
\begin{subequations}
\begin{align}
\label{Instant_MSE_bound_3_1_1}
p&=\mathbb{P}\Big(\sum_{i=\breve{n}+1}^ {\breve{n}+\check{n}} U_i \geq \check{n}t \big| \mathcal{E}_{\breve{n}+1}\Big)\mathbb{P}(\mathcal{E}_{\breve{n}+1})+\mathbb{P}\Big(\sum_{i=\breve{n}+1}^ {\breve{n}+\check{n}} U_i \geq \check{n}t \big| \mathcal{E}_{\breve{n}+1}^c\Big)\mathbb{P}(\mathcal{E}_{\breve{n}+1}^c)   \\
\label{Instant_MSE_bound_3_1_2}
&\stackrel{(a)}{\geq} \mathbb{P}\Big(\sum_{i=\breve{n}+1}^ {\breve{n}+\check{n}} U_i \geq \check{n}t \big| \mathcal{E}_{\breve{n}+1}\Big)\mathbb{P}(\mathcal{E}_{\breve{n}+1}), 
\end{align}
\end{subequations}
where $(a)$ is due to the axiomatic definition of probability (see probability axioms \cite[p. 9]{DPJN08}) that $0\leq \mathbb{P}\big(\sum_{i=\breve{n}+1}^ {\breve{n}+\check{n}} U_i \geq \check{n}t \big| \mathcal{E}_{\breve{n}+1}^c\big), \mathbb{P}(\mathcal{E}_{\breve{n}+1}^c) \leq 1$  and $0\leq \mathbb{P}\big(\sum_{i=\breve{n}+1}^ {\breve{n}+\check{n}} U_i \geq \check{n}t \big| \mathcal{E}_{\breve{n}+1}^c\big)\mathbb{P}(\mathcal{E}_{\breve{n}+1}^c)\leq 1$.    

To bound the first probabilistic expression that comprises the RHS of (\ref{Instant_MSE_bound_3_1_2}), we will apply total probability theorem (see Theorem \ref{Total_Prob_Thm}) w.r.t. $\mathcal{E}_{\breve{n}+2}$ and $\mathcal{E}_{\breve{n}+2}^c$. Doing so via Theorem \ref{Total_Prob_Thm} while letting $U_T=\sum_{i=\breve{n}+1}^{\breve{n}+\check{n}} U_i$,    
\begin{subequations}
\begin{align}
\label{Instant_MSE_bound_3_1_3_0}
\mathbb{P}\big( U_T  \geq \check{n}t \big| \mathcal{E}_{\breve{n}+1}\big) &=\mathbb{P}\big(U_T \geq \check{n}t \big|  \mathcal{E}_{\breve{n}+1}, \mathcal{E}_{\breve{n}+2}   \big)\mathbb{P}(\mathcal{E}_{\breve{n}+2})+ 
\mathbb{P}\big(U_T \geq \check{n}t \big|  \mathcal{E}_{\breve{n}+1}, \mathcal{E}_{\breve{n}+2}^c   \big) \mathbb{P}(\mathcal{E}_{\breve{n}+2}^c)   \\
\label{Instant_MSE_bound_3_1_3}
& \stackrel{(a)}{\geq} \mathbb{P}\Big(\sum_{i=\breve{n}+1}^ {\breve{n}+\check{n}} U_i \geq \check{n}t \big|  \mathcal{E}_{\breve{n}+1}, \mathcal{E}_{\breve{n}+2}  \Big)\mathbb{P}(\mathcal{E}_{\breve{n}+2}),
\end{align}
\end{subequations}
where $(a)$ is due to the fact that $0\leq\mathbb{P}\big( U_T \geq \check{n}t \big|  \mathcal{E}_{\breve{n}+1}, \mathcal{E}_{\breve{n}+2}^c  \big)\mathbb{P}(\mathcal{E}_{\breve{n}+2}^c)\leq 1$. Substituting the inequality in (\ref{Instant_MSE_bound_3_1_3}) into the RHS of (\ref{Instant_MSE_bound_3_1_2}) leads to       
\begin{equation}
\label{Instant_MSE_bound_3_1_4}
p\geq \mathbb{P}\Big(\sum_{i=\breve{n}+1}^ {\breve{n}+\check{n}} U_i \geq \check{n}t \big|  \mathcal{E}_{\breve{n}+1}, \mathcal{E}_{\breve{n}+2}   \Big)\prod_{j=\breve{n}+1}^{\breve{n}+2} \mathbb{P}(\mathcal{E}_j). 
\end{equation}
Conditioning consecutively w.r.t. $\mathcal{E}_{\breve{n}+3}, \ldots, \mathcal{E}_{\breve{n}+\check{n}}$ (and $\mathcal{E}_{\breve{n}+3}^c, \ldots, \mathcal{E}_{\breve{n}+\check{n}}^c$, respectively), it follows immediately from (\ref{Instant_MSE_bound_3_1_4}) that        
\begin{equation}
\label{Instant_MSE_bound_3_1_5}
p\geq \mathbb{P}\Big(\sum_{i=\breve{n}+1}^ {\breve{n}+\check{n}} U_i \geq \check{n}t \big|  \mathcal{E}_{\breve{n}+1}, \ldots, \mathcal{E}_{\breve{n}+\check{n}}   \Big)\prod_{j=\breve{n}+1}^{\breve{n}+\check{n}} \mathbb{P}(\mathcal{E}_j). 
\end{equation}
When events $\mathcal{E}_{\breve{n}+1}, \ldots, \mathcal{E}_{\breve{n}+\check{n}}$ happen consecutively, it follows via (\ref{RV_U_i_def}) and (\ref{Event_i_def}) for all $i\in \{\breve{n}+1, \breve{n}+2, \ldots, \breve{n}+\check{n}\}$ that    
\begin{equation}
\label{RV_U_i_cond_1}
(U_i | \mathcal{E}_i)= \check{t} \geq t\geq 0, \hspace{2mm} \forall i\in[\check{n}]. 
\end{equation}
As a result,      
\begin{subequations}
\begin{align}
\label{Sum_RV_U_i_cond_1}
\mathbb{P}\Big(\sum_{i=\breve{n}+1}^ {\breve{n}+\check{n}} U_i \geq \check{n}t \big| \big\{ \mathcal{E}_{\breve{n}+1}, \ldots, \mathcal{E}_{\breve{n}+\check{n}} \big\}   \Big)=&\mathbb{P}\Big(\sum_{i=\breve{n}+1}^ {\breve{n}+\check{n}} \check{t} \geq \check{n}t \Big)    \\
\label{Sum_RV_U_i_cond_2}
=&\mathbb{P}\big(\check{n}\check{t} \geq \check{n}t \big)   \\
\label{Sum_RV_U_i_cond_3}
=&\mathbb{P}\big(\check{t} \geq t \big) \stackrel{(a)}{=} 1,      
\end{align}
\end{subequations} 
where $(a)$ is due to the inequality on the RHS of (\ref{RV_U_i_cond_1}).         

Plugging (\ref{Sum_RV_U_i_cond_3}) into (\ref{Instant_MSE_bound_3_1_5}) then results in    
\begin{equation}
\label{Instant_MSE_bound_3_1_6}
p\geq\prod_{j=\breve{n}+1}^{\breve{n}+\check{n}} \mathbb{P}(\mathcal{E}_j)=\prod_{i=\breve{n}+1}^{\breve{n}+\check{n}} \mathbb{P}(\mathcal{E}_i). 
\end{equation}
Meanwhile, substituting (\ref{Instant_MSE_bound_3_1_6}) into the LHS of (\ref{Instant_MSE_bound_1}) produces 
\begin{equation}
\label{Instant_MSE_bound_4}
\mathbb{P}\bigg( F(\check{\bm{W}}) \geq \frac{t}{2N^2} \bigg) \geq\prod_{i=\breve{n}+1}^{\breve{n}+\check{n}} \mathbb{P}(\mathcal{E}_i).  
\end{equation}
For all $i\in \{\breve{n}+1, \breve{n}+2, \ldots, \breve{n}+\check{n}\}$, the RHS of (\ref{Instant_MSE_bound_4}) can be bounded from below by bounding the probability of an event $\mathcal{E}_i$ -- as defined in (\ref{Event_i_def}) -- from below. Thus, it follows through (\ref{Event_i_def}) that        
\begin{subequations}
\begin{align}
\label{Instant_MSE_bound_3_1_0}
\mathbb{P}(\mathcal{E}_i)&=\mathbb{P}\big(\|\bm{B}\check{\bm{x}}_{i,K}-\check{\bm{y}}_i^{*} \|^2 \geq t\big)\\
\label{Instant_MSE_bound_3_1}
&=\mathbb{P}\big( \|\bm{B}\check{\bm{x}}_{i,K}-\check{\bm{y}}_i^{*} \| \geq \sqrt{t} \big).	
\end{align}	
\end{subequations}
To bound the RHS of (\ref{Instant_MSE_bound_4}) from below, we are going to bound the RHS (\ref{Instant_MSE_bound_3_1}) from below, as developed in the sequel.   

To begin with, we state the reverse triangle inequality \cite[eq. (5.1.6)]{CDM00}: 
\begin{equation}
\label{Rev_triangle_inequality}
\| \bm{B}\check{\bm{x}}_{i,K}-\check{\bm{y}}_i^{*} \| \geq \big| \| \bm{B}\check{\bm{x}}_{i,K} \| - \| \check{\bm{y}}_i^{*} \| \big|.   
\end{equation}
Deploying (\ref{Rev_triangle_inequality}) in (\ref{Instant_MSE_bound_3_1}),   
\begin{subequations}
	\begin{align}
	\label{Prob_Exp_instant_MSE_bound_4}
	\mathbb{P}(\mathcal{E}_i)&\geq\mathbb{P}\big( \big| \| \bm{B}\check{\bm{x}}_{i,K} \| - \| \check{\bm{y}}_i^{*} \| \big|  \geq \sqrt{t} \big)     \\
	\label{Prob_Exp_instant_MSE_bound_5}
	&\stackrel{(a)}{\geq}\mathbb{P}\big( \| \bm{B}\check{\bm{x}}_{i,K} \| - \| \check{\bm{y}}_i^{*} \|   \geq \sqrt{t} \big)    \\
	\label{Prob_Exp_instant_MSE_bound_6}
	&=\mathbb{P}\big(\| \bm{B}\check{\bm{x}}_{i,K} \| \geq \sqrt{t}+ \| \check{\bm{y}}_i^{*} \| \big),  
	\end{align}	
\end{subequations}
where $(a)$ is due to the relation $|a|\geq a$ for $a\in\mathbb{R}$. Applying total probability theorem (see Theorem \ref{Total_Prob_Thm}) to the RHS of (\ref{Prob_Exp_instant_MSE_bound_6}),       
\begin{multline}
\label{Prob_Exp_instant_MSE_bound_7}
\mathbb{P}(\mathcal{E}_i)\geq \mathbb{P}\big( \|\bm{B}\check{\bm{x}}_{i,K} \| \geq \sqrt{t}+ \|\check{\bm{y}}_i^{*} \| \big| \|\check{\bm{y}}_i^{*} \|\geq\sqrt{t}\big)\mathbb{P}\big( \|\check{\bm{y}}_i^{*} \|\geq\sqrt{t}\big)\\ +
\mathbb{P}\big( \|\bm{B}\check{\bm{x}}_{i,K} \| \geq \sqrt{t}+ \|\check{\bm{y}}_i^{*} \| \big| \|\check{\bm{y}}_i^{*} \|<\sqrt{t}\big)\mathbb{P}\big( \| \check{\bm{y}}_i^{*} \|<\sqrt{t}\big).
\end{multline} 
To continue, we are going to bound $\mathbb{P}\big(\| \check{\bm{y}}_i^{*} \| \geq \sqrt{t} \big)$ and simplify (\ref{Prob_Exp_instant_MSE_bound_7}). To this end, employing Markov's inequality (see Proposition \ref{Prop_Markov_inequality}),    
\begin{equation}
\label{bm_y_i_bound_1}
\mathbb{P}\big(\| \check{\bm{y}}_i^{*} \| \geq \sqrt{t} \big)\leq \frac{\mathbb{E}\{ \| \check{\bm{y}}_i^{*} \| \} }{\sqrt{t}}.
\end{equation}  
Applying Jensen's inequality (see Proposition \ref{Prop_Jensen_inequality}) w.r.t. the concavity of the square-root function, 
\begin{equation}
\label{expec_bm_y_i_square}
\mathbb{E}\big\{ \| \check{\bm{y}}_i^{*} \|\big\}= \mathbb{E}\big\{ \sqrt{ \| \check{\bm{y}}_i^{*} \|^2 } \big\} \leq \sqrt{ \mathbb{E}\big\{ \| \check{\bm{y}}_i^{*} \|^2 \big\} }\stackrel{(a)}{=} \sqrt{N},
\end{equation}
where $(a)$ follows from (\ref{Expect_norm_check_y_i_3}). Substituting (\ref{expec_bm_y_i_square}) into (\ref{bm_y_i_bound_1}),    
\begin{equation}
\label{bm_y_i_bound_2}
\mathbb{P}\big(\| \check{\bm{y}}_i^{*} \| \geq \sqrt{t} \big)\leq \sqrt{N/t}.
\end{equation} 
To proceed, we are going to set the following condition: 
\begin{condition}
\label{Cond_t_N}
$t>>N$. 	
\end{condition}

Under Condition \ref{Cond_t_N}, these probabilistic consequences follow: $0\leq\mathbb{P}\big(\| \check{\bm{y}}_i^{*} \| \geq \sqrt{t} \big)\leq 0$ and $\mathbb{P}\big( \| \check{\bm{y}}_i^{*} \|<\sqrt{t}\big)=1-\mathbb{P}\big( \| \check{\bm{y}}_i^{*} \| \geq \sqrt{t}\big)=1$. Hence, assuming Condition \ref{Cond_t_N} is valid, substituting these probability values into the RHS of (\ref{Prob_Exp_instant_MSE_bound_7}) gives   
\begin{equation}
\label{Prob_Exp_instant_MSE_bound_8}
\mathbb{P}(\mathcal{E}_i)\geq \mathbb{P}\big( \|\bm{B}\check{\bm{x}}_{i,K} \| \geq \sqrt{t}+ \|\check{\bm{y}}_i^{*} \| \big| \|\check{\bm{y}}_i^{*} \|<\sqrt{t}\big). 
\end{equation}

Under Condition \ref{Cond_t_N}, conditioning the probability in the RHS of (\ref{Prob_Exp_instant_MSE_bound_8}) -- via total probability theorem (see Theorem \ref{Total_Prob_Thm}) -- w.r.t. the events $\|\check{\bm{y}}_i^{*} \| \leq \sqrt{N}$ and $\|\check{\bm{y}}_i^{*} \| > \sqrt{N}$ results in (\ref{Prob_Exp_instant_MSE_bound_8_1}), as shown at the top of this page.       
\begin{figure*}[!t] 
\begin{multline}
\label{Prob_Exp_instant_MSE_bound_8_1}
\mathbb{P}(\mathcal{E}_i)\geq \mathbb{P}\big( \|\bm{B}\check{\bm{x}}_{i,K} \| \geq \sqrt{t}+ \|\check{\bm{y}}_i^{*} \| \big| \|\check{\bm{y}}_i^{*} \|\leq\sqrt{N}<<\sqrt{t}\big)\mathbb{P}\big( \|\check{\bm{y}}_i^{*} \|\leq\sqrt{N}\big)+\\ \mathbb{P}\big( \|\bm{B}\check{\bm{x}}_{i,K} \| \geq \sqrt{t}+ \|\check{\bm{y}}_i^{*} \| \big| \sqrt{N}<\|\check{\bm{y}}_i^{*} \|<\sqrt{t} \big) \mathbb{P}\big( \|\check{\bm{y}}_i^{*} \|>\sqrt{N}\big). 
\end{multline}
\hrulefill
\end{figure*}
Because of the axiomatic definition of probability,   
\begin{subequations} 
\begin{align}
\label{Prob_Exp_instant_MSE_bound_8_2}
0 \leq& \mathbb{P}\big( \|\bm{B}\check{\bm{x}}_{i,K} \| \geq \sqrt{t}+ \|\check{\bm{y}}_i^{*} \| \big| \sqrt{N}<\|\check{\bm{y}}_i^{*} \|<\sqrt{t}\big) \leq 1     \\
\label{Prob_Exp_instant_MSE_bound_8_3}
0 \leq & \mathbb{P}\big( \|\check{\bm{y}}_i^{*} \|>\sqrt{N}\big) \leq 1.
\end{align}
\end{subequations} 
Employing the lower bounds in (\ref{Prob_Exp_instant_MSE_bound_8_2}) and (\ref{Prob_Exp_instant_MSE_bound_8_3}) in the RHS of (\ref{Prob_Exp_instant_MSE_bound_8_1}) under the validity of Condition \ref{Cond_t_N},     
\begin{equation}
\label{Prob_Exp_instant_MSE_bound_8_4}
\mathbb{P}(\mathcal{E}_i)\geq \xi_i \mathbb{P}\big( \|\bm{B}\check{\bm{x}}_{i,K} \| \geq \sqrt{t}+ \|\check{\bm{y}}_i^{*} \| \big| \|\check{\bm{y}}_i^{*} \|\leq\sqrt{N}\big), 
\end{equation}
where 
\begin{equation}
\label{chi_expression}
\xi_i=\mathbb{P}\big( \|\check{\bm{y}}_i^{*} \|\leq\sqrt{N}\big).
\end{equation}
With regard to the conditioning of (\ref{Prob_Exp_instant_MSE_bound_8_4}), the optimal value $\|\check{\bm{y}}_i^{*} \|$ can take is $\sqrt{N}$. For this optimal value, the RHS of (\ref{Prob_Exp_instant_MSE_bound_8_4}) would attain the lowest possible probabilistic bound. Thus,   
\begin{subequations}
	\begin{align}
	\label{Prob_Exp_instant_MSE_bound_9}
	\mathbb{P}(\mathcal{E}_i)\geq &\xi_i\mathbb{P}\big( \|\bm{B}\check{\bm{x}}_{i,K} \| \geq \sqrt{t}+ \sqrt{N}\big)  \\
	\label{Prob_Exp_instant_MSE_bound_10}
	\stackrel{(a)}{\geq} & \xi_i\mathbb{P}\big(\sigma_{\textnormal{min}}(\bm{B}) \|\check{\bm{x}}_{i,K} \| \geq \sqrt{t}+ \sqrt{N}\big),
	\end{align}
\end{subequations}           
where $(a)$ follows from \cite[eq. (4.5)]{vershynin_2018} -- see also Definition \ref{Ext_singular_values_def} and (\ref{Euclidean_norm_constraint}) -- that $\|\bm{B}\check{\bm{x}}_{i,K}\| \geq\sigma_{\textnormal{min}}(\bm{B}) \|\check{\bm{x}}_{i,K} \|$ and (\ref{Prob_Exp_instant_MSE_bound_10}) is valid under Condition \ref{Cond_t_N}.    

Applying \cite[eq. (4.5)]{vershynin_2018} -- see also (\ref{Euclidean_norm_constraint}) -- to (\ref{bm_x_i_K_definition}) successively to bound $\|\check{\bm{x}}_{i,K} \|$ from below leads to 
\begin{subequations}
	\begin{align}
	\label{bm_B_x_i_k_norm_1_2}
	\|\check{\bm{x}}_{i,K} \| \geq & \sigma_{\textnormal{min}}(\check{\bm{\Sigma}}_{i,K}\check{\bm{W}}_K) \|\check{\bm{x}}_{i,K-1} \|    \\
	\label{bm_B_x_i_k_norm_2_2} 
	\geq & \prod_{k=0}^{K-1}\sigma_{\textnormal{min}}\big(\check{\bm{\Sigma}}_{i,K-k}\check{\bm{W}}_{K-k}\big) \sigma_{\textnormal{min}}\big(\check{\bm{\Sigma}}_{i,0}\bm{A}\big) \|\check{\bm{x}}_i \|   \\
	\label{bm_B_x_i_k_norm_3_2}
	\stackrel{(a)}{=} & \prod_{k=0}^{K-1} \sigma_{\textnormal{min}}\big(\check{\bm{\Sigma}}_{i,K-k}\check{\bm{W}}_{K-k}\big) \sigma_{\textnormal{min}}\big(\check{\bm{\Sigma}}_{i,0}\bm{A}\big), 
	\end{align}
\end{subequations}      
where $(a)$ follows from the normalized FNN inputs that $\|\check{\bm{x}}_i \|=1$ (see Assumption \ref{normalization_assumption}). Thus, exploiting (\ref{bm_B_x_i_k_norm_3_2}) in the RHS of (\ref{Prob_Exp_instant_MSE_bound_10}),   
\begin{equation}
\label{Prob_Exp_instant_MSE_bound_11}
\mathbb{P}(\mathcal{E}_i) \geq \xi_i\mathbb{P}\Big( \sigma_{\textnormal{min}}(\bm{B}) \prod_{k=0}^{K-1} \sigma_{\textnormal{min}}\big(\check{\bm{\Sigma}}_{i,K-k}\check{\bm{W}}_{K-k}\big) \sigma_{\textnormal{min}}\big(\check{\bm{\Sigma}}_{i,0}\bm{A}\big)\geq \sqrt{t}+ \sqrt{N}\Big). 
\end{equation}
Diving both sides of the RHS of (\ref{Prob_Exp_instant_MSE_bound_11}) by $\prod_{k=0}^{K-1} \sigma_{\textnormal{min}}\big(\check{\bm{\Sigma}}_{i,K-k}\check{\bm{W}}_{K-k}\big) \sigma_{\textnormal{min}}\big(\check{\bm{\Sigma}}_{i,0}\bm{A}\big)$,   
\begin{equation}
\label{Prob_Exp_instant_MSE_bound_12}
\mathbb{P}(\mathcal{E}_i) \geq \xi_i \mathbb{P}\bigg( \sigma_{\textnormal{min}}(\bm{B})\geq \frac{\sqrt{t}+ \sqrt{N}}{\prod_{k=0}^{K-1}\sigma_{\textnormal{min}}\big(\check{\bm{\Sigma}}_{i,K-k}\check{\bm{W}}_{K-k}\big) \sigma_{\textnormal{min}}\big(\check{\bm{\Sigma}}_{i,0}\bm{A}\big)}\bigg).  
\end{equation}
Under the stated condition of Theorem \ref{Thm_probabilistic_MSE_lower_bound},   
\begin{equation}
\label{Assumption_C_sigma_1}
\prod_{k=0}^{K-1} \sigma_{\textnormal{min}}\big(\check{\bm{\Sigma}}_{i,K-k}\check{\bm{W}}_{K-k}\big) \sigma_{\textnormal{min}}\big(\check{\bm{\Sigma}}_{i,0}\bm{A}\big)\geq C_{\sigma}, 
\end{equation}
where $C_{\sigma} \in \mathbb{R}\backslash \{0\}$. By using the condition stated in (\ref{Assumption_C_sigma_1}), the RHS of (\ref{Prob_Exp_instant_MSE_bound_12}) simplifies to         
\begin{equation}
\label{Prob_Exp_instant_MSE_bound_13}
\mathbb{P}(\mathcal{E}_i)\geq \xi_i\mathbb{P}\big( \sigma_{\textnormal{min}}(\bm{B})\geq C_{\sigma}^{-1}(\sqrt{t}+ \sqrt{N}) \big).
\end{equation}  

To simplify the RHS of (\ref{Prob_Exp_instant_MSE_bound_13}) with regard to $\bm{B}\in\mathbb{R}^{d_y \times m}$ -- $m>d_y=2N^2$ per Assumption \ref{over-param_assumption} -- and $(\bm{B})_{i,j}\sim\mathcal{N}\big(0, \frac{1}{2N^2}\big)$, we are going to express $\sigma_{\textnormal{min}}(\bm{B})$ in terms of $\sigma_{\textnormal{min}}(\tilde{\bm{B}})$, where $\tilde{\bm{B}}\in\mathbb{R}^{m\times 2N^2}$ and $(\tilde{\bm{B}})_{i,j}\sim\mathcal{N}(0,1)$. Toward this goal, we first note that the singular values of $\bm{B}$ are the same as the singular values $\bm{B}^T$. In addition, we note that $\bm{B}^T$ and $\sqrt{(2N^2)^{-1} }\tilde{\bm{B}}$ have the same distribution. Accordingly, 
\begin{equation}
\label{min_sing_value_1_1}
\sigma_{\textnormal{min}}(\bm{B})=\sigma_{\textnormal{min}}(\bm{B}^T)=\sigma_{\textnormal{min}}\big(\sqrt{(2N^2)^{-1} }\tilde{\bm{B}}\big). 
\end{equation}
Employing the definition of Euclidean norm, $\sigma_{\textnormal{min}}(\alpha \tilde{\bm{B}})=\displaystyle\inf_{\|\bm{x}\|=1}\|\alpha\tilde{\bm{B}}\bm{x}\|=|\alpha|\inf_{\|\bm{x}\|=1}\|\tilde{\bm{B}}\bm{x}\|=|\alpha| \sigma_{\textnormal{min}}(\tilde{\bm{B}})$. Consequently, (\ref{min_sing_value_1_1}) simplifies to  
\begin{equation}
\label{min_sing_value_2_1}
\sigma_{\textnormal{min}}(\bm{B})=\sqrt{(2N^2)^{-1} }\sigma_{\textnormal{min}}\big(\tilde{\bm{B}}\big). 
\end{equation}
Meanwhile, deploying Gordon’s theorem for Gaussian matrices \cite[Theorem 2.6]{Rudelson_Extreme_SVals_10} (see Theorem \ref{Thm_Gordon_inequality}), it follows that    
\begin{subequations}
\begin{align}
\label{min_sing_value_3}
\mathbb{E}\{\sigma_{\textnormal{min}}\big(\tilde{\bm{B}}\big)\}&\geq \sqrt{m}-\sqrt{2N^2}	\\
\label{min_sing_value_3_1}
\mathbb{E}\{\sigma_{\textnormal{max}}\big(\tilde{\bm{B}}\big)\}&\leq \sqrt{m}+\sqrt{2N^2}.	
\end{align}
\end{subequations}
Substituting (\ref{min_sing_value_2_1}) into (\ref{Prob_Exp_instant_MSE_bound_13}),   
\begin{equation}
\label{Prob_Exp_instant_MSE_bound_14}
\mathbb{P}(\mathcal{E}_i)\geq \xi_i \mathbb{P}\big(\sigma_{\textnormal{min}}\big(\tilde{\bm{B}}\big)\geq C_{\sigma}^{-1}\sqrt{2N^2}(\sqrt{t}+ \sqrt{N})\big).
\end{equation}  

To continue, we require the Paley--Zygmund inequality \cite{Litvak_smallest_sing_value_05} (see Lemma \ref{PZ_identity}). Invoking Lemma \ref{PZ_identity} to simplify (\ref{Prob_Exp_instant_MSE_bound_14}), letting $f=\sigma_{\textnormal{min}}\big(\tilde{\bm{B}}\big)$, considering $\lambda=C_{\sigma}^{-1}\sqrt{2N^2}(\sqrt{t}+ \sqrt{N})$, and choosing $p=2$ (this gives $q=2$ per Lemma \ref{PZ_identity}) leads to     
\begin{equation}
\label{Prob_Exp_instant_MSE_bound_15}
\mathbb{P}(\mathcal{E}_i)\geq \xi_i\frac{\Big[\mathbb{E}\big\{ \sigma_{\textnormal{min}}^2\big(\tilde{\bm{B}}\big) \big\}-\big(C_{\sigma}^{-1}\sqrt{2N^2}(\sqrt{t}+ \sqrt{N})\big)^2\Big]^2}{\mathbb{E}\big\{ \sigma_{\textnormal{min}}^4\big(\tilde{\bm{B}}\big) \big\}}, 
\end{equation}
where -- as required by Lemma \ref{PZ_identity} -- $0\leq C_{\sigma}^{-1}\sqrt{2N^2}(\sqrt{t}+ \sqrt{N}) \leq \sqrt{\mathbb{E}\{ \sigma_{\textnormal{min}}^2\big(\tilde{\bm{B}}\big) \}}$, $\sigma_{\textnormal{min}}^2\big(\tilde{\bm{B}}\big)=[\sigma_{\textnormal{min}}\big(\tilde{\bm{B}}\big)]^2$, and $\sigma_{\textnormal{min}}^4\big(\tilde{\bm{B}}\big)=[\sigma_{\textnormal{min}}\big(\tilde{\bm{B}}\big)]^4$. 

To simplify the RHS of (\ref{Prob_Exp_instant_MSE_bound_15}), we are going to bound $\mathbb{E}\big\{ \sigma_{\textnormal{min}}^2\big(\tilde{\bm{B}}\big) \big\}$ from below and $\mathbb{E}\big\{ \sigma_{\textnormal{min}}^4\big(\tilde{\bm{B}}\big) \big\}$ from above. Toward the former end, employing Jensen's inequality (see Proposition \ref{Prop_Jensen_inequality}) with regard to the convexity of the squaring function,   
\begin{equation}
\label{expectation_bm_B_x_i_k_square_norm_1}
\mathbb{E}\{\sigma_{\textnormal{min}}^2\big(\tilde{\bm{B}}\big)\} =\mathbb{E}\{[\sigma_{\textnormal{min}}\big(\tilde{\bm{B}}\big)]^2 \} \geq  \big[ \mathbb{E}\{\sigma_{\textnormal{min}}\big(\tilde{\bm{B}}\big)\} \big]^2\stackrel{(a)}{\geq}  \big[\sqrt{m}-\sqrt{2N^2}\big]^2, 
\end{equation}
where $(a)$ follows from (\ref{min_sing_value_3}). We now continue to the derivation of the upper bound of $\mathbb{E}\big\{ \sigma_{\textnormal{min}}^4\big(\tilde{\bm{B}}\big) \big\}$. To this end, employing the integral identity (see Lemma \ref{Lemma_integral_identity}) with regard to the fact that $\sigma_{\textnormal{min}}^4\big(\tilde{\bm{B}}\big)=[\sigma_{\textnormal{min}}\big(\tilde{\bm{B}}\big)]^4$ is non-negative,   
\begin{subequations}
	\begin{align}
	\label{exp_sigma_min_4_1}   
	\mathbb{E}\big\{ \sigma_{\textnormal{min}}^4\big(\tilde{\bm{B}}\big) \big\}=&\int_{0}^{\infty}\mathbb{P}\big(\sigma_{\textnormal{min}}^4\big(\tilde{\bm{B}}\big)> t\big)dt \\
	\label{exp_sigma_min_4_2} 
	=& \int_{0}^{\infty}\mathbb{P}\big(\sigma_{\textnormal{min}}\big(\tilde{\bm{B}}\big)> t^{1/4}\big)dt \\
	\label{exp_sigma_min_4_3} 
	\stackrel{(a)}{\leq} & \int_{0}^{\infty}\mathbb{P}\big(\sigma_{\textnormal{max}}\big(\tilde{\bm{B}}\big)> t^{1/4}\big)dt, 
	\end{align}
\end{subequations}
where $(a)$ follows from the fact that $\sigma_{\textnormal{min}}\big(\tilde{\bm{B}}\big)\leq \sigma_{\textnormal{max}}\big(\tilde{\bm{B}}\big)$. Subtracting $\mathbb{E}\big\{\sigma_{\textnormal{max}}\big(\tilde{\bm{B}}\big)\big\}$ from both sides of the inequality that comprises the RHS of (\ref{exp_sigma_min_4_3}) leads to 
\begin{subequations}
\begin{align}
\label{exp_sigma_min_4_4_0}   
\mathbb{E}\big\{ \sigma_{\textnormal{min}}^4\big(\tilde{\bm{B}}\big) \big\} & \leq  \int_{0}^{\infty}\mathbb{P}\big(\sigma_{\textnormal{max}}\big(\tilde{\bm{B}}\big)-\mathbb{E}\big\{\sigma_{\textnormal{max}}\big(\tilde{\bm{B}}\big)\big\}>t^{1/4} -\mathbb{E}\big\{\sigma_{\textnormal{max}}\big(\tilde{\bm{B}}\big)\big\} \big)dt \\ 
\label{exp_sigma_min_4_4}
& \stackrel{(a)}{\leq} \int_{0}^{\infty}\mathbb{P}\big(\sigma_{\textnormal{max}}\big(\tilde{\bm{B}}\big)-\mathbb{E}\big\{\sigma_{\textnormal{max}}\big(\tilde{\bm{B}}\big)\big\} >t^{1/4}-(\sqrt{m}+\sqrt{2N^2}) \big)dt,
\end{align}
\end{subequations}
where $(a)$ follows from the upper bound stated in (\ref{min_sing_value_3_1}).   

To simplify the RHS of (\ref{exp_sigma_min_4_4}), we proceed with integration by substitution. To this end, let
\begin{equation}
\label{tilde_t_1}
\tilde{t}=t^{1/4}-(\sqrt{m}+\sqrt{2N^2}) \Leftrightarrow t=[\tilde{t}+(\sqrt{m}+\sqrt{2N^2})]^4. 
\end{equation} 
Applying the laws of differential calculus to (\ref{tilde_t_1}), 
\begin{equation}
\label{diff_tilde_t_1}
d\tilde{t}=(1/4)t^{-3/4} dt \Leftrightarrow dt=4t^{3/4}d\tilde{t}.
\end{equation}
Substituting the RHS of (\ref{tilde_t_1}) into the RHS of (\ref{diff_tilde_t_1}): 
\begin{equation}
\label{diff_tilde_t_2}
dt=4[\tilde{t}+(\sqrt{m}+\sqrt{2N^2})]^3d\tilde{t}.
\end{equation} 
Deploying the LHS of (\ref{tilde_t_1}) and (\ref{diff_tilde_t_2}) into the upper bound expression of (\ref{exp_sigma_min_4_4}) leads to     
\begin{multline}
\label{exp_sigma_min_4_5_1}
\mathbb{E}\big\{ \sigma_{\textnormal{min}}^4\big(\tilde{\bm{B}}\big) \big\}\leq   \displaystyle\int_{-(\sqrt{m}+\sqrt{2N^2})}^{\infty}4[\tilde{t}+(\sqrt{m}+\sqrt{2N^2})]^3  \mathbb{P}\big(\sigma_{\textnormal{max}}\big(\tilde{\bm{B}}\big)-\mathbb{E}\big\{\sigma_{\textnormal{max}}\big(\tilde{\bm{B}}\big)\big\} >\tilde{t}\big)d\tilde{t}.
\end{multline}
Since $\mathbb{P}\big(\sigma_{\textnormal{max}}\big(\tilde{\bm{B}}\big)-\mathbb{E}\big\{\sigma_{\textnormal{max}}\big(\tilde{\bm{B}}\big)\big\} >\tilde{t}\big)\leq \mathbb{P}\big(\sigma_{\textnormal{max}}\big(\tilde{\bm{B}}\big)-\mathbb{E}\big\{\sigma_{\textnormal{max}}\big(\tilde{\bm{B}}\big)\big\} \geq\tilde{t}\big)$, the RHS of (\ref{exp_sigma_min_4_5_1}) can be further bounded as       
\begin{multline}
\label{exp_sigma_min_4_5}
\mathbb{E}\big\{ \sigma_{\textnormal{min}}^4\big(\tilde{\bm{B}}\big) \big\}\leq   \int_{-(\sqrt{m}+\sqrt{2N^2})}^{\infty}4[\tilde{t}+(\sqrt{m}+\sqrt{2N^2})]^3  \mathbb{P}\big(\sigma_{\textnormal{max}}\big(\tilde{\bm{B}}\big)-\mathbb{E}\big\{\sigma_{\textnormal{max}}\big(\tilde{\bm{B}}\big)\big\} \geq\tilde{t}\big)d\tilde{t}.
\end{multline}
The RHS of (\ref{exp_sigma_min_4_5}) can be simplified by employing the Gaussian concentration inequality stated in Theorem \ref{Thm_Gaussian_conc_inequality} while considering the scenario that $\tilde{t}>0$. To enforce the constraint $\tilde{t}>0$, it follows through (\ref{tilde_t_1}) that $t^{1/4}-(\sqrt{m}+\sqrt{2N^2})>0$ and thus $t>(\sqrt{m}+\sqrt{2N^2})^4$. Meanwhile, from Condition \ref{Cond_t_N} which we exploited to simplify (\ref{Prob_Exp_instant_MSE_bound_7}) into (\ref{Prob_Exp_instant_MSE_bound_8}), $t>>N$. Consequently, it can be asserted via Assumption \ref{over-param_assumption} -- the over-parameterization assumption -- that the following condition subsumes Condition \ref{Cond_t_N}:    
\begin{condition}
\label{cond_t_m_N}
\begin{equation}
\label{t_constraint_upper}
t>\overbrace{(\sqrt{m}+\sqrt{2N^2})^4}^{=t_l}=(\sqrt{m}+\sqrt{2}N)^4.
\end{equation}	
\end{condition}
    
To bound the RHS of (\ref{exp_sigma_min_4_5}) under Condition \ref{cond_t_m_N}, we shall underscore the fact that $\sigma_{\textnormal{max}}\big(\tilde{\bm{B}}\big)$ is a 1-Lipshitz function (see Definition \ref{L_Lipschitz_func_def}) when $\tilde{\bm{B}}$ is rearranged into a vector in $\mathbb{R}^{2mN^2}$ \cite{Rudelson_Extreme_SVals_10,Boucheron_Con_eqs_13}. Therefore, whenever $\tilde{t}>0$ (equivalently, $t>(\sqrt{m}+\sqrt{2N^2})^4$), the Gaussian concentration inequality (see Theorem \ref{Thm_Gaussian_conc_inequality}) ascertains that  
\begin{equation}
\label{sigma_max_Gaussian_conc_inequality}
\mathbb{P}\big(\sigma_{\textnormal{max}}\big(\tilde{\bm{B}}\big)-\mathbb{E}\big\{\sigma_{\textnormal{max}}\big(\tilde{\bm{B}}\big)\big\} \geq\tilde{t}\big) \leq e^{-\tilde{t}^2/2}.
\end{equation}
As a result, substituting (\ref{sigma_max_Gaussian_conc_inequality}) into the RHS of (\ref{exp_sigma_min_4_5}) leads to     
\begin{equation}
\label{exp_sigma_min_4_6}
\mathbb{E}\big\{ \sigma_{\textnormal{min}}^4\big(\tilde{\bm{B}}\big) \big\}\leq   4\int_{-(\sqrt{m}+\sqrt{2N^2})}^{\infty}[\tilde{t}+(\sqrt{m}+\sqrt{2N^2})]^3 e^{-\tilde{t}^2/2}d\tilde{t}.
\end{equation} 
If we now let $x=\tilde{t}$ and $c=\sqrt{m}+\sqrt{2N^2}$, then $d\tilde{t}=dx$. Employing these parameters, (\ref{exp_sigma_min_4_6}) can be alternatively expressed as     
\begin{equation}
\label{exp_sigma_min_4_7}
\mathbb{E}\big\{ \sigma_{\textnormal{min}}^4\big(\tilde{\bm{B}}\big) \big\}\leq   4\overbrace{\int_{-c}^{\infty}(x+c)^3 e^{-x^2/2}dx}^{=I}.
\end{equation} 

To continue, we will prove the following lemma.    
\begin{lemma}
\label{Lem_I}	
For $c=\sqrt{m}+\sqrt{2}N$ and $I$ defined in (\ref{exp_sigma_min_4_7}), 
\begin{equation}
\label{I_simplified}
I=(2+c^2)e^{-c^2/2}+  c\sqrt{\frac{\pi}{2}}\Big[\textnormal{erf}\big(c/\sqrt{2}\big)+1\Big] \times (3+c^2).
\end{equation}
\proof We start by stating the binomial theorem (binomial expansion) \cite[eq. 1.111, p. 25]{ISGI07}:     
\begin{equation}
\label{Binomial_theorem}
(a+x)^n=\sum_{k=0}^n {n \choose k} x^k a^{n-k}.
\end{equation}
Employing (\ref{Binomial_theorem}), 
\begin{equation}
\label{x_plus_c_3}
(x+c)^3=x^3+3cx^2+3c^2x+c^3.
\end{equation} 
Substituting (\ref{x_plus_c_3}) into the $I$ expression given by (\ref{exp_sigma_min_4_7}), 
\begin{multline}
\label{I_2}
I=\overbrace{\int_{-c}^{\infty} x^3e^{-x^2/2}dx}^{=I_0} + 3c\overbrace{\int_{-c}^{\infty}x^2e^{-x^2/2}dx}^{=I_1}  +  3c^2 \overbrace{\int_{-c}^{\infty}xe^{-x^2/2}dx}^{=I_2}  + c^3 \overbrace{\int_{-c}^{\infty} e^{-x^2/2}dx}^{=I_3}. 
\end{multline} 
In the sequel, we are going to simplify $I_0$, $I_1$, $I_2$, and $I_3$, as equated in (\ref{I_2}). To this end, we will employ the following integration identities \cite[p. 336-337]{ISGI07} which will help us make our simplifications brief.         
\begin{equation}
\label{Int_identity_1}
\int_{0}^u e^{-q^2x^2}dx =\frac{\sqrt{\pi}}{2q}\Phi(qu), \hspace{1mm} q>0 \hspace{1mm}\text{\cite[eq. (3.321.2)]{ISGI07}}
\end{equation}
\begin{equation}
\label{Int_identity_2}
\int_{0}^\infty e^{-q^2x^2}dx =\frac{\sqrt{\pi}}{2q}, \hspace{1mm} q>0 \hspace{1mm}\text{\cite[eq. (3.321.3)]{ISGI07}}
\end{equation}
\begin{equation}
\label{Int_identity_3}
\int_{0}^u xe^{-q^2x^2}dx=\frac{1}{2q^2}\big[1- e^{-q^2u^2}\big] \hspace{1mm}\text{\cite[eq. (3.321.4)]{ISGI07}}
\end{equation}
\begin{equation}
\label{Int_identity_4}
\int_{0}^u x^2e^{-q^2x^2}dx=\frac{1}{2q^3}\Big[\frac{\sqrt{\pi}}{2}\Phi(qu)-qu e^{-q^2u^2} \Big] \hspace{1mm}\text{\cite[eq. (3.321.5)]{ISGI07}}
\end{equation}
\begin{equation}
\label{Int_identity_5}
\int_{0}^u x^3e^{-q^2x^2}dx=\frac{1}{2q^4} \Big[1-(1+q^2u^2)e^{-q^2u^2} \Big] \hspace{1mm}\text{\cite[eq. (3.321.6)]{ISGI07}}
\end{equation}
\begin{equation}
\label{Int_identity_6}
\int_{0}^\infty x^m e^{-\beta x^n} dx=\frac{\Gamma(\gamma)}{n\beta^\gamma} \hspace{1mm}\text{\cite[eq. (3.326.2)]{ISGI07}}, 
\end{equation}
where (\ref{Int_identity_6}) is valid for $\gamma=\frac{m+1}{n}$, $\textnormal{Re}\{\beta\}>0$, $\textnormal{Re}\{m\}>0$, and $\textnormal{Re}\{n\}>0$.     

To continue, we begin by simplifying $I_3$ as   
\begin{subequations}
\begin{align}
\label{I_3_1}
I_3=&\int_{-c}^{\infty} e^{-x^2/2}dx \\
\label{I_3_2}
=& \int_{-c}^{0} e^{-x^2/2}dx+\int_{0}^{\infty} e^{-x^2/2}dx  \\
\label{I_3_3}
\stackrel{(a)}{=} &\int_{0}^{c} e^{-t^2/2}dt+\int_{0}^{\infty} e^{-x^2/2}dx   \\
\label{I_3_4}
\stackrel{(b)}{=} & \sqrt{\frac{\pi}{2}}\Phi\big(c/\sqrt{2}\big)+\sqrt{\frac{\pi}{2}}= \sqrt{\frac{\pi}{2}} \Big[\Phi(c/\sqrt{2})+1 \Big]  \\
\label{I_3_5}
\stackrel{(c)}{=} & \sqrt{\frac{\pi}{2}} \Big[\textnormal{erf}\big(c/\sqrt{2}\big)+1 \Big], 
\end{align}
\end{subequations}   
where $(a)$ follows from integration by substitution w.r.t. $t=-x$; and $(b)$ follows from (\ref{Int_identity_1}) and (\ref{Int_identity_2}); and $(c)$ follows from (\ref{erf_func_def}).           

We then proceed with the simplification of $I_2$ as 
\begin{subequations}
\begin{align}
\label{I_2_1}
I_2&= \int_{-c}^{\infty}xe^{-x^2/2}dx     \\
\label{I_2_2}
&=\int_{-c}^{0}xe^{-x^2/2}dx+\int_{0}^{\infty}xe^{-x^2/2}dx      \\
\label{I_2_3}
&\stackrel{(a)}{=}\int_{0}^{\infty}xe^{-x^2/2}dx - \int_{0}^{-c}xe^{-x^2/2}dx     \\
\label{I_2_4}
&\stackrel{(b)}{=} \int_{0}^{\infty}xe^{-x^2/2}dx - \int_{0}^{c}te^{-t^2/2}dt   \\ 
\label{I_2_5}
&\stackrel{(c)}{=} \Gamma(1) - (1-e^{-c^2/2})  \\
\label{I_2_5_1}
&\stackrel{(d)}{=} 0!-(1-e^{-c^2/2})   \\
\label{I_2_6} 
&\stackrel{(e)}{=}1-(1-e^{-c^2/2})=e^{-c^2/2}, 
\end{align}		
\end{subequations}	
where $(a)$ is due to the property of integration; $(b)$ follows through integration by substitution w.r.t. $t=-x$; $(c)$ follows from (\ref{Int_identity_6}) and (\ref{Int_identity_3}), respectively; $(d)$ is because of Definition \ref{Gamma_func_def_all_cases}; and the result in $(e)$ can also be obtained directly from (\ref{I_2_1}) by employing integration by substitution w.r.t. $u=e^{-x^2/2}$.        

Let us now continue to the simplification of $I_1$ as follows:    
\begin{subequations}
\begin{align}
\label{I_1_1}
I_1&=\int_{-c}^{\infty}x^2e^{-x^2/2}dx         \\
\label{I_1_2}
&=\int_{-c}^{0}x^2e^{-x^2/2}dx+\int_{0}^{\infty}x^2e^{-x^2/2}dx			\\
\label{I_1_3}
&\stackrel{(a)}{=} \int_{0}^{c} t^2e^{-t^2/2}dt+\int_{0}^{\infty}x^2e^{-x^2/2}dx	 			\\
\label{I_1_4}	
&\stackrel{(b)}{=} \sqrt{\frac{\pi}{2}}\Phi\big(c/\sqrt{2}\big)-ce^{-c^2/2}+\sqrt{2}\Gamma(3/2)   \\
\label{I_1_5}	
&\stackrel{(c)}{=} \sqrt{\frac{\pi}{2}}\textnormal{erf}\big(c/\sqrt{2}\big)-ce^{-c^2/2}+\sqrt{2}\Gamma(3/2),     			
\end{align}		
\end{subequations}
where $(a)$ follows from integration by substitution w.r.t. $x=-t$; $(b)$ follows from (\ref{Int_identity_4}) and (\ref{Int_identity_6}), respectively; and $(c)$ follows from (\ref{erf_func_def}). Meanwhile, for odd $n$, $n!!$ (see Definition \ref{Double_factorial_def}) can be equated as $n!!=\pi^{-\frac{1}{2}}2^{\frac{1}{2}n+\frac{1}{2}}\Gamma\big(\frac{1}{2}n+1\big)$ \cite[eq. (5.4.2), p. 137]{OLDC10}. Thus, for $n=1$,   
\begin{equation}
\label{gamma_fun_simplification_1}
1!!=1=\pi^{-\frac{1}{2}} \times 2 \times \Gamma\Big(\frac{1}{2}+1\Big) \Leftrightarrow  \Gamma(3/2)=\frac{\sqrt{\pi}}{2}.   
\end{equation}      	
Substituting the RHS of (\ref{gamma_fun_simplification_1}) into the RHS of (\ref{I_1_5}) then gives
\begin{equation}
\label{I_1_6}
I_1=\sqrt{\frac{\pi}{2}}\Big[\textnormal{erf}\big(c/\sqrt{2}\big)+1\Big]-ce^{-c^2/2}.
\end{equation}  	
It is to be noted that the result in (\ref{I_1_6}) can also be obtained directly from (\ref{I_1_1}) by employing integration by parts w.r.t. $u=x$ and $dv=xe^{-x^2/2}dx$.  

Finally, we continue to the simplification of $I_0$ as progressed below: 
\begin{subequations}
\begin{align}
\label{I_0_1}
I_0&=\int_{-c}^{\infty} x^3e^{-x^2/2}dx     \\
\label{I_0_2}
&= \int_{-c}^{0} x^3e^{-x^2/2}dx+\int_{0}^{\infty} x^3e^{-x^2/2}dx         \\
\label{I_0_3}
&\stackrel{(a)}{=}\int_{0}^{\infty} x^3e^{-x^2/2}dx -  \int_{0}^{-c} x^3e^{-x^2/2}dx    \\
\label{I_0_4}
&\stackrel{(b)}{=}\int_{0}^{\infty} x^3e^{-x^2/2}dx -  \int_{0}^{c} t^3e^{-t^2/2}dt         \\
\label{I_0_5}
&\stackrel{(c)}{=} 2-2\big[1-(1+c^2/2) e^{-c^2/2} \big]     \\
\label{I_0_6}
&=2(1+c^2/2) e^{-c^2/2},
\end{align}		
\end{subequations}
where $(a)$ is due to the property of integration; $(b)$ follows through integration by substitution w.r.t. $t=-x$; and $(c)$ follows from (\ref{Int_identity_6}) and (\ref{Int_identity_5}), respectively.    

To finalize the lemma, it is now time to plug the derived integrals into (\ref{I_2}). Hence, substituting (\ref{I_3_5}), (\ref{I_2_6}), (\ref{I_1_6}), and (\ref{I_0_6}) into (\ref{I_2}) gives
\begin{multline}
\label{I_3}
I=2(1+c^2/2) e^{-c^2/2} + 3c\times \sqrt{\frac{\pi}{2}}\Big[\textnormal{erf}\big(c/\sqrt{2}\big)+1\Big]  -3c\times ce^{-c^2/2}  +   3c^2\times e^{-c^2/2} \\ + c^3 \times \sqrt{\frac{\pi}{2}} \Big[\textnormal{erf}\big(c/\sqrt{2}\big)+1 \Big]. 
\end{multline}    
\begin{equation}
\label{I_4}
I=(2+c^2)e^{-c^2/2}+  c\sqrt{\frac{\pi}{2}}\Big[\textnormal{erf}\big(c/\sqrt{2}\big)+1\Big] \times  (3+c^2).
\end{equation}
The result in (\ref{I_4}) is equal to (\ref{I_simplified}). This completes the proof of Lemma \ref{Lem_I}.          \QED
\end{lemma}

To continue bounding (\ref{exp_sigma_min_4_7}) via Lemma \ref{Lem_I}, we are going to exploit the following two bounds. First, it is inferred for $x>>1$ via (\ref{erf_func_def}) that 
\begin{equation}
\label{erf_x_approx_1}
\textnormal{erf}(x)\leq \textnormal{erf}(\infty)=1.  
\end{equation}
Second, for $c=\sqrt{m}+\sqrt{2}N$ ($c$ would be very large since $m$ is very large) and hence $c>>1$, 
\begin{equation}
\label{exp_x_^2_approx_1}
e^{-c^2/2}\approx 0 \hspace{2mm} \textnormal{and} \hspace{2mm} (3+c^2)\leq 2c^2.  
\end{equation} 
Deploying (\ref{erf_x_approx_1}) and (\ref{exp_x_^2_approx_1}) in the RHS of (\ref{I_4}),    
\begin{subequations}
\begin{align}
\label{I_bound_1}
I\leq &c\sqrt{2\pi}\times 2c^2 \leq 6c^3    \\
\label{I_bound_2}
&\stackrel{(a)}{=}6\big(\sqrt{m}+\sqrt{2N^2}\big)^3,  
\end{align} 
\end{subequations}
where $(a)$ is because $c=\sqrt{m}+\sqrt{2N^2}$.   
 
Meanwhile, deploying the bound in (\ref{I_bound_2}) in the RHS of (\ref{exp_sigma_min_4_7}) gives 
\begin{equation}
\label{exp_sigma_min_4_8_1}
\mathbb{E}\big\{ \sigma_{\textnormal{min}}^4\big(\tilde{\bm{B}}\big) \big\}\leq   24\big(\sqrt{m}+\sqrt{2N^2}\big)^3.
\end{equation}
Considering the scenario $\big(\sqrt{m}+\sqrt{2N^2}\big)\geq 24$ which is the case w.r.t. $m$ chosen as very large under the considered over-parameterization per Assumption \ref{over-param_assumption}, the LHS of (\ref{exp_sigma_min_4_8_1}) can be further bounded as      
\begin{equation}
\label{exp_sigma_min_4_8}
\mathbb{E}\big\{ \sigma_{\textnormal{min}}^4\big(\tilde{\bm{B}}\big) \big\} \leq \big(\sqrt{m}+\sqrt{2N^2}\big)^4.
\end{equation}
           
To further simplify (\ref{Prob_Exp_instant_MSE_bound_15}), we first need to simplify the underneath necessary condition emanating from the Paley--Zygmund inequality: 
\begin{equation}
\label{C_sigma_cond_1}
C_{\sigma}^{-1}\sqrt{2N^2}(\sqrt{t}+ \sqrt{N}) \leq \sqrt{\mathbb{E}\{ \sigma_{\textnormal{min}}^2\big(\tilde{\bm{B}}\big) \}}.
\end{equation}
To bound the RHS of (\ref{C_sigma_cond_1}), we need to bound $\mathbb{E}\{ \sigma_{\textnormal{min}}^2\big(\tilde{\bm{B}}\big) \}$. To this end, applying Jensen's inequality (see Proposition \ref{Prop_Jensen_inequality}) w.r.t. the concavity of the square-root function,   
\begin{subequations}
\begin{align}
\label{bound_1}
\mathbb{E}\{ \sigma_{\textnormal{min}}^2\big(\tilde{\bm{B}}\big) \}&=\mathbb{E}\bigg\{ \sqrt{\sigma_{\textnormal{min}}^4\big(\tilde{\bm{B}}\big)} \bigg\}\leq \sqrt{\mathbb{E}\{ \sigma_{\textnormal{min}}^4\big(\tilde{\bm{B}}\big) \} }      \\
\label{bound_2}
&\stackrel{(a)}{\leq} \big(\sqrt{m}+\sqrt{2N^2}\big)^2,  
\end{align} 
\end{subequations}
where $(a)$ follows from (\ref{exp_sigma_min_4_8}). Employing (\ref{bound_2}) in the RHS of (\ref{C_sigma_cond_1}),   
\begin{equation}
\label{C_sigma_cond_2}
(\sqrt{t}+ \sqrt{N}) \leq \frac{C_{\sigma}}{\sqrt{2N^2}} \sqrt{\big(\sqrt{m}+\sqrt{2N^2}\big)^2}.
\end{equation}      
\begin{equation}
\label{C_sigma_cond_3}
\sqrt{t} \leq \frac{C_{\sigma}}{\sqrt{2N^2}} \big(\sqrt{m}+\sqrt{2N^2}\big)-\sqrt{N}.
\end{equation}
Squaring (\ref{C_sigma_cond_3}) leads to a condition:   
\begin{condition}
\label{cond_t_u}
\begin{equation}
\label{C_sigma_cond_4}
t \leq \overbrace{\bigg[\frac{C_{\sigma}}{\sqrt{2N^2}} \big(\sqrt{m}+\sqrt{2N^2}\big)-\sqrt{N}\bigg]^2}^{=t_u}.
\end{equation}
\end{condition}
Merging the conditions in (\ref{t_constraint_upper}) and (\ref{C_sigma_cond_4}) leads to a condition: 
\begin{condition}
\label{cond_t_l_t_u}
\begin{equation}
\label{merged_condition}
t_l< t \leq t_u,  
\end{equation}
where $t_l$ and $t_u$ are defined in (\ref{t_constraint_upper}) and (\ref{C_sigma_cond_4}), respectively.
\end{condition}
  
It is thus inferred via (\ref{merged_condition}) that $t_u>t_l$ and hence  
\begin{subequations}
	\begin{align}
		\label{C_sigma_cond_5}
		\bigg[\frac{C_{\sigma}}{\sqrt{2N^2}} \big(\sqrt{m}+\sqrt{2N^2}\big)-\sqrt{N}\bigg]^2 & >\big[(\sqrt{m}+\sqrt{2N^2})^2\big]^2   \\
		\label{C_sigma_cond_6}
		\bigg[\frac{C_{\sigma}}{\sqrt{2N^2}} \big(\sqrt{m}+\sqrt{2N^2}\big)-\sqrt{N}\bigg] & >\big[(\sqrt{m}+\sqrt{2N^2})^2\big].
	\end{align}
\end{subequations}
Meanwhile, it directly follows through (\ref{C_sigma_cond_6}) that   
\begin{equation}
\label{C_sigma_cond_7}
\frac{C_{\sigma}}{\sqrt{2N^2}}>(\sqrt{m}+\sqrt{2N^2})+\frac{\sqrt{N}}{\sqrt{m}+\sqrt{2N^2}}.
\end{equation}
As a result, (\ref{C_sigma_cond_7}) simplifies to an inequality 
\begin{equation}
\label{C_sigma_cond_8}
C_{\sigma}>\overbrace{\sqrt{2}N\Big[(\sqrt{m}+\sqrt{2}N)+\frac{\sqrt{N}}{\sqrt{m}+\sqrt{2}N}\Big]}^{=C_{\sigma}(m, N)}.
\end{equation}
Consolidating the conditions expressed in (\ref{Assumption_C_sigma_1}) and (\ref{C_sigma_cond_8}), the overall condition on $C_{\sigma}$ becomes:   
\begin{condition}
\label{cond_C_sigma}
\begin{equation}
\label{Merged_condition_C_sigma}
\prod_{k=0}^{K-1} \sigma_{\textnormal{min}}\big(\check{\bm{\Sigma}}_{i,K-k}\check{\bm{W}}_{K-k}\big) \sigma_{\textnormal{min}}\big(\check{\bm{\Sigma}}_{i,0}\bm{A}\big)\geq C_{\sigma} > C_{\sigma}(m, N), 
\end{equation}		
\end{condition} 
where $C_{\sigma}(m, N)$ is equated in the RHS of (\ref{C_sigma_cond_8}). Therefore, when the conditions stated in (\ref{t_constraint_upper}) and (\ref{Merged_condition_C_sigma}) are satisfied, substituting the bounds in (\ref{expectation_bm_B_x_i_k_square_norm_1}) and (\ref{exp_sigma_min_4_8}) into the RHS of (\ref{Prob_Exp_instant_MSE_bound_15}) gives        
\begin{equation}
\label{Prob_Exp_instant_MSE_bound_16}
\mathbb{P}(\mathcal{E}_i)\geq \xi_i\frac{\Big[(\sqrt{m}-\sqrt{2N^2})^2-\big(C_{\sigma}^{-1}\sqrt{2N^2}(\sqrt{t}+ \sqrt{N})\big)^2\Big]^2}{\big[\big(\sqrt{m}+\sqrt{2N^2}\big)^2\big]^2}. 
\end{equation}
Factoring out the RHS of (\ref{Prob_Exp_instant_MSE_bound_16}) gives      
\begin{equation}
\label{Prob_Exp_instant_MSE_bound_17}
\mathbb{P}(\mathcal{E}_i)\geq \xi_i \big( f_1^2- f_2^2  \big)^2, 
\end{equation}
where $f_1=\frac{\sqrt{m}-\sqrt{2}N}{\sqrt{m}+\sqrt{2}N}$ and $f_2=\frac{\sqrt{2}NC_{\sigma}^{-1}(\sqrt{t}+ \sqrt{N})}{\sqrt{m}+\sqrt{2}N}$. 
\begin{remark}
\label{Rem_prob_lower_bound}
As $\sqrt{m}-\sqrt{2}N<\sqrt{m}+\sqrt{2}N$, $f_1<1$. From (\ref{C_sigma_cond_3}), $\sqrt{2}NC_{\sigma}^{-1}(\sqrt{t}+ \sqrt{N})\leq\sqrt{m}+\sqrt{2}N$ and hence $f_2\leq 1$. Since $\xi_i$ denotes probability as in (\ref{chi_expression}), $0\leq \xi_i=\mathbb{P}\big( \|\check{\bm{y}}_i^{*} \|\leq\sqrt{N}\big)\leq 1$. Therefore, it follows immediately from (\ref{Prob_Exp_instant_MSE_bound_17}) that $0\leq \mathbb{P}(\mathcal{E}_i) \leq 1$ which is in accordance with the axiomatic definition of probability.        
\end{remark}

Meanwhile, replacing (\ref{Prob_Exp_instant_MSE_bound_17}) into the RHS of (\ref{Instant_MSE_bound_4}) results in      
\begin{equation}
\label{Prob_Exp_instant_MSE_bound_18}
\mathbb{P}\bigg( F(\check{\bm{W}}) \geq \frac{t}{2N^2} \bigg) \geq \big( f_1^2- f_2^2 \big)^{2\check{n}} \prod_{i=\breve{n}+1}^{\breve{n}+\check{n}} \xi_i.
\end{equation}
To proceed to the final result that comprises Theorem \ref{Thm_probabilistic_MSE_lower_bound}, we prove the following lemma.         

\begin{lemma}
\label{Lem_dep_Gaussian_norm}
For $\xi_i=\mathbb{P}\big( \|\check{\bm{y}}_i^{*} \|\leq\sqrt{N}\big)$ as in (\ref{chi_expression}), 
\begin{equation}
\label{Prob_xi}
\xi_i\geq 2^{-2L(Q+1)}\bigg[1+ \textnormal{erf}\bigg( \frac{1}{\sqrt{2(Q+1)}} \bigg)\bigg].
\end{equation}

\proof Emanating from the CE-BEM expressed via eq. (\ref{BEM_equation}), we first underline the fact that all the elements of $\check{\bm{y}}_i^{*}$ -- which are the channel labels as in (\ref{test_label_vector_1}) -- are not independent Gaussian distributed RVs with a unit variance. As a result, the existing concentration inequality derived for a Gaussian norm of independent standard normal RVs \cite[p. 126-127]{Boucheron_Con_eqs_13} cannot be applied. To overcome this challenge, we start with the developments of Lemma \ref{On_channel_energy} and then proceed to drive a concentration bound that would bound the LHS of (\ref{Prob_xi}) from below.   

Employing (\ref{norm_check_y_i_2_7}) and (\ref{BEM_equation_simpli_2}),   
\begin{subequations}
\begin{align}
\label{norm_check_y_i_sq_1}
\|\check{\bm{y}}_i^{*}\|^2&=\sum_{n=0}^{N-1}\sum_{l=0}^{L-1} h^H[i\tilde{N}+n; l] h[i\tilde{N}+n; l]    \\
\label{norm_check_y_i_sq_2}
&=\sum_{n=0}^{N-1}\sum_{l=0}^{L-1} \bm{v}_{i\tilde{N}+n}^H\bm{h}_{i\tilde{N}+n,l}^H \bm{h}_{i\tilde{N}+n,l}\bm{v}_{i\tilde{N}+n}, 
\end{align}	
\end{subequations}
where $\bm{v}_{i\tilde{N}+n}$ and $\bm{h}_{i\tilde{N}+n,l}$ are given by (\ref{BEM_coefficient_relation_3}) and (\ref{BEM_coefficient_relation_2_1}), respectively. From (\ref{BEM_coefficient_relation_2_1}), it is evident that $\bm{h}_{i\tilde{N}+n,l}$ does not depend on $\tilde{N}$ and $n$. As a result, we can equate that   
\begin{equation}
\label{BEM_coefficient_relation_let_1}
\bm{h}_{i,l}=\bm{h}_{i\tilde{N}+n,l}=\big[ h_0[i; l], \ldots, h_Q[i; l] \big],  
\end{equation}       
where $h_q[i; l] \sim \mathcal{CN}(0, \sigma_{q,l}^2)$. Using (\ref{BEM_coefficient_relation_let_1}), (\ref{norm_check_y_i_sq_2}) can be expressed as    
\begin{equation}
\label{norm_check_y_i_sq_3}
\|\check{\bm{y}}_i^{*}\|^2=\sum_{n=0}^{N-1}\sum_{l=0}^{L-1} \bm{v}_{i\tilde{N}+n}^H\bm{h}_{i,l}^H \bm{h}_{i,l}\bm{v}_{i\tilde{N}+n}.
\end{equation}  
 
If we now consider a matrix $\check{\bm{Y}}_i^{*}\in\mathbb{C}^{N\times L}$ defined as  
\begin{equation}
\label{bm_Y_check_def}
\check{\bm{Y}}_i^{*}=
\begin{bmatrix}
\bm{h}_{i,0}\bm{v}_{i\tilde{N}} & \bm{h}_{i,1}\bm{v}_{i\tilde{N}} & \ldots & \bm{h}_{i,L-1}\bm{v}_{i\tilde{N}}\\
\bm{h}_{i,0}\bm{v}_{i\tilde{N}+1} & \bm{h}_{i,1}\bm{v}_{i\tilde{N}+1} & \ldots & \bm{h}_{i,L-1}\bm{v}_{i\tilde{N}+1}\\
\vdots & \vdots & \vdots & \vdots \\
\bm{h}_{i,0}\bm{v}_{i\tilde{N}+N-1} & \bm{h}_{i,1}\bm{v}_{i\tilde{N}+N-1} & \ldots & \bm{h}_{i,L-1}\bm{v}_{i\tilde{N}+N-1} 
\end{bmatrix}
\end{equation}       
then 
\begin{equation}
\label{bm_Y_check_Frob}
\|\check{\bm{y}}_i^{*}\|^2=\|\check{\bm{Y}}_i^{*}\|_F^2.
\end{equation}    
Deploying (\ref{bm_Y_check_Frob}) and (\ref{chi_expression}), the following probability relations follow: 
\begin{subequations}
\begin{align}  
\label{prob_xi_1}
\xi_i=\mathbb{P}\big( \|\check{\bm{y}}_i^{*} \|\leq\sqrt{N}\big)&=\mathbb{P}\big( \|\check{\bm{y}}_i^{*} \|^2\leq N\big)     \\
\label{prob_xi_2}
&\stackrel{(a)}{=}\mathbb{P}\big( \|\check{\bm{Y}}_i^{*}\|_F^2\leq N\big),
\end{align}	
\end{subequations}
where $(a)$ is due to (\ref{bm_Y_check_Frob}). Using (\ref{bm_Y_check_def}) w.r.t. all $n\in\{0, \ldots, N-1\}$ and all $l\in\{0, \ldots, L-1\}$,    
\begin{subequations}
\begin{align}   
\label{bm_Y_check_elem_sq_1}  
\big|\big(\check{\bm{Y}}_i^{*}\big)_{n,l}\big|^2&=\big|\bm{h}_{i,l}\bm{v}_{i\tilde{N}+n}\big|^2=\bm{v}_{i\tilde{N}+n}^H\bm{h}_{i,l}^H \bm{h}_{i,l}\bm{v}_{i\tilde{N}+n}  \\  
\label{bm_Y_check_elem_sq_2}
&\stackrel{(a)}{=}\textnormal{tr}\big(\bm{v}_{i\tilde{N}+n}^H\bm{h}_{i,l}^H    \bm{h}_{i,l}\bm{v}_{i\tilde{N}+n}\big) \\
\label{bm_Y_check_elem_sq_3}
&\stackrel{(b)}{=} \textnormal{tr}\big(\bm{v}_{i\tilde{N}+n}\bm{v}_{i\tilde{N}+n}^H\bm{h}_{i,l}^H \bm{h}_{i,l}\big)    \\
\label{bm_Y_check_elem_sq_4}
&\stackrel{(b)}{\leq} \textnormal{tr}\big(\bm{v}_{i\tilde{N}+n}\bm{v}_{i\tilde{N}+n}^H\big)    \textnormal{tr}\big(\bm{h}_{i,l}^H \bm{h}_{i,l}\big)
\end{align}	
\end{subequations} 
where $(a)$ follows from the fact that for a $1\times 1$ matrix $\bm{A}=a$, $\textnormal{tr}\big(\bm{A})=a$; $(b)$ follows from the identity that $\textnormal{tr}(\bm{AB})=\textnormal{tr}(\bm{BA})$ \cite{Magnus_07}; and $(c)$ follows from the identity $\textnormal{tr}(\bm{AB})\leq\textnormal{tr}(\bm{A})\textnormal{tr}(\bm{B})$ \cite[eq. (2.3)]{UZT10} since both $\bm{A}=\bm{v}_{i\tilde{N}+n}\bm{v}_{i\tilde{N}+n}^H$ and $\bm{B}=\bm{h}_{i,l}^H \bm{h}_{i,l}$ are Hermitian and positive semidefinite matrices (see Definition \ref{Pos_semi_definite_def}).       

From (\ref{BEM_coefficient_relation_3}), $(\bm{v}_{i\tilde{N}+n})_{j}^H(\bm{v}_{i\tilde{N}+n})_{j}=1$. As a result, 
\begin{equation}
\label{trace_constraint_1}
\textnormal{tr}\big(\bm{v}_{i\tilde{N}+n}\bm{v}_{i\tilde{N}+n}^H\big)=Q+1.
\end{equation} 
Employing (\ref{BEM_coefficient_relation_let_1}),       
\begin{equation}
\label{trace_constraint_2}
\textnormal{tr}\big(\bm{h}_{i,l}^H \bm{h}_{i,l}\big)=\sum_{q=0}^Q \big| h_q[i; l] \big|^2.  
\end{equation}
Substituting (\ref{trace_constraint_1}) and (\ref{trace_constraint_2}) into the RHS of (\ref{bm_Y_check_elem_sq_4}), 
\begin{equation}
\label{bm_Y_check_elem_sq_5}
\big|\big(\check{\bm{Y}}_i^{*}\big)_{n,l}\big|^2 \leq (Q+1) \sum_{q=0}^Q \big| h_q[i; l] \big|^2. 
\end{equation}    
Employing the definition of Frobenius norm (see Definition \ref{Frob_and_spec_norma}) w.r.t. (\ref{bm_Y_check_def}),      
\begin{subequations}
\begin{align}
\label{bm_Y_check_Frob_sq_1}
\|\check{\bm{Y}}_i^{*}\|_F^2&=\sum_{n=0}^{N-1}\sum_{l=0}^{L-1} \big|\big(\check{\bm{Y}}_i^{*}\big)_{n,l}\big|^2   \\
\label{bm_Y_check_Frob_sq_2}
&\stackrel{(a)}{\leq} N(Q+1) \sum_{l=0}^{L-1} \sum_{q=0}^Q \big| h_q[i; l] \big|^2  \\
\label{bm_Y_check_Frob_sq_3}
&\stackrel{(b)}{=} N(Q+1) \sum_{l=0}^{L-1} \sum_{q=0}^Q \big( \big[X_{q,l}^{(1)}\big]^2+\big[X_{q,l}^{(2)}\big]^2 \big), 
\end{align}
\end{subequations}
where $(a)$ follows from (\ref{bm_Y_check_elem_sq_5}); $(b)$ follows from the fact that $h_q[i; l]=\textnormal{Re}\{h_q[i; l]\}+\jmath \textnormal{Im}\{h_q[i; l]\}$ such that $X_{q,l}^{(1)}=\textnormal{Re}\{h_q[i; l]\} \sim \mathcal{N} (0, \sigma_{q,l}^2/2)$ and $X_{q,l}^{(2)}=\textnormal{Im}\{h_q[i; l]\}\sim \mathcal{N} (0, \sigma_{q,l}^2/2)$.     

Substituting the bound in (\ref{bm_Y_check_Frob_sq_3}) into the RHS of (\ref{prob_xi_2}) leads to
\begin{subequations}
\begin{align}
\label{prob_xi_3}
\xi_i\geq&\mathbb{P}\bigg(\sum_{l=0}^{L-1} \sum_{q=0}^Q \sum_{r=1}^2 \big[X_{q,l}^{(r)}\big]^2 \leq \frac{1}{Q+1} \bigg)   \\
\label{prob_xi_4}
=&\mathbb{P}\bigg(\sqrt{\sum_{l=0}^{L-1} \sum_{q=0}^Q \sum_{r=1}^2 \big[X_{q,l}^{(r)}\big]^2} \leq \frac{1}{\sqrt{Q+1}} \bigg), 
\end{align}	
\end{subequations}
where 
\begin{equation}
\label{distrib_real_imag}
X_{q,l}^{(r)}\sim\mathcal{N} (0, \sigma_{q,l}^2/2), \hspace{2mm} \forall r\in[2],
\end{equation}
and the variances are constrained per (\ref{sigma_q_l_def})-(\ref{gamma_def}) -- also via Assumption \ref{CE_BEM_coeffcients_normalization} -- such that     
\begin{equation}
\label{var_constraint}
\sum_{l=0}^{L-1} \sum_{q=0}^Q \sigma_{q,l}^2=1.
\end{equation}
To continue, we are going to bound the square root expression in the RHS of (\ref{prob_xi_4}). To this end, we follow with the proof of the underneath lemma. 

\begin{lemma}
\label{Lem_sqrt_sum}
For $a_1, \ldots, a_n\geq 0$, 
\begin{equation}
\label{inequality_sq_root_sum_4}
\sqrt{\small\sum_{j=1}^n a_j} \leq \sum_{j=1}^n \sqrt{a_j}.
\end{equation}

\proof Let us begin with two variables $a_1, a_2\geq 0$. Using binomial expansion,   
\begin{subequations} 
\begin{align}
\label{inequality_sq_root_sum_1}
(\sqrt{a_1}+\sqrt{a_2})^2&=a_1+a_2+2\sqrt{a_1a_2}  \\
\label{inequality_sq_root_sum_2}
\sqrt{a_1}+\sqrt{a_2}&=\sqrt{a_1+a_2+2\sqrt{a_1a_2}}  \\
\label{inequality_sq_root_sum_3}
\sqrt{a_1}+\sqrt{a_2}& \stackrel{(a)}{\geq} \sqrt{a_1+a_2}, 
\end{align}
\end{subequations}
where $(a)$ follows from the fact that $a_1a_2\geq 0$ as long as $a_1, a_2\geq 0$. If we now employ (\ref{inequality_sq_root_sum_3}) w.r.t. $\tilde{a}_1=a_1+a_2\geq 0$ and $a_3\geq 0$,   
\begin{subequations} 
\begin{align}   
\label{inequality_sq_root_sum_3_1}
\sqrt{\tilde{a}_1+a_3} \leq & \sqrt{\tilde{a}_1}+\sqrt{a_3}   \\
\label{inequality_sq_root_sum_3_2}
\sqrt{a_1+a_2+a_3} \leq & \sqrt{a_1+a_2}+\sqrt{a_3}   \\
\label{inequality_sq_root_sum_3_3}
\sqrt{a_1+a_2+a_3} &\stackrel{(a)}{\leq} \sqrt{a_1}+\sqrt{a_2}+\sqrt{a_3}, 
\end{align}
\end{subequations}
where $(a)$ is due to (\ref{inequality_sq_root_sum_3}). Continuing the same procedure w.r.t. $a_3, \ldots, a_n\geq 0$, (\ref{inequality_sq_root_sum_3_3}) can be generalized for $n$ terms. This produces (\ref{inequality_sq_root_sum_4}) and concludes the proof of Lemma \ref{Lem_sqrt_sum}.    \QED  
\end{lemma}
  
Deploying the inequality in (\ref{inequality_sq_root_sum_4}) in the RHS of (\ref{prob_xi_4}),    
\begin{equation}
\label{prob_xi_5}
\xi_i\geq \mathbb{P}\bigg(\sum_{l=0}^{L-1} \sum_{q=0}^Q \sum_{r=1}^2 \big| X_{q,l}^{(r)} \big| \leq \frac{1}{\sqrt{Q+1}} \bigg),
\end{equation}   
where $\sqrt{\big[X_{q,l}^{(r)}\big]^2}=\big| X_{q,l}^{(r)} \big|$ is employed. For a RV $Z\sim \mathcal{N}(0, \sigma^2)$, $\mathbb{P}(Z\geq 0)=\mathbb{P}(Z\leq 0)=1/2$ regardless of the value of $\sigma^2$. Thus, with regard to (\ref{distrib_real_imag}), $\big| X_{q,l}^{(r)} \big|=X_{q,l}^{(r)}$ or $\big| X_{q,l}^{(r)} \big|=-X_{q,l}^{(r)}$ with probabilities 1/2 each. In this respect, if we let $\theta_{q,l}^{(r)}$ be a RV exhibiting a symmetric Bernoulli distribution (see Definition \ref{Symm_Bernoulli_distribution}), it follows that 
\begin{equation}
\label{Abs_RV_X_q,l,r}
\big| X_{q,l}^{(r)} \big|=\theta_{q,l}^{(r)}X_{q,l}^{(r)}, \hspace{3mm} \theta_{q,l}^{(r)} \in \{-1, 1\}. 
\end{equation}   
Substituting (\ref{Abs_RV_X_q,l,r}) into the RHS of (\ref{prob_xi_5}) then leads to 
\begin{equation}
\label{prob_xi_6}
\xi_i\geq \mathbb{P}\bigg(\sum_{l=0}^{L-1} \sum_{q=0}^Q \sum_{r=1}^2  \theta_{q,l}^{(r)}X_{q,l}^{(r)} \leq \frac{1}{\sqrt{Q+1}} \Big| \theta_{q,l}^{(r)} \in \{-1, 1\}   \bigg).
\end{equation} 
Regarding the RHS of (\ref{prob_xi_6}), notice that there are $2L(Q+1)$ Gaussian distributed RVs and symmetric Bernoulli distributed (or Rademacher distributed) RVs that form a sum of $2L(Q+1)$ products.       

If we not consider a random Rademacher product given by $\prod_{l'=0}^{L-1} \prod_{q'=0}^{Q} \prod_{r'=1}^{2} \theta_{q',l'}^{(r')}$, it will have a value of +1 or -1. As a result, total probability theorem (see Theorem \ref{Total_Prob_Thm}) w.r.t. the disjoint events $\prod_{l'=0}^{L-1} \prod_{q'=0}^{Q} \prod_{r'=1}^{2} \theta_{q',l'}^{(r')}=1$ or $\prod_{l'=0}^{L-1} \prod_{q'=0}^{Q} \prod_{r'=1}^{2} \theta_{q',l'}^{(r')}=-1$ is applied to the RHS of (\ref{prob_xi_6}) to give (\ref{prob_xi_7}), as shown at the top of this page.           
\begin{figure*}[!t]
\begin{equation}
\label{prob_xi_7}
\xi_i\geq \sum_{s\in \{-1, 1\}} \mathbb{P}\Bigg(\sum_{l=0}^{L-1} \sum_{q=0}^Q \sum_{r=1}^2  \theta_{q,l}^{(r)}X_{q,l}^{(r)} \leq \frac{1}{\sqrt{Q+1}} \Bigg| \prod_{l'=0}^{L-1} \prod_{q'=0}^{Q} \prod_{r'=1}^{2} \theta_{q,l}^{(r)}=s   \bigg)\mathbb{P}\bigg(\prod_{l'=0}^{L-1} \prod_{q'=0}^{Q} \prod_{r'=1}^{2} \theta_{q',l'}^{(r')}=s \bigg).
\end{equation}
\hrulefill
\end{figure*}

As the respective probability w.r.t. $s=-1$ in the summand of the RHS of (\ref{prob_xi_7}) is greater than or equal to zero, the RHS of (\ref{prob_xi_7}) is bounded from below by the RHS of (\ref{prob_xi_8}), also asserting the lower bound for $\xi_i$.      

\begin{figure*}[!t]
	\begin{equation}
	\label{prob_xi_8}
	\xi_i\geq \mathbb{P}\Bigg(\sum_{l=0}^{L-1} \sum_{q=0}^Q \sum_{r=1}^2  \theta_{q,l}^{(r)}X_{q,l}^{(r)} \leq \frac{1}{\sqrt{Q+1}} \Bigg| \prod_{l'=0}^{L-1} \prod_{q'=0}^{Q} \prod_{r'=1}^{2} \theta_{q',l'}^{(r')}=1   \bigg)\mathbb{P}\bigg(\prod_{l'=0}^{L-1} \prod_{q'=0}^{Q} \prod_{r'=1}^{2} \theta_{q',l'}^{(r')}=1 \bigg).
	\end{equation}
	\hrulefill
\end{figure*}
Resuming our analysis from (\ref{prob_xi_8}), let us define an event $\bar{\mathcal{E}}$ as   
\begin{equation}
\label{Event_E_bar}
\bar{\mathcal{E}} \eqdef \bigg(\prod_{l'=0}^{L-1} \prod_{q'=0}^{Q} \prod_{r'=1}^{2} \theta_{q',l'}^{(r')}=1\bigg) 
\end{equation} 
Employing (\ref{Event_E_bar}) in the RHS of (\ref{prob_xi_8}), 
\begin{equation}
\label{prob_xi_9}
\xi_i\geq \mathbb{P}\Big(\sum_{l=0}^{L-1} \sum_{q=0}^Q \sum_{r=1}^2  \theta_{q,l}^{(r)}X_{q,l}^{(r)} \leq \frac{1}{\sqrt{Q+1}} \Big| \bar{\mathcal{E}}  \Big)\mathbb{P}\big(\bar{\mathcal{E}} \big).
\end{equation}  
Moreover, because $2L(Q+1)$ will be an even number for $L, Q\in\mathbb{Z}^{+}$ in line with the presumption of the considered CE-BEM, $\prod_{l=0}^{L-1} \prod_{q=0}^{Q} \prod_{r=1}^{2} \theta_{q,l}^{(r)}=1$ if $\theta_{0,0}^{(1)}=\theta_{0,0}^{(2)}=\ldots=\theta_{Q,L-1}^{(1)}=\theta_{Q,L-1}^{(2)}=1$; $\theta_{0,0}^{(1)}=\theta_{0,0}^{(2)}=\ldots=\theta_{Q,L-1}^{(1)}=\theta_{Q,L-1}^{(2)}=-1$; or all other combinations of $\theta_{q',l'}^{(r')}$s whose product is equal to one. Thus, the event in (\ref{Event_E_bar}) boils down to 
\begin{equation}
\label{Event_E_bar_simp_1}
\bar{\mathcal{E}} =\bar{\mathcal{E}}_{-1} \cup \bar{\mathcal{E}}_{1} \cup \bigcup_{v\in\{-1,1\}} \bar{\mathcal{E}}_{v}^c.
\end{equation}
where for $v\in \{-1,1\}$ 
\begin{subequations}
	\begin{align}
		\label{Event_E_bar_-1_1}
		\bar{\mathcal{E}}_{v} & \eqdef \Big( \theta_{0,0}^{(1)}=\theta_{0,0}^{(2)}=\ldots=\theta_{Q,L-1}^{(1)}=\theta_{Q,L-1}^{(2)}=v \Big)   \\
		\label{Event_E_bar_complement_-1_1}
		\bar{\mathcal{E}}_{v}^c & \eqdef \Big( \Big\{ \theta_{0,0}^{(1)}, \theta_{0,0}^{(2)}, \ldots, \theta_{Q,L-1}^{(1)}, \theta_{Q,L-1}^{(2)} \Big\} \neq \Big\{ v, v, \ldots, v, v \Big\}   \Big| \prod_{l'=0}^{L-1} \prod_{q'=0}^{Q} \prod_{r'=1}^{2} \theta_{q',l'}^{(r')}=1 \Big).
	\end{align}
\end{subequations}

Deploying (\ref{Event_E_bar_simp_1}) in the RHS of (\ref{prob_xi_9}) and bounding from below,     
\begin{equation}
\label{prob_xi_10}
\xi_i\geq \sum_{v\in\{-1,1\}}\mathbb{P}\bigg(\sum_{l=0}^{L-1} \sum_{q=0}^Q \sum_{r=1}^2  \theta_{q,l}^{(r)}X_{q,l}^{(r)} \leq \frac{1}{\sqrt{Q+1}} \bigg| \bar{\mathcal{E}}_{v}  
 \bigg)\mathbb{P}\big(\bar{\mathcal{E}}_v \big). 
\end{equation}  
Since $\theta_{0,0}^{(1)}, \theta_{0,0}^{(2)}, \ldots, \theta_{Q,L-1}^{(2)}$ are independent symmetric Bernoulli distributed RVs that take either 1 or -1 with probability 1/2 (see Definition \ref{Symm_Bernoulli_distribution}), it is inferred through (\ref{Event_E_bar_-1_1}) that    
\begin{subequations} 
\begin{align}
\label{Pob_Event_E_bar_simp_2}
\mathbb{P}\big(\bar{\mathcal{E}}_{-1}\big)&=\prod_{l'=0}^{L-1} \prod_{q'=0}^{Q} \prod_{r'=1}^{2} \mathbb{P}\big(\theta_{q',l'}^{(r')}=-1\big)=\bigg(\frac{1}{2}\bigg)^{2L(Q+1)}   \\
\label{Pob_Event_E_bar_simp_3}
\mathbb{P}\big(\bar{\mathcal{E}}_{1}\big)&=\prod_{l'=0}^{L-1} \prod_{q'=0}^{Q} \prod_{r'=1}^{2} \mathbb{P}\big(\theta_{q',l'}^{(r')}=1\big)=\bigg(\frac{1}{2}\bigg)^{2L(Q+1)}.
\end{align}
\end{subequations} 
Deploying (\ref{Event_E_bar_-1_1}), (\ref{Pob_Event_E_bar_simp_2}), and (\ref{Pob_Event_E_bar_simp_3}) in the RHS of (\ref{prob_xi_10}) gives       
\begin{equation}
\label{prob_xi_11}
\xi_i\geq 2^{-2L(Q+1)} \big(p_1  + p_1\big),
\end{equation}
where     
\begin{subequations}
\begin{align}
\label{p1_def}
p_1&= \mathbb{P}\Big(Y \leq \frac{1}{\sqrt{Q+1}} \Big)   \\ 
\label{p2_def}
p_2 & = \mathbb{P}\Big(-Y \leq \frac{1}{\sqrt{Q+1}}\Big)=\mathbb{P}\Big(Y \geq -\frac{1}{\sqrt{Q+1}}\Big)   \\ 
\label{Y_RV_def}
Y& = \sum_{l=0}^{L-1} \sum_{q=0}^Q \sum_{r=1}^2 X_{q,l}^{(r)}.
\end{align}
\end{subequations}
Referring to (\ref{distrib_real_imag}), a RV $Y$ per (\ref{Y_RV_def}) is the sum of $2L(Q+1)$ Gaussian distributed RVs and hence a Gaussian RV. As a result, $Y\sim \mathcal{N}(0, \sigma_Y^2)$ and       
\begin{subequations}
\begin{align}
\label{Y_RV_var_calc_1}
\sigma_Y^2=\sum_{l=0}^{L-1} \sum_{q=0}^Q \sum_{r=1}^2 \mathbb{E}\big\{  \big[X_{q,l}^{(r)}\big]^2 \big\}&\stackrel{(a)}{=}\sum_{l=0}^{L-1} \sum_{q=0}^Q \sum_{r=1}^2 \sigma_{q,l}^2/2    \\
\label{Y_RV_var_calc_2}
&=\sum_{l=0}^{L-1} \sum_{q=0}^Q \sigma_{q,l}^2 \stackrel{(b)}{=} 1, 
\end{align}
\end{subequations}
where $(a)$ follows from (\ref{distrib_real_imag}) and $(b)$ follows from (\ref{var_constraint}). Accordingly, $Y\sim \mathcal{N}(0, 1)$ and hence $Y$ is a standard normal RV.         

With regard to a standard normal RV $Y\sim \mathcal{N}(0, 1)$, the probabilities in the RHSs of (\ref{p1_def}) and (\ref{p2_def}) can be obtained by integrating the standard Gaussian's probability density function (PDF) $f_{Y}(y)$ given by \cite{vershynin_2018}          
\begin{equation}
\label{Standa_norm_PDF}
f_{Y}(y)=\frac{1}{\sqrt{2\pi}} e^{-y^2/2}.
\end{equation}
Therefore, using (\ref{Standa_norm_PDF}) and (\ref{p1_def}), 
\begin{subequations}
\begin{align}
\label{p_1_calc_1}
p_1&= \frac{1}{\sqrt{2\pi}} \int_{-\infty}^{1/\sqrt{Q+1}}e^{-y^2/2} dy         \\
\label{p_1_calc_2}
&=\frac{1}{\sqrt{2\pi}} \bigg[ \int_{-\infty}^{0}e^{-y^2/2} dy+ \int_{0}^{1/\sqrt{Q+1}}e^{-y^2/2} dy    \bigg]       \\
\label{p_1_calc_3}
&\stackrel{(a)}{=} \frac{1}{\sqrt{2\pi}} \bigg[ \int_{0}^{\infty}e^{-t^2/2} dt+ \int_{0}^{1/\sqrt{Q+1}}e^{-y^2/2} dy    \bigg]     \\
\label{p_1_calc_4}
&\stackrel{(b)}{=} \frac{1}{\sqrt{2\pi}} \bigg[\sqrt{\frac{\pi}{2}}+\sqrt{\frac{\pi}{2}} \Phi\bigg(\frac{1}{\sqrt{2(Q+1)}} \bigg)   \bigg], \\
\label{p_1_calc_5}
&\stackrel{(c)}{=} \frac{1}{2} \bigg[1+ \textnormal{erf}\bigg(\frac{1}{\sqrt{2(Q+1)}} \bigg)\bigg].
\end{align}
\end{subequations}  
where $(a)$ follows through integration by substitution w.r.t. $t=-y$; $(b)$ follows from (\ref{Int_identity_2}) and (\ref{Int_identity_1}), respectively; and $(c)$ follows from (\ref{Error_func_def}).      

Similarly, deploying (\ref{Standa_norm_PDF}) and (\ref{p2_def}) would lead to:    
\begin{subequations}
\begin{align}
\label{p_2_calc_1}
p_2&=\frac{1}{\sqrt{2\pi}} \int_{-\frac{1}{\sqrt{Q+1}}}^{\infty}e^{-y^2/2} dy     \\
\label{p_2_calc_2}
&= \frac{1}{\sqrt{2\pi}} \bigg[\int_{-\frac{1}{\sqrt{Q+1}}}^{0}e^{-y^2/2} dy+\int_{0}^{\infty}e^{-y^2/2} dy  \bigg]  \\
\label{p_2_calc_3}
&\stackrel{(a)}{=} \frac{1}{\sqrt{2\pi}} \bigg[\int_{0}^{\frac{1}{\sqrt{Q+1}}}e^{-t^2/2} dt+\int_{0}^{\infty}e^{-y^2/2} dy  \bigg]  \\
\label{p_2_calc_4}
&\stackrel{(b)}{=} \frac{1}{2} \bigg[1+ \textnormal{erf}\bigg(\frac{1}{\sqrt{2(Q+1)}} \bigg)\bigg].    
\end{align}
\end{subequations}
where $(a)$ follows through integration by substitution w.r.t. $t=-y$; $(b)$ follows from the simplification of (\ref{p_1_calc_3}) into (\ref{p_1_calc_5}), as the integrals in (\ref{p_1_calc_3}) and (\ref{p_2_calc_3}) are identical.     

Moreover, substituting (\ref{p_1_calc_5}) and (\ref{p_2_calc_4}) into the RHS of (\ref{prob_xi_11}) results in   
\begin{equation}
\label{prob_xi_12}
\xi_i\geq 2^{-2L(Q+1)}\bigg[1+ \textnormal{erf}\bigg( \frac{1}{\sqrt{2(Q+1)}} \bigg) \bigg],
\end{equation}   
This completes the proof of Lemma \ref{Lem_dep_Gaussian_norm}.       \QED
\end{lemma}

At last, substituting (\ref{prob_xi_12}) into the RHS of (\ref{Prob_Exp_instant_MSE_bound_18}) gives    
\begin{equation}
\label{Prob_Exp_instant_MSE_bound_19}
\mathbb{P}\bigg( F(\check{\bm{W}}) \geq \frac{t}{2N^2} \bigg) \geq \Bigg(\frac{\big( f_1^2- f_2^2 \big)^{2}}{2^{2L(Q+1)}} \bigg[1+ \textnormal{erf}\bigg( \frac{1}{\sqrt{2(Q+1)}} \bigg) \bigg]\Bigg)^{\check{n}}. 
\end{equation}
The expression in (\ref{Prob_Exp_instant_MSE_bound_19}) is exactly the one in (\ref{Prob_Exp_instant_MSE_bound}). This completes the proof of Theorem \ref{Thm_probabilistic_MSE_lower_bound}.   \QEDclosed

\section{Proof of Corollary \ref{Coro_min_lower_probab_bound}}
\label{proof_Coro_min_lower_probab_bound}
As expressed in Theorem \ref{Thm_probabilistic_MSE_lower_bound}, $f_1$ and $\big[1+ \textnormal{erf}\big( \big(\sqrt{2(Q+1)}\big)^{-1} \big) \big]$ are always constants. Hence, the probability in the LHS of (\ref{Prob_Exp_instant_MSE_bound}) depends on $f_2$ which, in turn, depends on $C_{\sigma}$. This parameter -- as in (\ref{Assumption_C_sigma_1}) -- which is the lower bound of the product of the singular values of random matrices will control the lower bound of the probability on the LHS of (\ref{Prob_Exp_instant_MSE_bound}). Thus, 
\begin{subequations}
	\begin{align}
	\label{min_proba_bound_exp_1}
	\min \mathbb{P}\bigg( F(\check{\bm{W}}) \geq \frac{t}{2N^2} \bigg)&=\min_{f_2} \mathbb{P}\bigg( F(\check{\bm{W}}) \geq \frac{t}{2N^2} \bigg)     \\
	\label{min_proba_bound_exp_2}
	&\stackrel{(a)}{\geq} \min_{f_2} \big[ f_1^2- f_2^2  \big]^2,  
	\end{align}	
\end{subequations}
where $(a)$ follows since all multiplying constants of the RHS of (\ref{Prob_Exp_instant_MSE_bound}) will not affect the minimum value of the RHS of (\ref{Prob_Exp_instant_MSE_bound}). Since the RHS of (\ref{min_proba_bound_exp_2}) is a convex function of $f_2$, its minimum occurs at $\frac{d}{df_2} \big[ f_1^2- f_2^2  \big]^2=0$:   
\begin{equation}
\label{derivative_minimum}
2(f_1-f_2)(f_1+f_2)\times -2f_2=0 \Leftrightarrow f_1=f_2.
\end{equation} 
Thus, replacing $f_1=f_2$ into (\ref{min_proba_bound_exp_2}) delivers (\ref{min_proba_bound_exp}). This completes the proof of Corollary \ref{Coro_min_lower_probab_bound}.     \QEDclosed     

\section{Proof of Lemma \ref{Lem_Conc_Gau_norm_diff_variance}}
\label{proof_Lem_Conc_Gau_norm_diff_variance} 
For $\bm{x}=[X_1, \ldots, X_{2n}]^T\in\mathbb{R}^{2n}$ with $X_i\sim\mathcal{N}(0, \sigma_i^2)$, $\|\bm{x}\|- \mathbb{E}\big\{ \|\bm{x}\| \big\} \leq \|\bm{x}\|$. As a result,   
\begin{subequations}
\begin{align} 
\label{Lem_Conc_Gau_norm_diff_variance_2}
\mathbb{P}\big( \|\bm{x}\|- \mathbb{E}\big\{ \|\bm{x}\| \big\}  > t\big)&\leq \mathbb{P}\big( \|\bm{x}\| > t\big)    \\
\label{Lem_Conc_Gau_norm_diff_variance_3}
&=1-\mathbb{P}\big( \|\bm{x}\| \leq t\big).        
\end{align}	
\end{subequations}
If we can bound $\mathbb{P}\big( \|\bm{x}\| \leq t\big)$ from below, we can bound (\ref{Lem_Conc_Gau_norm_diff_variance_3}) and hence (\ref{Lem_Conc_Gau_norm_diff_variance_2}) from above. We are thus going to bound -- in what follows -- $\mathbb{P}\big( \|\bm{x}\| \leq t\big)$ from below.     

By definition, $\|\bm{x}\|=\sqrt{\sum_{i=1}^{2n} X_i^2}$. Employing (\ref{inequality_sq_root_sum_4}), $\|\bm{x}\|=\sqrt{\sum_{i=1}^{2n} X_i^2}\leq \sum_{i=1}^{2n} \sqrt{X_i^2}$. Consequently,   
\begin{subequations}
\begin{align}
\label{Lem_Conc_Gau_norm_diff_variance_5}
\mathbb{P}\big( \|\bm{x}\| \leq t\big)&\geq \mathbb{P}\bigg( \sum_{i=1}^{2n} \sqrt{X_i^2} \leq t\bigg)       \\
\label{Lem_Conc_Gau_norm_diff_variance_6}
&\stackrel{(a)}{=} \mathbb{P}\bigg( \sum_{i=1}^{2n} |X_i| \leq t\bigg), 
\end{align}	
\end{subequations} 
where $(a)$ is due to the identity $\sqrt{X_i^2}=|X_i|$. Meanwhile, following the arguments that led to (\ref{Abs_RV_X_q,l,r}), $|X_i|$ can also be expressed in terms of $X_i$ and a symmetric Bernoulli (or Rademacher) distributed RV (see Definition \ref{Symm_Bernoulli_distribution}) $\theta_i\in \{-1, 1\}$ as           
\begin{equation}
\label{Abs_RV_X_i}
|X_i|=\theta_i X_i, \hspace{3mm} \theta_i \in \{-1, 1\},  
\end{equation}
where 
\begin{equation}
\label{RV_X_i_dist}
X_i\sim\mathcal{N}(0, \sigma_i^2) \hspace{2mm} \textnormal{and} \hspace{2mm} \sigma^2=\sum_{i=1}^{2n} \sigma_i^2.  
\end{equation}
Substituting (\ref{Abs_RV_X_i}) into the RHS of (\ref{Lem_Conc_Gau_norm_diff_variance_6}),   
\begin{equation}
\label{Lem_Conc_Gau_norm_diff_variance_7}
\mathbb{P}\big( \|\bm{x}\| \leq t\big) \geq \mathbb{P}\bigg( \sum_{i=1}^{2n} \theta_i X_i   \leq t\big| \theta_i \in \{-1, 1\} \bigg).
\end{equation}   

If we now consider a random Rademacher product given by $\prod_{j=1}^{2n} \theta_j$, it will take a value of +1 or -1. Consequently, total probability theorem (see Theorem \ref{Total_Prob_Thm}) w.r.t. the disjoint events $\prod_{j=1}^{2n} \theta_j=1$ and $\prod_{j=1}^{2n} \theta_j=-1$ can be applied to the RHS of (\ref{Lem_Conc_Gau_norm_diff_variance_7}) to give  
\begin{equation}
\label{Lem_Conc_Gau_norm_diff_variance_8}
\mathbb{P}\big( \|\bm{x}\| \leq t\big) \geq \sum_{u\in\{-1, 1\}}\mathbb{P}\bigg( \sum_{i=1}^{2n} \theta_i X_i \leq t\bigg| \prod_{j=1}^{2n} \theta_j=u\bigg)  \mathbb{P}\bigg(\prod_{j=1}^{2n} \theta_j=u\bigg).
\end{equation}
Since the second summand of (\ref{Lem_Conc_Gau_norm_diff_variance_8}) w.r.t. $u=-1$ takes a probability between 0 and 1, the RHS of (\ref{Lem_Conc_Gau_norm_diff_variance_8}) can be bounded from below as    
\begin{equation}
\label{Lem_Conc_Gau_norm_diff_variance_9}
\mathbb{P}\big( \|\bm{x}\| \leq t\big) \geq \mathbb{P}\bigg( \sum_{i=1}^{2n} \theta_i X_i \leq t\bigg| \prod_{j=1}^{2n} \theta_j=1\bigg)\mathbb{P}\bigg(\prod_{j=1}^{2n} \theta_j=1\bigg).
\end{equation} 
For the purpose of our ensuing analysis, let us define an event $\bar{\mathcal{E}}$ as   
\begin{equation}
\label{Event_E_bar_for_x}
\bar{\mathcal{E}} \eqdef \bigg(\prod_{j=1}^{2n} \theta_j=1\bigg) 
\end{equation} 
Employing (\ref{Event_E_bar_for_x}) into (\ref{Lem_Conc_Gau_norm_diff_variance_9}), 
\begin{equation}
\label{Lem_Conc_Gau_norm_diff_variance_10}
\mathbb{P}\big( \|\bm{x}\| \leq t\big)\geq \mathbb{P}\Big(\sum_{i=1}^{2n}  \theta_i X_i  \leq t \big| \bar{\mathcal{E}}\Big)\mathbb{P}\big(\bar{\mathcal{E}} \big).
\end{equation}  

Since $2n$ will be an even number for any $n\in\mathbb{N}$, $\prod_{j=1}^{2n} \theta_j=1$ if $\theta_{1}=\theta_{2}=\ldots=\theta_{2n}=1$; $\theta_{1}=\theta_{2}=\ldots=\theta_{2n}=-1$; or all other combinations of $\theta_j$s whose product is equal to one. Thus, the event in (\ref{Event_E_bar_for_x}) can be expressed through the union of disjoint events as
\begin{equation}
	\label{Event_E_bar_for_x_simp_1}
	\bar{\mathcal{E}}=\bar{\mathcal{E}}_{-1} \cup \bar{\mathcal{E}}_{1} \cup \bigcup_{u\in\{-1,1\}} \bar{\mathcal{E}}_{u}^c, 
\end{equation}
where for $u\in\{-1,1\}$       
\begin{subequations}
	\begin{align}
		\label{Event_E_bar_for_x_-1_1}
		\bar{\mathcal{E}}_{u} & \eqdef \Big( \theta_{1}=\theta_{2}=\ldots=\theta_{2n-1}=\theta_{2n}=u \Big)   \\
		\label{Event_E_bar_complement_for_x_-1_1} 
		\bar{\mathcal{E}}_{u}^c  & \eqdef \Big( \big\{ \theta_1, \theta_2, \ldots, \theta_{2n-1}, \theta_{2n} \big\}  \neq \big\{ u, u, \ldots, u, u \big\} \Big| \prod_{j=1}^{2n} \theta_j=1 \Big).
	\end{align}
\end{subequations}
Using (\ref{Event_E_bar_for_x_simp_1}) in the RHS of (\ref{Lem_Conc_Gau_norm_diff_variance_10}) and bounding from below,   
\begin{equation}
\label{Lem_Conc_Gau_norm_diff_variance_11}
\mathbb{P}\big( \|\bm{x}\| \leq t\big)\geq \sum_{u\in\{-1,1\}}\mathbb{P}\bigg(\sum_{i=1}^{2n} \theta_i X_i   
\leq t \Big| \bar{\mathcal{E}}_{u}\bigg)\mathbb{P}\big(\bar{\mathcal{E}}_u \big). 
\end{equation} 
Because $\theta_{1}, \theta_{2}, \ldots, \theta_{2n}$ are independent symmetric Bernoulli distributed RVs that would take either 1 or -1 with probability 1/2 (see Definition \ref{Symm_Bernoulli_distribution}), it follows from (\ref{Event_E_bar_for_x_-1_1}) that   
\begin{subequations} 
	\begin{align}
	\label{Event_E_bar_for_x_simp_2}
	\mathbb{P}\big(\bar{\mathcal{E}}_{-1}\big)&=\prod_{i=0}^{2n} \mathbb{P}\big(\theta_i=-1\big)=\bigg(\frac{1}{2}\bigg)^{2n}=2^{-2n}   \\
	\label{Event_E_bar_for_x_simp_3}
	\mathbb{P}\big(\bar{\mathcal{E}}_{1}\big)&=\prod_{i=1}^{2n} \mathbb{P}\big(\theta_i=1\big)=\bigg(\frac{1}{2}\bigg)^{2n}=2^{-2n}.
	\end{align}
\end{subequations} 
Substituting (\ref{Event_E_bar_for_x_-1_1}), (\ref{Event_E_bar_for_x_simp_2}), and (\ref{Event_E_bar_for_x_simp_3}) into the RHS of (\ref{Lem_Conc_Gau_norm_diff_variance_11}) leads to          
\begin{equation}
\label{Lem_Conc_Gau_norm_diff_variance_12}
\mathbb{P}\big( \|\bm{x}\| \leq t\big)\geq 2^{-2n} \big[\mathbb{P}\big(Z   
\leq t \big)  + \mathbb{P}\big( -Z   
\leq t \big)\big],
\end{equation}
where     
\begin{equation}
\label{Z_RV_def}
Z =\sum_{i=1}^{2n} X_i.
\end{equation}
With regard to (\ref{RV_X_i_dist}), a RV $Z$ defined in (\ref{Z_RV_def}) would be the sum of zero mean Gaussian RVs. Thus, $Z\sim\mathcal{N}(0, \sigma_z^2)$ for     
\begin{equation}
\label{Var_calc_RV_Z}
\sigma_z^2=\sum_{i=1}^{2n} \mathbb{E}\big\{X_i^2\big\}=\sum_{i=1}^{2n}\sigma_i^2=\sigma^2.
\end{equation}   
Following (\ref{Var_calc_RV_Z}), $Z\sim\mathcal{N}(0, \sigma^2)$. As a result, the two probabilities in the RHS of (\ref{Lem_Conc_Gau_norm_diff_variance_12}) would be the same. Consequently,    
\begin{equation}
\label{Lem_Conc_Gau_norm_diff_variance_13}
\mathbb{P}\big( \|\bm{x}\| \leq t\big)\geq 2^{1-2n} \mathbb{P}\big(Z\leq t \big).
\end{equation}

To simplify the RHS of (\ref{Lem_Conc_Gau_norm_diff_variance_13}), we integrate w.r.t. the Gaussian PDF of $Z$ (a zero mean RV) given by \cite{Proakis_5th_ed_08}       
\begin{equation}
\label{Standa_PDF_Z}
f_{Z}(z)=\frac{1}{\sigma\sqrt{2\pi}} e^{-z^2/(2\sigma^2)}.
\end{equation}
Accordingly,   
\begin{equation}
\label{Prob_Z_leq_t_1}
\mathbb{P}\big(Z\leq t \big)=\frac{1}{\sigma\sqrt{2\pi}}\int_{-\infty}^t  e^{-z^2/(2\sigma^2)} dz. 
\end{equation}
If we now let $x=z/\sigma$, $dx=dz/\sigma$ and hence $dz=\sigma dx$. Applying these parameters into the RHS of (\ref{Prob_Z_leq_t_1}), integration by substitution leads to      
\begin{subequations}
\begin{align}
\label{Prob_Z_leq_t_2}
\mathbb{P}\big(Z\leq t \big)=&\frac{1}{\sqrt{2\pi}}\int_{-\infty}^{\frac{t}{\sigma}}  e^{-x^2/2} dx   \\
\label{Prob_Z_leq_t_3}
=&\frac{1}{\sqrt{2\pi}} \bigg[ \int_{-\infty}^{0}  e^{-x^2/2} dx + \int_{0}^{\frac{t}{\sigma}}  e^{-x^2/2} dx  \bigg]     \\
\label{Prob_Z_leq_t_4}
\stackrel{(a)}{=}&\frac{1}{\sqrt{2\pi}} \bigg[ \int_{0}^{\infty}  e^{-u^2/2} du + \int_{0}^{\frac{t}{\sigma}}  e^{-x^2/2} dx  \bigg]    \\
\label{Prob_Z_leq_t_5}  
\stackrel{(b)}{=}&\frac{1}{\sqrt{2\pi}} \bigg[ \sqrt{\frac{\pi}{2}} + \sqrt{\frac{\pi}{2}} \Phi\bigg(\frac{t}{\sigma\sqrt{2}} \bigg)  \bigg]      \\
\label{Prob_Z_leq_t_6}
\stackrel{(c)}{=}&\frac{1}{2} \bigg[ 1 +\textnormal{erf}\bigg(\frac{t}{\sigma\sqrt{2}} \bigg)  \bigg],
\end{align}
\end{subequations}  
where $(a)$ follows from integration by substitution w.r.t. $u=-x$; $(b)$ follows from (\ref{Int_identity_2}) and (\ref{Int_identity_1}), respectively; and $(c)$ follows from (\ref{erf_func_def}).    

Meanwhile, plugging (\ref{Prob_Z_leq_t_6}) into the RHS of (\ref{Lem_Conc_Gau_norm_diff_variance_13}),    
\begin{equation}
\label{Lem_Conc_Gau_norm_diff_variance_14}
\mathbb{P}\big( \|\bm{x}\| \leq t\big)\geq 2^{-2n} \bigg[ 1 +\textnormal{erf}\bigg(\frac{t}{\sigma\sqrt{2}} \bigg)  \bigg].
\end{equation}
Hence, 
\begin{equation}
\label{Lem_Conc_Gau_norm_diff_variance_15}
-\mathbb{P}\big( \|\bm{x}\| \leq t\big)\leq -2^{-2n} \bigg[ 1 +\textnormal{erf}\bigg(\frac{t}{\sigma\sqrt{2}} \bigg)  \bigg].
\end{equation}
From (\ref{Lem_Conc_Gau_norm_diff_variance_2}) and (\ref{Lem_Conc_Gau_norm_diff_variance_3}), 
\begin{equation}
\label{Lem_Conc_Gau_norm_diff_variance_16}
\mathbb{P}\big( \|\bm{x}\|- \mathbb{E}\big\{ \|\bm{x}\| \big\}  > t\big)\leq 1+\big(-\mathbb{P}\big( \|\bm{x}\| \leq t\big)\big).
\end{equation}
Exploiting the RHS of (\ref{Lem_Conc_Gau_norm_diff_variance_15}) into the RHS of (\ref{Lem_Conc_Gau_norm_diff_variance_16}),   
\begin{equation}
\label{Lem_Conc_Gau_norm_diff_variance_17}
\mathbb{P}\big( \|\bm{x}\|- \mathbb{E}\big\{ \|\bm{x}\| \big\}  > t\big)\leq 1-2^{-2n} \bigg[ 1 +\textnormal{erf}\bigg(\frac{t}{\sigma\sqrt{2}} \bigg)  \bigg].
\end{equation}
This is exactly (\ref{Lem_Conc_Gau_norm_diff_variance_1}) and hence the end of Lemma \ref{Lem_Conc_Gau_norm_diff_variance}'s proof.     \QEDclosed

\section*{Acknowledgments}
The authors acknowledge the U.S. Department of Commerce and the National Institute of Standards and Technology (NIST) for funding this work. The first author acknowledges Prof. Ali H. Sayed (EPFL, Switzerland) for a fruitful discussion. The first author also acknowledges Dr. Raied Caromi (NIST, MD, USA) for his technical support -- during the initial empirical experimentation of this work -- regarding \textsc{Keras} and \textsc{ENKI}.  

\section*{Disclaimer} 
The identification of any commercial product or trade name does not imply endorsement or recommendation by the National Institute of Standards and Technology, nor is it intended to imply that the materials or equipment identified are necessarily the best available for the purpose.


\end{document}